\documentclass [superscriptaddress,reprint,amsmath,amssymb,showkeys]{revtex4-2}
\usepackage{graphicx}
\usepackage{dcolumn}
\usepackage{bm}
\usepackage{color}
\usepackage{float}
\usepackage[utf8]{inputenc}
\usepackage[mathlines]{lineno}
\usepackage{comment}
\usepackage{amsmath}
\usepackage{xr}
\usepackage{xr-hyper}
\usepackage{hyperref}
\usepackage{soul,xcolor}
\soulregister\cite7
\soulregister\ref7
\setstcolor{blue}
\newcommand{\CPB}{CsPbBr$_3$}
\newcommand{\STO}{SrTiO$_3$}
\newcommand{\BTO}{BaTiO$_3$}

\newcommand{\nd}{2$^{nd}$}

\renewcommand{\st}{1$^{st}$} 
\newcommand{\BZ}{Brillouin zone}

\newcommand{\nit}{\noindent}
\renewcommand{\tt}[1]{\text{{#1}}} 

\renewcommand {\vec}  [1]  {\ensuremath{\boldsymbol{#1}}}
\newcommand {\nvec}  [1]  {\ensuremath{\boldsymbol{\bar{#1}}}}
\newcommand   {\avg}  [1] {\ensuremath{\left\langle#1\right\rangle}}

\newcommand{\bra}[1]{\ensuremath{ \left\langle #1 \right| }}		
\newcommand{\ket}[1]{\ensuremath{ \left| #1 \right\rangle }}
\newcommand{\braket}[2]{\ensuremath{ \left\langle #1 \middle| #2 \right\rangle }}
\newcommand{\braxket}[3]{\ensuremath{ \left\langle #1 \middle| #2 \middle| #3 \right\rangle }}

\makeatletter
\newcommand*{\addFileDependency}[1]
{
  \typeout{(#1)}
  \@addtofilelist{#1}
  \IfFileExists{#1}{}{\typeout{No file #1.}}
}
\makeatother

\begin{document}

\title{The Disorder Origin of Raman Scattering In Perovskites Single Crystals}

\author{Matan Menahem}
\thanks{These authors contributed equally to this work}
\affiliation{Department of Chemical and Biological Physics, Weizmann Institute of Science, Rehovot 76100, Israel}
\author{Nimrod Benshalom}
\thanks{These authors contributed equally to this work}
\affiliation{Department of Chemical and Biological Physics, Weizmann Institute of Science, Rehovot 76100, Israel}
\author{Maor Asher}
\affiliation{Department of Chemical and Biological Physics, Weizmann Institute of Science, Rehovot 76100, Israel}
\author{Sigalit Aharon}
\affiliation{Department of Molecular Chemistry and Material Science, Weizmann Institute of Science, Rehovot 76100, Israel}
\author{Roman Korobko}
\affiliation{Department of Chemical and Biological Physics, Weizmann Institute of Science, Rehovot 76100, Israel}
\author{Sam Safran}
\affiliation{Department of Chemical and Biological Physics, Weizmann Institute of Science, Rehovot 76100, Israel}
\author{Olle Hellman}
\affiliation{Department of Molecular Chemistry and Material Science, Weizmann Institute of Science, Rehovot 76100, Israel}
\author{Omer Yaffe}
\email{omer.yaffe@weizmann.ac.il}
\affiliation{Department of Chemical and Biological Physics, Weizmann Institute of Science, Rehovot 76100, Israel}

\date{\today}

\begin{abstract}

The anharmonic lattice dynamics of oxide and halide perovskites play a crucial role in their mechanical and optical properties.
Raman spectroscopy is one of the key methods used to study these structural dynamics. 
However, despite decades of research, existing interpretations cannot explain the temperature dependence of the observed Raman spectra. 
We demonstrate the non-monotonic evolution with temperature of the scattering intensity and present a model for \nd -order Raman scattering that accounts for this unique trend. 
By invoking a low-frequency anharmonic feature, we are able to reproduce the Raman spectral line-shapes and integrated intensity temperature dependence.
Numerical simulations support our interpretation of this low-frequency mode as a transition between two minima of a double-well potential surface.
The model can be applied to other dynamically disordered crystal phases, providing a better understanding of the structural dynamics, leading to favorable electronic, optical, and mechanical properties in functional materials.

\end{abstract}

\maketitle

\section*{Introduction}

Lattice dynamics and structural phase transitions in perovskite crystals have been extensively studied for decades.~\cite{Kamba2021,Cowley2006,Miyata2017} 
The high coordination number for cations in the perovskite structure (12) offers a range of bonding strengths that leads to strongly anharmonic ionic thermal motion.~\cite{Mitzi1999,Lanigan-Atkins2021} 
Importantly, such anharmonic motion enables desirable and tunable electro-mechanical and optoelectronic responses.~\cite{Wright2016,Munson2018} 

Raman scattering spectroscopy is a powerful tool in the study of vibrational properties of materials, and was extensively used to investigate the structural dynamics of perovskites.~\cite{Migoni1976,Fontana1991,Menahem2021,SharmaMAPI2,Cohen2022}   
However, the interpretation of their Raman spectrum and its evolution with temperature has proven very challenging. 
The anharmonic character leads to inadequacy of the Raman selection rules, as they are dictated by the average symmetry of the crystals.~\cite{Bhagavantam1931,Cohen2022,Cowley1968,Saksena1940}
This is most apparent in the cubic phase ($Pm\bar{3}m$) of perovskite crystals, which is \st -order Raman inactive and yet, typically displays a large Raman cross-section.~\cite{Fontana2020,Perry1967}   
Despite the many similarities in their crystal structure and observed spectra, existing interpretations for the Raman scattering of oxide and halide perovskites are markedly different.
Moreover, these models fail to capture or explain several important experimental observations.  

In the case of oxide perovskites, it is widely accepted that the Raman spectrum in the cubic phase is dominated by two-phonon Raman scattering,~\cite{Cowley1964b,Fontana2020,Perry1967,Fontana1990} associated with phonons at high symmetry points in the \BZ.~\cite{Born1947,Johnson1964,Burstein1965}
However, the intensity of the scattered light is unusually strong and does not scale with temperature as two-phonon scattering dictates.~\cite{Fontana1972,Huller1969,Khatib1989,Stachiotti1993} 

The Raman spectra of halide perovskites are dominated by an over-damped peak centered at zero-frequency (termed 'central peak').~\cite{Guo2017b,Gao2021}
It was suggested that the central peak is related to polar fluctuations~\cite{Fontana1990,Stachiotti1993,YaffePRL2017} between short-lived distorted structures,~\cite{BechtelPRM2019,Beecher2016,Whalley2016b,Zhu2019} giving rise to \st -order Raman scattering.
While it explains the observed Raman and is backed up by molecular dynamics simulations,~\cite{YaffePRL2017,Gao2021} it contradicts the Raman selection rules derived from the average symmetry.
Moreover, this picture is at odds with two additional experimental observation. 
The first is that the halide perovskites have a very sharp absorption edge (i.e. low Urbach energy) in the visible range, which implies a highly ordered crystal structure.~\cite{Gehrmann2019,DeWolf2014}
The second is that the 'central peak' is observed in Raman scattering but not in THz-absorption measurements.~\cite{Maeng2019,Xia2021,VorakiatJPhysChemLett2016}

\begin{figure*}[ht]
    \centering
    \includegraphics[width = 17 cm]{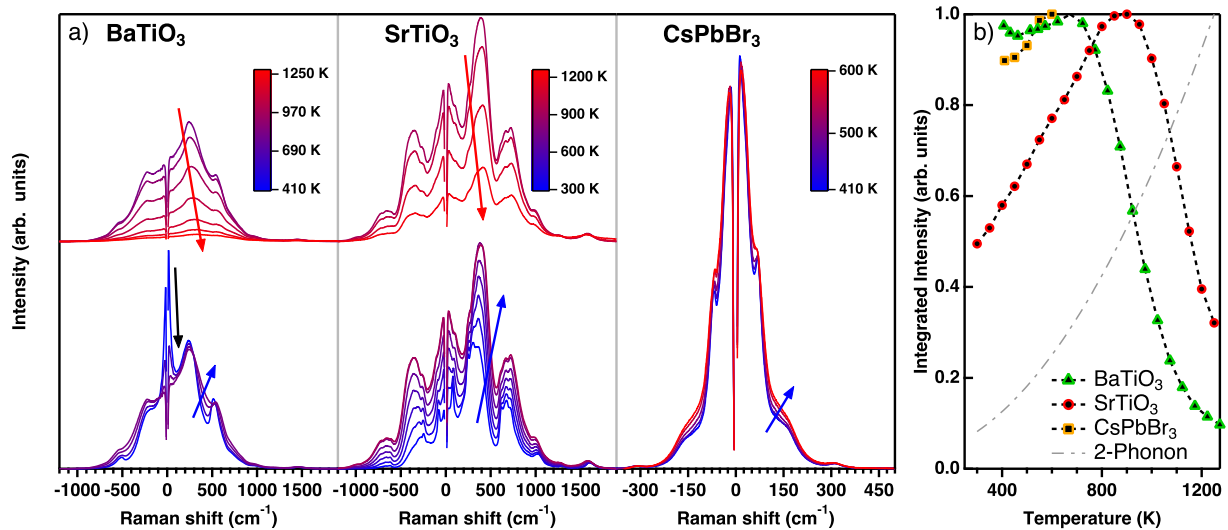}
    \caption{\textbf{Temperature-dependent Raman spectra of cubic BaTiO$_{\vec{3}}$, SrTiO$_{\vec{3}}$ and CsPbBr$_{\vec{3}}$}. a) Spectra (from left to right) of \BTO, \STO\ and \CPB. The spectra of \BTO\ and \STO\ are divided to two scattering regimes, showing the non-monotonic temperature dependence. Arrows emphasize trends with increasing temperature. b) Normalized Stokes integrated intensity of \BTO\ (green), \STO\ (red) and \CPB\ (yellow). In grey is the expected temperature dependence from conventional two phonon scattering. The deviation from the expected temperature dependence is evident.}
    \label{fig:TRaman_Data}
\end{figure*}

In this study we propose a unified description for the Raman scattering of both oxide and halide perovskites that explains its unique temperature dependence, as well as resolves the problems described above.
Our model attempts to capture the symmetry broken structures using a picture consistent with the average symmetry and its constraints.

First, we measure the temperature evolution of the Raman spectra of \BTO, \STO, and \CPB\ crystals, observing an unusual trend with temperature, in strike contrast to harmonic Raman theory.
Next, motivated by the double-well (DW) picture of an atomic potential surface in cubic halide perovskites,~\cite{Yang2020} we simulate the dynamics of a single particle in 1D potential and find it leads to excess Raman scattering intensity.
To reconcile our experimental observations with the simulated results, while maintaining the cubic symmetry, we present a \nd-order Raman scattering model for dynamically disordered crystals.
The core concept of our model is a localized low-frequency fluctuation that embodies all anharmonic dynamics, and participates in light scattering.
By allowing the potential surface to change with temperature, our model is able to explain the Raman spectra of oxide and halide perovskites.

\section*{Results and Discussions}

The Raman spectra of \BTO, \STO\ and \CPB\ at selected temperatures in the cubic phase are shown in Fig.~1(a).
We collected the Raman spectra in a back-scattering configuration, with special care taken to isolate the effect of temperature on the intensity, as described in the Methods.
The data of \BTO\ and \CPB\ begin after the tetragonal-cubic phase transition (405~K).
The data of \STO\ begin at 300~K since the phase transition is at 105~K.~\cite{He2020}
The full data set, including the spectra of \STO\ below 300~K, is depicted in Fig.~S1 and S2 in the Supplemental Material.
Figure~1(b) shows the Stokes integrated intensity as function of temperature for \BTO\ (green), \STO\ (red) and \CPB\ (yellow) along with the expected trend for 2-phonon scattering (gray).
In particular, we observe a non-monotonic trend with temperature, with a clear maximum for \STO\ and \BTO\ and subsequent decrease in scattering intensity.
We therefore divide the spectra of for \STO\ and \BTO\ in Fig.~1(a) into two regimes according to the trend in the Stokes integrated intensity, below and above the maximum intensity temperature ($T_m$).

The effect of temperature on the spectral line-shape and integrated intensity in the oxide perovskites is different above and below $T_m$: \STO\ exhibits peak broadening and increase in integrated intensity; in \BTO\ a sharp central peak appears at the phase transition~\cite{Laabidi1991} and decreases with temperature (black arrow in Fig.~1(a)), causing only slight changes to the integrated intensity (polarized spectra of the \BTO\ central peak are depicted in Fig.~S3 in the SM). 
The spectra of \CPB\ exhibit a similar trend to \BTO, with no decrease in the central peak intensity, and only $10\%$ increase in integrated intensity.
Due to the similar crystal structures and Raman spectra, we speculate a decreasing integrated intensity would have also been observed in \CPB, had the crystals remained intact above 700~K.
The decrease in integrated intensity is in stark contrast to the harmonic theory of Raman scattering.~\cite{Cardona1982}
Conventionally, the Stokes integrated intensity is expected to increase with temperature due the to thermal population of vibrations.

Contemporary explanations for the unconventional properties of halide perovskites~\cite{BigReview} are based on a DW potential model.
Frozen phonon calculations have shown the existence of a DW potential around instability points in the \BZ~\cite{Beecher2016,Marronnier2017,Yang2020} as well as a 2D 'saddle' potential related to the phase diagram.~\cite{BechtelPRM2019}
To test the effect of a DW potential on the Raman scattering and whether it can account for the decrease in integrated intensity, we numerically simulated the Raman spectra of a single particle in a DW potential.

\begin{figure*}[ht]
\centering
\includegraphics[width=17 cm]{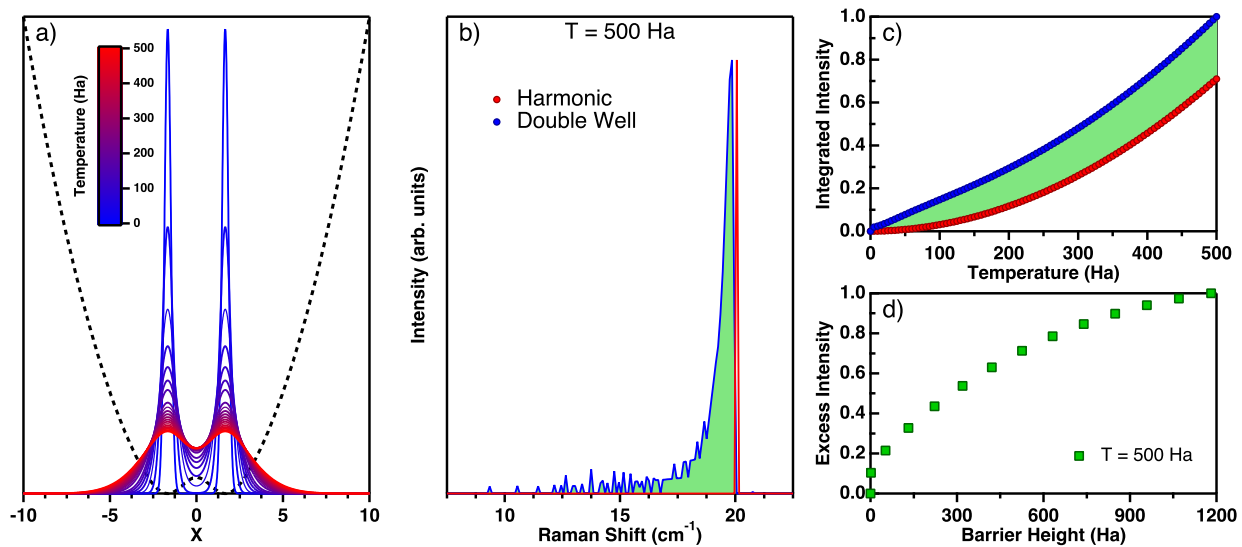}
\caption{\label{fig:cubic_sim} \textbf{Numerical simulation results for a particle in a potential well}. a) Probability density for the location of the particle at various temperatures in a DW potential. b) Simulated \nd -order Raman spectra of a particle in a double-well (blue) and parabolic (red) potential at $T = 500~\tt{Ha}$. c) Integrated intensity of \nd -order inelastic spectra as function of temperature for a double-well (blue) and parabolic (red) potential. d) Excess intensity at $T = 500~\tt{Ha}$ as function of the barrier height ($H_{\tt{Barrier}}=V(0)-V(W)$) showing positive monotonic dependence. Green shading emphasizes the excess intensity.}
\end{figure*}

\subsection*{Numerical Simulation}
We numerically solve the time-independent Schr$\ddot{\tt{o}}$dinger equation for a single particle in a 1D DW potential:
\begin{equation}
    \left(\frac{-\hbar^{2}}{2m}\frac{\partial^{2}}{\partial x^{2}}+V\left(x\right)\right)\psi\left(x\right)=E\psi\left(x\right)~,
    \label{eq:schrodinger_real_space}
\end{equation}
\nit where $E$ and $\psi(x)$ are the real eigenvalues and eigenstates, respectively.
Because the cubic phase is Raman inactive, the decrease in integrated intensity suggests the effective potential approaches a harmonic curve at high temperatures.
We therefore use the following DW potential form:
\begin{subequations}
\begin{equation}
    V(x) = Ax^2 + B \mathcal{G}(x,0,\sigma)
    \label{eq:numer_potential}
\end{equation}
\begin{equation}
    \mathcal{G}(x,\mu,\sigma) = (2\pi\sigma^2)^{\tt{-}\frac{1}{2}} e^{\frac{-(x-\mu)^2}{2\sigma^2}}
    \label{eq:gaussian}
\end{equation}
\end{subequations}
\nit where $\mathcal{G}(x,\mu,\sigma)$ is a Gaussian barrier with $A,B,\sigma \geq 0$.

Given a set of eigenstates, $\psi_n(x) \equiv \braket{x}{n}$, the temperature dependent probability density of the particle is given by
\begin{equation}
    \mathcal{P}(x) = \sum_{n}\frac{e^{-E_{n}\beta}}{Z_{T}}\left|\psi_{n}\left(x\right)\right|^{2}~,
\label{eq:sim_prob}
\end{equation}
with $E_n$ the eigenvalue and $Z_T = \sum_{n}e^{-E_{n}\beta}$ the canonical partition function. 
Isolating the effect of lattice dynamics, the \nd -order inelastic scattering spectra were calculated from the four-point correlation function for particle position (the subscript $2$ denotes \nd -order scattering - see Sec. S3 in the SM for full derivation):
\begin{equation}
\begin{split}
    \mathcal{S}_2(\omega) &\propto \int \tt{d}t~ e^{-i\omega t} \avg{ x(t)x(t)x(0)x(0)}_{T}\\
    & \propto 
     \sum_{n}\sum_{m} \frac{e^{-E_{n}\beta}}{Z_{T}} \left| \braxket{m}{x^2}{n} \right|^{2} \delta \left( \omega_{nm}-\omega \right)~,
\end{split}
\label{eq:sim_spectrum}
\end{equation}
\nit with the temperature-dependent integrated intensity given by:
\begin{equation}
    I = \int \mathcal{S}_2(\omega)~\tt{d}\omega~,
\label{eq:sim_ii}
\end{equation}
\nit where  $\omega_{nm} \equiv (E_n - E_m)/\hbar$, $\beta = (k_BT)^{-1}$ and $\hbar$ and $k_B$ are the Planck and Boltzmann constants.
The same calculation was performed for a particle in a perfect harmonic potential ($B=0$) for comparison.

Figure~2(a) shows the probability density calculated at various temperatures for the case of a DW potential.
At low temperatures the particle is confined to one of the wells, with a sharp Gaussian probability density around the well's minima ($x=\pm W$) and almost none at the barrier.
As temperature rises, the probability density broadens symmetrically, increasing in the barrier.
As expected, inside the harmonic potential the particle is localized at the bottom of the well ($x=0$), and the probability density broadens with increasing temperature (see Fig.~S6 in the SM).

Figures~2(b) and 2(c) show the simulated \nd -order Raman spectra at $T = 500~\tt{Ha}$ and the temperature dependent integrated intensity, calculated by Eq.~(4) and (5), respectively, for the DW (blue) and parabolic (red) potential.
The effect of the potential barrier is apparent - an excess in scattering intensity compared to the harmonic case (green shading).
Moreover, the largest relative difference arises at low temperatures. 
The full simulated spectra are depicted in Fig.~S5 in the SM.


Our harmonic spectra agree with the analytical solution, showing a single sharp peak at the overtone frequency (see Sec.~S3 in the SM).    
Comparison of the harmonic and DW spectra show the excess intensity stems from low-frequency features in the DW simulation, absent in the harmonic case.
These originate in transitions between nearly degenerate localized states that are forbidden in the harmonic case.
This may be understood as spontaneous symmetry breaking, where localization of the atomic states relaxes the harmonic selection rules for the matrix element $\braxket{m}{x^2}{n}$.~\cite{Mayteevarunyoo2008}
This effect persists at high temperatures, far above the barrier height (Fig.~S6 in the SM).
However, had that been the whole story, we would expect the integrated intensity to monotonically increase with temperature as shown in Fig.~2(b).
Figure~2(d) shows a monotonic dependence of the excess intensity at $T=500 ~\tt{Ha}$ on the barrier height.
Thus, a decreasing barrier can explain the observed temperature dependence.
This explains why other static-potential theories with symmetry breaking~\cite{Wang2021,Zhao2020,Zhao2021,Zhao2022} are insufficient to explain the observed temperature dependence.


To apply and test these lessons on the experimental spectra we propose a phenomenological model for \nd -order Raman scattering, with the effect of a DW potential realized by a low-frequency feature, corresponding to symmetry breaking fluctuations. 
The effective temperature dependence of the potential barrier is introduced through a fitting parameter $\eta(T)$ that allows the system to gradually reduce to harmonic dynamics, restoring the cubic selection rules at high temperatures.

\subsection*{Description of The Model}

%
%
As in the numerical simulation, we conserve the cubic point group constraints and assume the crystal is \st-order Raman inactive, such that the leading term becomes two-phonon Raman scattering.
Formally, \nd -order Raman scattering involves only transitions that conserve momentum, \textit{i.e.} any combination of two modes, at $\vec{q}$ and $\vec{q}'=-\vec{q} \equiv \nvec{q}$.~\cite{Cardona1982}
As the Fourier transform of a four-point correlation function, the \nd -order Stokes cross-section may be approximated as the convolution:~\cite{Benshalom2022}
%
\begin{equation}
\begin{split}
    \mathcal{S}_2(\omega,0) \propto 
    \sum_{j,j',\vec{q}} \chi_{j,j'}^{\vec{q}} & \int\tt{d}\Omega~ \mathcal{J}_j(\Omega,\vec{q})[n(\Omega,T) + 1]\\
    \times~&\mathcal{J}_{j'}(\omega-\Omega,\nvec{q})[n(\omega-\Omega,T) + 1] \\
    = 
    \sum_{j,j',\vec{q}} \chi_{j,j'}^{\vec{q}} & ~ \mathcal{J}_j(T,\omega,\vec{q}) \ast \mathcal{J}_{j'}(T,\omega,\nvec{q})~,
\end{split}
\label{eq:S2w}
\end{equation}
%
\nit where $\chi_{j,j'}^{\vec{q}}$ are the \nd-order susceptibility mode derivatives, $n(\omega,T)$ is the Bose-Einstein occupation factor (see Eq.~(S8) in the SM) and we use the shorthand $f(T,\omega,\vec{q}) \equiv f(\omega,\vec{q})[n(\omega,T) + 1]$.
The spectral function $\mathcal{J}_j(\omega,\vec{q})$  is defined as:
\begin{equation}
\begin{split}
    \mathcal{J}_j(\omega,\vec{q}) = [n(\omega,&T)+1]^{-1} \\
    & \times \int \tt{d}t~e^{-i \omega t} \avg{ U_{j}(\vec{q},t) U_{j}(\vec{q},0)}_T~,
\end{split}
\label{eq:SpectFunc}
\end{equation}
%
\nit the Fourier transform of the thermally averaged ($\avg{\cdots}_{T}$) auto-correlation function of atomic displacement $U_{j}(\vec{q},t)$ for mode $j$.
The spectral function for a crystal of N$_{\tt{uc}}$ unit cells must satisfy:~\cite{Kwok1968}
\begin{subequations}
\label{eq:Jcon}
    \begin{equation}
        \int_{0}^{\infty}\tt{d}\Omega~ \mathcal{J}_j(\Omega,\vec{q}) = 1
    \label{eq:j1}
    \end{equation}
    \begin{equation}
        \frac{1}{\tt{N}_{\tt{uc}}} \sum_{j,\vec{q}}\mathcal{J}_j(\omega,\vec{q}) = g(\omega)
    \label{eq:jvdos}
    \end{equation}
\end{subequations}
\nit where $g(\omega)$ is the vibrational density of states (vDOS).

To introduce the spontaneous symmetry breaking created by the DW potential, as indicated by simulation, we invoke a low-frequency feature (LFF), $\mathcal{D}(\omega,\vec{q})$, in the spectral function.
The LFF is not included in $g(T,\omega)$ and is completely decoupled from phononic vibrations in the sense that there is no energy transfer between the two modalities.
This LFF can represent a low-frequency vibration that breaks the global symmetry, a localized fluctuation, or act as an effective description of multiple oscillations.~\cite{Andrews1982,Krumhansl1975,Porter2009,Shirane1993}
A similar approach was introduced by Fontana and Lambert~\cite{Fontana1972} for the Raman spectra of \BTO, by Safran \textit{et al.}~\cite{Safran1977} for magnetic phase transitions, and by Dultz and others~\cite{Dultz1973,Dultz1976,Briganti1981,Sanyal1982} for disorder-induced light scattering.
We thus divide the spectral function to:
\begin{equation}
     \sum_{j,\vec{q}}\mathcal{J}_j(\omega,\vec{q}) = \eta \sum_{\vec{q}} \mathcal{D}(\omega,\vec{q}) + (1-\eta) \sum_{j,\vec{q}}\mathcal{P}_j(\omega,\vec{q})~,
\label{eq:JDP}
\end{equation}
\nit where $\mathcal{P}_j(\omega,\vec{q})$ represents the phononic part in the spectral function, and $0 \leq \eta \leq 1$ is the LFF component fraction.
The LFF component is normalized by
\begin{equation}
\label{eq:Dcon}
     \frac{1}{\tt{N}_{\tt{uc}}} \int_{0}^{\infty}\tt{d}\Omega~ \sum_{\vec{q}} \mathcal{D}(\Omega,\vec{q}) = 3N~,
\end{equation}
where $N$ is the number of atoms per unit cell, such that conditions~(8) for the full spectral function hold.

We further assume that $\mathcal{D}(\omega,\vec{q})$ is a separable function:
\begin{equation}
    \mathcal{D}(\omega,\vec{q}) = D(\omega)C(\vec{q})~.
\label{eq:DwCq}
\end{equation}
Neutron scattering and statistical mechanics indicate that $C(\vec{q})$ has a Lorentzian line-shape, centered at $\vec{q}_0$.~\cite{Cowley1996,Papon2002,DiAntonio1993,Vugmeister1999,Shirane1993}
According to the Orstein-Zernike theory in one dimension~\cite{OZTheory}, it would have the form:
\begin{equation}
\begin{split}
    C(q) &= \frac{T/T_c}{a^2(\vec{q}-\vec{q_0})^2 + (T-T_c)/T_c} \\
    &= \frac{\xi^{2}(\xi^{2}-1)^{-1}}{a^2(\vec{q}-\vec{q_0})^2 + (\xi^{2}-1)^{-1}}~,
\end{split}
\label{eq:Cq}
\end{equation}
\nit where $a$ is the lattice constant and $\xi$ is the temperature-dependent correlation length in units of unit-cells, defined as $\xi^2 = T/(T-T_c) \geq 1$.
As temperature approaches the phase transition ($T \rightarrow T_c$), the correlation length $\xi$ diverges, and the fluctuation leads the phase transition into the ordered, lower temperature phase.
$\vec{q}_0$ is the high-temperature phase fluctuation wave-vector, and it defines the symmetry breaking across the two phases.~\cite{Safran1976,Schmutz1979}
For example, in ferro-distortive phase transition $\vec{q}_0 = 0$,~\cite{Huller1969} whereas in anti-ferro-distortive phase transition $\vec{q}_0$ would be the \BZ\ edge.~\cite{Lanigan-Atkins2021,Fleury1968,Shirane1969}

Plugging Eq.~(9) and (11) back into Eq.~(6) and approximating $\chi_{j, j'}^{ \vec{q}} \equiv 1$, we find the cross-section is composed of three parts:
\begin{equation}
\begin{split}
    \mathcal{S}_2(\omega,0) \propto 
    \eta^2 \left [ \sum_{\vec{q}} C(\vec{q})C(\nvec{q}) \right ] D(T,\omega) \ast D(T,\omega)\\
    +
    2 \eta (1-\eta) \left [ \sum_{j,\vec{q}} C(\vec{q}) \mathcal{P}_{j}(T,\omega,\nvec{q})\right ] \ast D(T,\omega)& \\
    +
    (1-\eta)^2 \sum_{j,j',\vec{q}} \mathcal{P}_j(T,\omega,\vec{q}) \ast \mathcal{P}_{j'}(T,\omega,\nvec{q})&~,
\end{split}
\label{eq:S2_disorder}
\end{equation}
%
the first represents 2-LFF scattering, \textit{i.e.}, the creation of two LFF excitations in the case of Stokes scattering.
The second involves 1-LFF and 1-phonon excitation.
The final third term is equivalent to the standard expression for 2-phonon scattering~\cite{Cardona1982}.
Although detrimental for spectral accuracy, the removal of $\chi_{j, j'}^{\vec{q}}$-dependency is not as crude as might first appear;
selection rules for \nd-order scattering are usually so lax as to allow most phonon pairs to participate, even in high symmetry structures,~\cite{Burstein1965} and it has long been known that the 2-phonon vDOS can provide a rough estimation of the \nd-order Raman spectrum.~\cite{Y.Yu2010}
Most importantly, as we demonstrate below, this simplified version is sufficient to capture and explain the non-monotonic temperature dependence of the integrated intensity.

Assuming the frequency of $\mathcal{D}(T,\omega,\vec{q})$ is much lower than phonon frequencies, we define the vDOS as $\tt{N}_{\tt{uc}}^{-1}\sum_{j,\vec{q}} \mathcal{P}_j(T,\omega,\vec{q}) = g(T,\omega)$.
Simplifying further by setting $C(\vec{q}) \equiv 1$, the cross-section becomes
\begin{equation}
\begin{split}
    \mathcal{S}_2(\omega,0) \propto 
    \eta^2~ D(T,\omega) \ast D(T,\omega&) \\
    +
    2\eta (1 - \eta)~ g(T,\omega) \ast D&(T,\omega) \\
    +
     (1-\eta)^2~ g(T,\omega&) \ast g(T,\omega)~,
\end{split}
\label{eq:S2_final}
\end{equation}
where we used Eq.~(8b) to find that $\mathcal{S}_2(\omega,0)$ is given by the convolutions of $g(T,\omega)$ with itself and with $D(T,\omega)$.

\begin{figure*}[ht]
\centering
\includegraphics[width=17 cm]{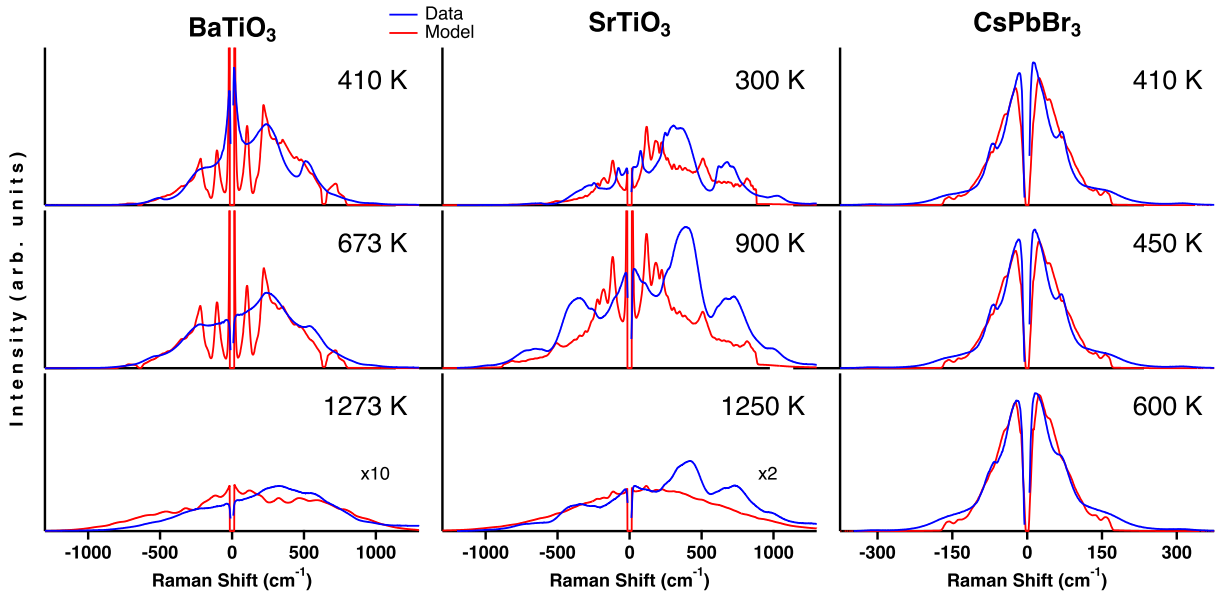}
\caption{\label{fig:cubic_fit} \textbf{Fit results} of Eq.~(14) to the data of \BTO, \STO\ and \CPB\ at three selected temperatures. Blue and red traces are the data and model, respectively.}
\end{figure*}

\subsection*{Fitting to the Model}

Assuming $g(T,\omega)$ is known and $D(T,\omega)$ is a damped Lorentz oscillator,~\cite{YaffePRL2017} Eq.~(14) has 4 remaining fitting parameters: (1) $A$ - global proportionality factor of $\mathcal{S}_2(\omega,0)$, (2) $\eta$ - weight of the LFF in the spectral function, (3) $\omega_D$ - frequency of the LFF, (4) $\Gamma_D$ - damping coefficient of the LFF.

Figure~3 shows the fit results for \BTO, \STO\ and \CPB\ at selected temperatures.
We used experimental densities of states~\cite{Choudhury2010,Khatib1989,Lanigan-Atkins2021} from inelastic neutron scattering at 300~K to calculate $g(T,\omega)$. 
The fitted function $D(T,\omega)$ and $g(T,\omega)$ were normalized according to Eq.(8) and (10).

We begin the fitting process by matching the highest temperature integrated intensity to the model with $\eta=0$, \textit{i.e.}, to conventional 2-phonon scattering.
This procedure fixes the proportionality factor, which is set throughout all temperatures.
Next, we use the Lorentz parameters $\omega_D$, $\Gamma_D$, and the LFF weight component $\eta$, to fit the integrated intensity as function of temperature.
As the fitted plots in Fig.~3 show, by adding $D(T,\omega)$ to the spectral function we are able to reproduce the non-monotonic temperature evolution of the Raman spectra.
Discrepancies in the spectral shape are the result of our simplifying approximations $C(\vec{q}), \chi_{j,j'}^{\vec{q}} \equiv 1$, and complete decoupling between phonons and the LFF.~\cite{Scalabrin1977,Shapiro1972,Petzelt1987} 

We emphasize that it is difficult to attach physical meaning to the fitted parameters without additional constraints on either $D(T,\omega)$ or $\eta$.
The final values and the integrated intensities are depicted in Fig.~S7 and S7 in the SM.
The number of parameters may be reduced further by establishing the origin of $D(\omega)$ and independently estimating $\omega_D$ and $\Gamma_D$ or the relation $\Gamma_D=\Gamma(\omega_D)$.

Nonetheless, close examination of the fitting process revealed the possible different mechanisms governing the temperature dependence.
A slow increase in intensity correlates with an increase in the frequency of the LFF ($\omega_D$), while a decrease in intensity correlates with a decrease in LFF contributions ($\eta$).
Our interpretation of the Raman spectra as \nd-order scattering enhanced by a LFF also resolves some of the difficulties with competing interpretations mentioned above.
It reproduces the non-monotonic temperature dependence without challenging the selection rules derived for the cubic average structure.
Moreover, because its low frequency and localization in space, the LFF should not couple efficiently to electrons.
Therefore, it remains below the frequency range of THz-absorption measurements, and does not affect the electronic band structure, allowing for a sharp absorption edge.

Our model also offers insight into another aspect of the different perovskite Raman spectra.
There is a striking difference between the sharp central peak of \BTO\ and the shallow central feature of \STO\ and \CPB.
It was already suggested that the central feature in \CPB\ is related to scattering by polar fluctuation.~\cite{YaffePRL2017,Gao2021}.
However, our calculations show that the central feature can be reproduced without the 2-LFF term in \STO\ and \CPB, but not in \BTO.
This suggests the central peak of \BTO\ is indeed directly induced by the low-frequency anharmonic motion leading the phase transition and stabilizing the cubic phase.
As temperature approaches $T_c$, the correlation length of the symmetry-breaking motion increases, leading to an effective tetragonal local structure even before the phase transition.
In both \CPB\ and \STO\ the phase transition is lead by vibrations with \vec{q}-vectors at the \BZ\ edge~\cite{Lanigan-Atkins2021,Fleury1968,Shirane1969}, but since \BTO\ goes into a ferroelectric phase, its transition is lead by a $\Gamma$-point vibration.~\cite{Huller1969}
It is therefore possible that as the correlation length grows, cubic selection rules in \BTO\ relax, leading to \st -order scattering and a sharp central peak.

\section*{Conclusions}

We presented numerical simulations of a 1D quantum particle, showing the effect of a DW potential on the Raman scattering of disordered crystals.
The DW induces low-frequency features interpreted as transitions between potential wells, causing an excess in scattering intensity, which can explain the abnormally high Raman cross-section in cubic perovskites.
The simulation results link the decrease in scattering intensity in higher temperatures to a temperature-dependent potential surface, which evolves from a multi-well to harmonic-like potential.
Motivated by this insight, we presented a phenomenological \nd -order Raman scattering model that conserves the cubic symmetry of the crystal structure.
By introducing a low-frequency feature and incorporating it into the \nd -order scattering formalism, we successfully reproduced the broad Raman spectra and non-monotonic temperature dependence.
The existence of this low-frequency feature is supported by various observations, such as over-damped modes and critical inelastic neutron scattering,~\cite{Shapiro1972,Weadock2020,Cowley2006,Bruce1980,LeguyNatComm2015,Lanigan-Atkins2021} central peak in Raman scattering,~\cite{Fontana1972,YaffePRL2017,Gao2021,SharmaMAPI2,Perry1967,Fontana2020,Laabidi1991} soft modes,~\cite{Scott1974,Guo2017c} short phonon coherence length~\cite{Songvilay2019,Mayers2018,Gold-Parker2018} and a high dielectric function at low-frequencies.~\cite{Filippone2020,Svirskas2020,Petzelt1987,Kamba2021}

Our interpretation is disencumbered by contradictions with Raman selection rules or complementary measurements.
The model similarly applies to oxide and halide perovskites, emphasizing the similarities between the crystal families and their interaction with light.\
It can be applied to other dynamically-disordered crystal phases, providing better intuition of the effect of structural dynamics on the electronic, optical, and mechanical properties of functional materials. 

\subsection*{Acknowledgements}

The authors would like to thank Dr. Iddo Pinkas for help in designing the experimental setup, and Dr.\ Lior Segev (WIS) for developing the Raman software.
O.Y. acknowledges funding from: ISF(1861/17), BSF (grant No.2016650), ERC (850041 - ANHARMONIC).

\section*{\label{sec_methods} Methods}

\subsection*{Raman~Scattering}

Raman scattering measurements were conducted in a home-built back-scattering system using a 2.54~eV (for \BTO\ and \STO) CW sapphire pumped-diode laser (Coherent Inc., USA) and a 1.58~eV CW pump-diode laser (Toptica Inc., USA).
The samples were heated in a closed temperature control system (Linkam Scientific) under continues Nitrogen flow.
Measures taken to isolate the effect of temperature on the intensity involve: maintaining constant measurement conditions, careful refocusing of the laser before every measurement and repeated measurement cycles (heating and cooling) across the full temperature range and phase transitions.
The Stokes integrated intensity was calculated by numerically integrating over the experimental spectra in the Stokes scattering side ($\omega > 0$).

\clearpage
\onecolumngrid

\centering{{\Huge Supplemental Material}} \\ 
\centering{\textbf{The Disorder Origin of Raman Scattering In Perovskites Single Crystals}}

\renewcommand{\thepage}{S\arabic{page}}  
\renewcommand{\thesection}{S\arabic{section}}   
\renewcommand{\thesubsection}{S\arabic{section}.\alph{subsection}} 
\renewcommand{\thetable}{S\arabic{table}}   
\renewcommand{\thefigure}{S\arabic{figure}}
\renewcommand{\theequation}{S\arabic{equation}}
\setcounter{page}{1}
\setcounter{figure}{0}
\setcounter{equation}{0}

\section{Materials}

Single crystals of \BTO\ and \STO\ with (001) orientation were purchased from MTI cooporation (Richmond, CA, USA) and CrystTech GmbH (Berlin, Germany), respectively.
A single crystal of \CPB\ was synthesized according to the procedure described in Ref.~\cite{Rakita2016} and characterized by optical microscopy, x-ray diffraction, and Raman scattering (data not shown).

\section{\label{secSI_Raman}Raman Scattering}

Raman measurements were preformed in a system designed for polarization orientation (PO) Raman measurements.
The full scheme is described thoroughly in Ref.~\cite{Menahem2021,Asher2020,SharmaMAPI2,Benshalom2022,Cohen2022} and the chosen incident laser lines are described in the Methods section in the main text.

The unpolarized Raman spectra are obtained by averaging the parallel and perpendicular spectra at two perpendicular directions.  
This procedure is equivalent to measuring with an unpolarized laser source and no analyzer. 
The advantage of this process is that we require a scrambler to avoid the inherent polarization of the laser source.

The polarization dependent intensity for a given mode is given by the following equation:~\cite{Y.Yu2010}

\begin{equation}
    \mathcal{S} \propto |\hat{e}_{s} \cdot \vec{J} \cdot \vec{R} \cdot \vec{J} \cdot \hat{e}_{i}|^{2}
\label{eq:Raman_Intensity}
\end{equation}

Where $\hat{e}_{i}$ and $\hat{e}_{s}$ are the polarization vectors of the incident and scattered light, respectively. $\vec{R}$ is the Raman tensor, and $\vec{J}$ is the Jones matrix,~\cite{KarnertSchiRep2016,KranertPRL2016} introduced to account for birefringence effects due to anisotropy in the crystal structure.~\cite{FoxOPOS,GuoPRL2018,Li2020}  $\hat{e}_{i}$ and $\hat{e}_{s}$ are either parallel or perpendicular to one another, depending on the measurement configuration.

In principle, the Raman tensor is a $3 \times 3$ matrix, constrained by symmetry considerations.
The shape of the Raman tensor depends on the symmetry of the crystal, symmetry of the mode and the crystal face measured.
To generalize our claims, we consider scattering from a general Raman tensor, unconstrained by symmetry.
Rotation of the Raman tensor to different crystallographic orientations would mix the diagonal and off-diagonal elements of the tensor.
Therefore, we use the most general real-valued tensor:

\begin{equation}
 \vec{R}=\begin{pmatrix} a&c\\d&b \end{pmatrix}
\label{eq:Raman_tensor}
\end{equation}

Our polarization manipulation is confined to the $xy$ plane of the laboratory frame.
Therefore, we are blind to the $z$ components of $\vec{R}$ and may consider it as a $2 \times 2$ tensor.
In our laboratory frame, the polarization vectors are given by:

\begin{equation}
\hat{e}_i=\left( \begin{array}{c}
cos(\theta) \\ 
sin(\theta) \end{array}
\right) ~,~ \hat{e}_{S,||}^T=\left( \begin{array}{c}
cos(\theta) \\ 
sin(\theta) \end{array}
\right) ~,~ \hat{e}_{S,\perp}^T=\left( \begin{array}{c}
-sin(\theta) \\ 
 cos(\theta) \end{array}
\right)
\end{equation}

\nit and the Jones-matrix is given by:

\begin{equation}
\vec{J}=\left( \begin{array}{cc}
1 & 0 \\ 
0 & e^{i\left |\phi\right |} \end{array}\right)
\end{equation}

Plugging all definitions back to Eq.~(S1) and solve, we find:

\begin{subequations}

\begin{equation}
    \mathcal{S}_{\parallel}(\theta) \propto \left | a \cos ^{2}(\theta) + (c+d) e^{i \left| \phi \right |} \sin (\theta) \cos (\theta) + b e^{2 i \left| \phi \right |} \sin ^{2}(\theta) \right|^{2}
\end{equation}

\begin{equation}
    \mathcal{S}_{\perp}(\theta) \propto \left| d e^{i \left| \phi \right |} \cos ^{2}(\theta)-\left( a-b e^{2 i \left| \phi \right |} \right) \sin (\theta) \cos (\theta)- c e^{i \left| \phi \right |} \sin ^{2}(\theta) \right|^{2}
\end{equation}
\end{subequations}

\nit where $\mathcal{S}_{\parallel}(\theta)$ and $ \mathcal{S}_{\perp}(\theta)$ are the intensity in the parallel and perpendicular configurations, respectively.
The unpolarized spectra are obtained by summing:

\begin{equation}
\begin{split}
    \mathcal{S}_{\tt{unpol}}  &= \mathcal{S}_{\parallel}(\theta) + \mathcal{S}_{\perp}(\theta) +\mathcal{S}_{\parallel} \left( \theta + \frac{\pi}{2} \right) + \mathcal{S}_{\perp} \left( \theta + \frac{\pi}{2} \right)\\
    &= a^2 + b^2 + c^2 + d^2 \\
    &= \frac{1}{\pi} \int_0^{2\pi} \mathcal{S}_{\parallel}(\theta) + \mathcal{S}_{\perp}(\theta) ~\tt{d}\theta
\end{split}
\label{eq:Raman_unpolsum}
\end{equation}

\nit which is independent of $\theta$.
The generality of the Raman tensor and the restriction to the laboratory frame guarantee the result is applicable to all modes in the spectrum, and all orders of scattering. 
Therefore, we can apply Eq.~(S6) to the PO Raman spectra regardless of the choice of $\theta$.
To properly follow the intensity evolution with temperature, $\theta$ was kept constant throughout the temperature cycles, and all spectra were normalized to a single global value.


    \begin{figure}[t]
        \centering
        \includegraphics[width = 15 cm]{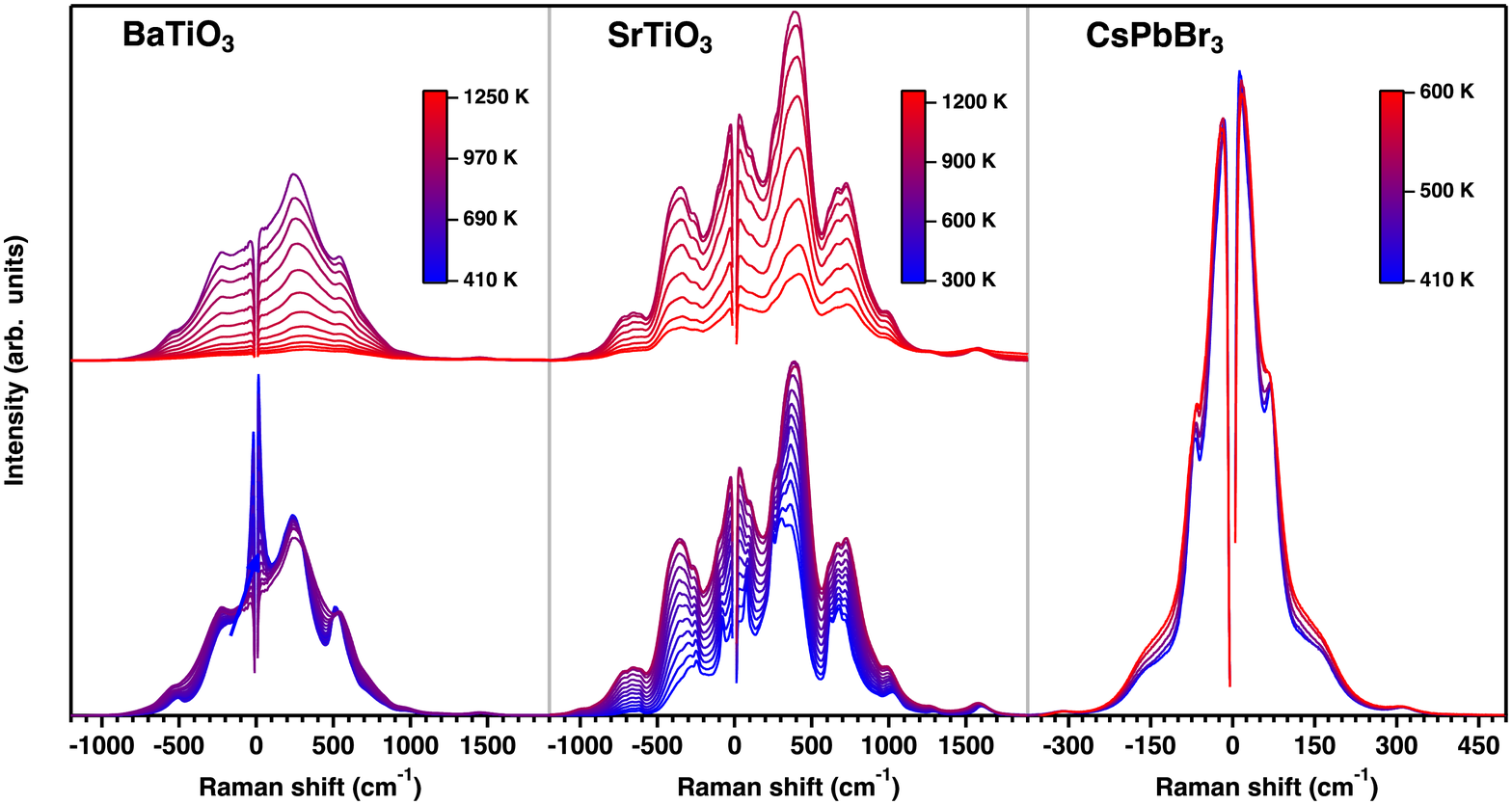}
        \caption{All Temperature-dependent unpolarized Raman scattering spectra of BaTiO$_{\vec{3}}$, SrTiO$_{\vec{3}}$ and CsPbBr$_{\vec{3}}$ in the cubic phase. The spectra of \BTO\ and \STO\ are divided to two scattering regimes, showing the non-monotonic temperature dependence.}
        \label{fig:SI_expRaman}
    \end{figure}

Figure~S1 presents the unpolarized Raman spectra of \BTO, \STO\ and \CPB.
The data in Fig.~1 in the main text were taken from this data set.
The spectra of \BTO\ and \STO\ are divided to two scattering regimes, as described in the main text.

\subsection{\label{secSI_STO_Cryo} \STO\ data below 300~K}

Unlike \BTO\ and \CPB, the phase transition temperature of \STO\ is below room temperature ($T_c = $ 105~K).
Therefore, to measure the Raman spectra of \STO\ in the range of $T_c \leq T \leq $300~K the measurement system had to be modified.
The \STO\ crystal was loaded into liquid N$_2$ cooled optical-cryostat (Janis Inc., USA), set up in the same experimental system, instead of the closed temperature control system (Linkam Scientific).
The same spot was measured for both data sets, below and above 300~K.

Figure~S2 shows the temperature dependent unpolarized Raman spectra of \STO\ between the tetragonal-cubic phase transition temperature and room temperature.
The spectra show the same trends with temperature as shown in Fig.~S1.
Comparing the spectra below (100~K, black trace) and above (110~K, blue trace) the phase transition, we see a very subtle effect on the Raman spectra.
Small peaks in the spectrum of the tetragonal phase, assigned to \st -order Raman scattering~\cite{Schaufele1967,Taylor1979} or localized modes~\cite{Nilsen1968}, disappear as we cross the phase transition to the cubic phase.

    \begin{figure}[t]
        \centering
        \includegraphics[width = 15 cm]{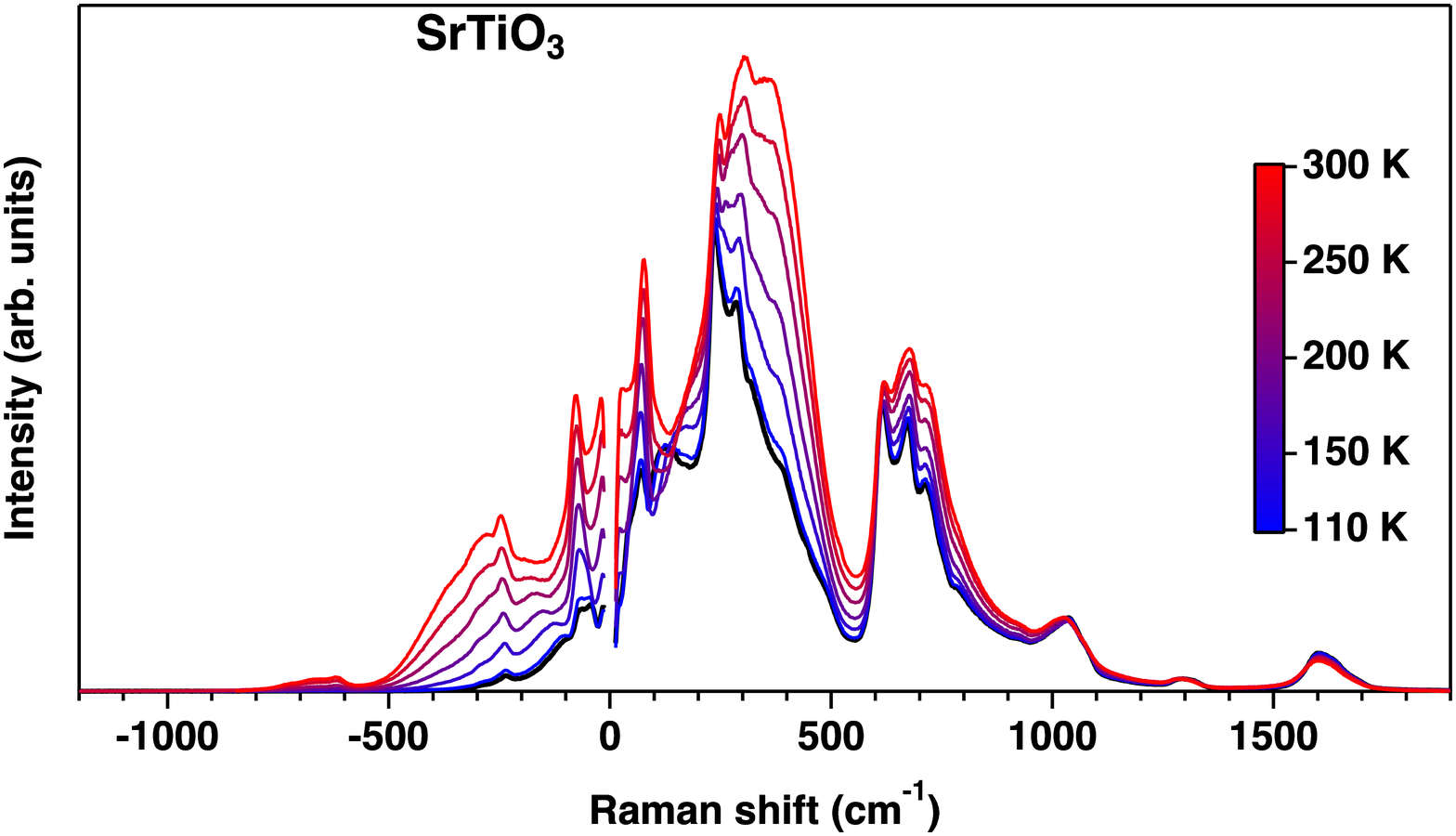}
        \caption{Temperature dependent Raman scattering spectra of cubic \STO\ below room temperature.
        The spectra show the same trends with temperature as shown in Fig.~S1 below room temperature down to the phase transition temperature ($T_c = $105~K).
        Raman spectrum at the tetragonal phase ($T=$ 100~K) is in black, showing the same broad features appearing in the cubic phase, together with small peaks assigned as \st -order Raman scattering peaks.
        The intensity of the \nd -order scattering is as high as the the \st -order scattering in the tetragonal phase.}
        \label{fig:SI_STO_LowT}
    \end{figure}

\subsection{\label{secSI_BTO_CP} \BTO\ Central Peak in PO}

When exciting \BTO\ single crystal in our PO Raman setup, we can selectively observe only the central peak.
This is done when the excitation polarization is parallel to one of the crystallographic axis, and collection of light scattered perpendicular to the incident polarization ($Z(XY)Z$ or $Z(YX)Z$ in Porto notation~\cite{PortoNotationCardona}).
This scheme allows for observation of modes with non-vanishing off-diagonal components in the Raman tensor.
Fortunately, this configuration emphasizes the central peak in \BTO\ by rejecting many modes around it.

    \begin{figure}[t]
        \centering
        \includegraphics[width = 15 cm]{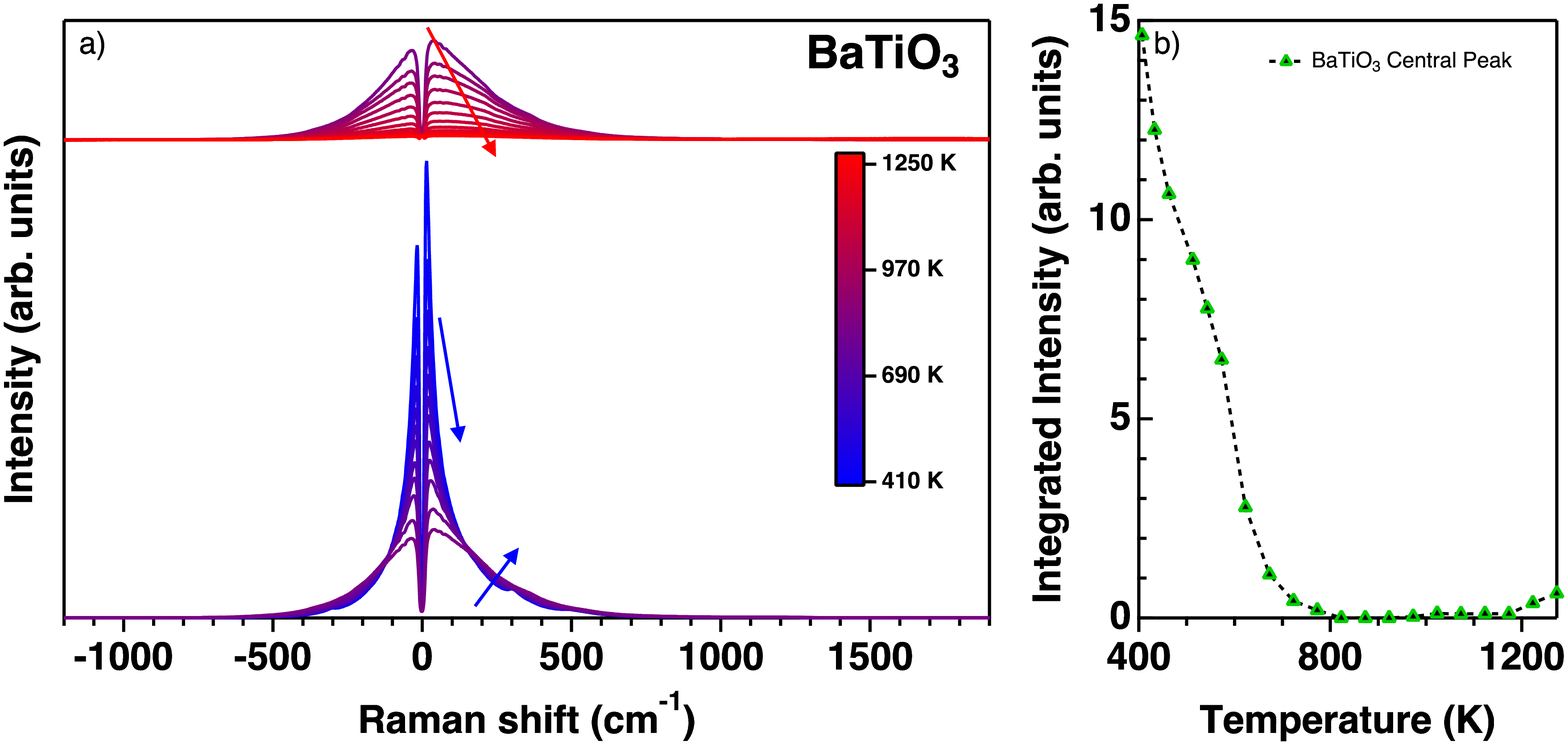}
        \caption{Temperature dependent polarized Raman scattering spectra of cubic \BTO\ showing mostly the central peak.
        Incident polarization is along one of the crystallographic axes, and scattered polarization is perpendicular ($Z(XY)Z$ or $Z(YX)Z$ in Porto notation).
        The spectra are divided to the two scattering regimes as described in the main text.}
        \label{fig:SI_BTO_CP}
    \end{figure}

Conventionally, the central peak is related to a long-amplitude relaxational motion with no restoring fore.~\cite{YaffePRL2017,Scalabrin1977,Laabidi1991}
Therefore, to extract the temperature dependence of the central peak in \BTO, the spectra were fitted to a Debye relaxor and a sum of damped Lorentz oscillators multiplied by the Bose-Einstein occupation number:

\begin{equation}
\begin{split}
    S(\omega) =[ n(\omega,&T) + 1]  \frac{ c_0 | \omega | \Gamma_0}{ \omega^2  + \Gamma_0^2} ~~+ \\
    [ &n(\omega,T) + 1] \sum_i  \frac{ c_i \omega \Gamma_i^3 }{\omega^2 \Gamma_i^2 + \left( \omega^2 - \Omega_i^2 \right) ^2}
\end{split}
\label{eq:lorentz}
\end{equation}

\nit where subscripts $0/i$ stand for the Debye and $i^{th}$ Lorentz oscillator, $\omega$ and $\Omega_i$ are the probing and mode-central frequency respectively, $\Gamma_{0/i}$ is the damping constant inversely proportional to the lifetime of the fluctuation, and $c_{0/i}$ is a unit-less fitting parameter.
The Bose-Einstein occupation number as function of $\omega$ and the absolute temperature $T$ is given by:

\begin{equation}
n(\omega,T)  = ( e^{\beta \hbar \omega} -1)^{-1}
\label{eq:nbed}
\end{equation}

\nit where $\beta = (k_B T)^{-1}$, $k_B$ and $\hbar$ are the Boltzmann and reduced Planck's constants.
Damped Lorentz oscillators were used to capture the rest of the spectral features, without assuming any more meaning to the fit results.
The integrated intensity of the central peak was extracted by integrating the Debye relaxor term as function of frequency.

Figure~S3 shows the temperature dependent polarized Raman spectra of \BTO\ in this configuration, showing mostly the central peak.
The spectra are divided into the two scattering regimes as in the Fig.~1 in the main text.
The decreasing intensity of the central peak is clearly evident.



\subsection{\label{secSI_classic}Comparing to 1- and 2-phonon scattering}

To compare the temperature dependence of the integrated intensity of \BTO, \STO\ and \CPB\ to the expected dependencies from \st - and \nd -order Raman scattering, we have to approximate the expected dependencies using the available data. 

The structure factor in \st -order Raman scattering at a given temperature has the form:

\begin{equation}
    \mathcal{S}_{1}(\omega,0)_{T_0} \propto \sum_{j} \chi_j \mathcal{J}_j(\omega,0)[n(\omega,T_0) + 1]
\label{eq:SI_Raman_S1}
\end{equation}

\nit where $\mathcal{S}_{m}(\omega,\vec{q})$ is the structure factor of $m^{\text{th}}$ order Raman scattering at angular-frequency $\omega$ and scattering wave-vector (and crystal momentum) $\vec{q}=0$, due to momentum conservation. 
$\chi_j$ is the Raman tensor of mode $j$ and $\mathcal{J}_j(\omega,\vec{q})$ is the spectral function, defined in Eq.~(7) in the main text.
The temperature dependence of \st -order Raman follows the Bose-Einstein distribution, as described in Eq.~(S8).
Conventionally, it is assumed that the spectral function does not depend on temperature.
Therefore, the expected temperature dependent integrated intensity can be approximated by:

\begin{equation}
    I(T) = \int_0^{\infty} \tt{d}\omega ~ \mathcal{S}_1(\omega,0)_{T_0} \frac{[n(T,\omega)+1]}{[n(T_0,\omega)+1]}
\label{eq:I1}
\end{equation}

The structure factor in \nd -order Raman scattering at a given temperature was taken from Eq.~(14) in the main text, with $\eta=0$:

\begin{equation}
    \mathcal{S}_2(\omega,0)_{T_0} = [n(T_0,\omega)+1]g(\omega) \ast [n(T_0,\omega)+1]g(\omega)
\end{equation}

\nit where we assume a temperature independent density of states.
Therefore, the expected temperature dependent integrated intensity was estimated by calculating $\mathcal{S}_2(\omega,0)_T$ for any given temperature and integrating:

\begin{equation}
    I(T) = \int_0^{\infty} \tt{d}\omega ~ \mathcal{S}_2(\omega,0)_{T}
\label{eq:I2}
\end{equation}

Figure~S4 shows the temperature dependent Stokes integrated intensity of the Raman spectra of \BTO, \STO\ and \CPB, compared to the calculated expected dependencies according to Eq.~(S10) and (S11).
The calculated dependencies were normalized to cross the maximum intensity point.
Clearly, the experimental dependencies of \BTO\ and \CPB\ do not follow the expected behavior.
In the case of \STO, in the range of $T_c < T \leq T_m$ the integrated intensity increases with temperature like \st -order Raman scattering, which is forbidden by the selection rules for the cubic perovskite structure.
Moreover, the decreasing integrated intensity cannot be reconciled with the calculated intensities, which are monotonic with temperature.

   \begin{figure}[t]
        \centering
        \includegraphics[width = 15 cm]{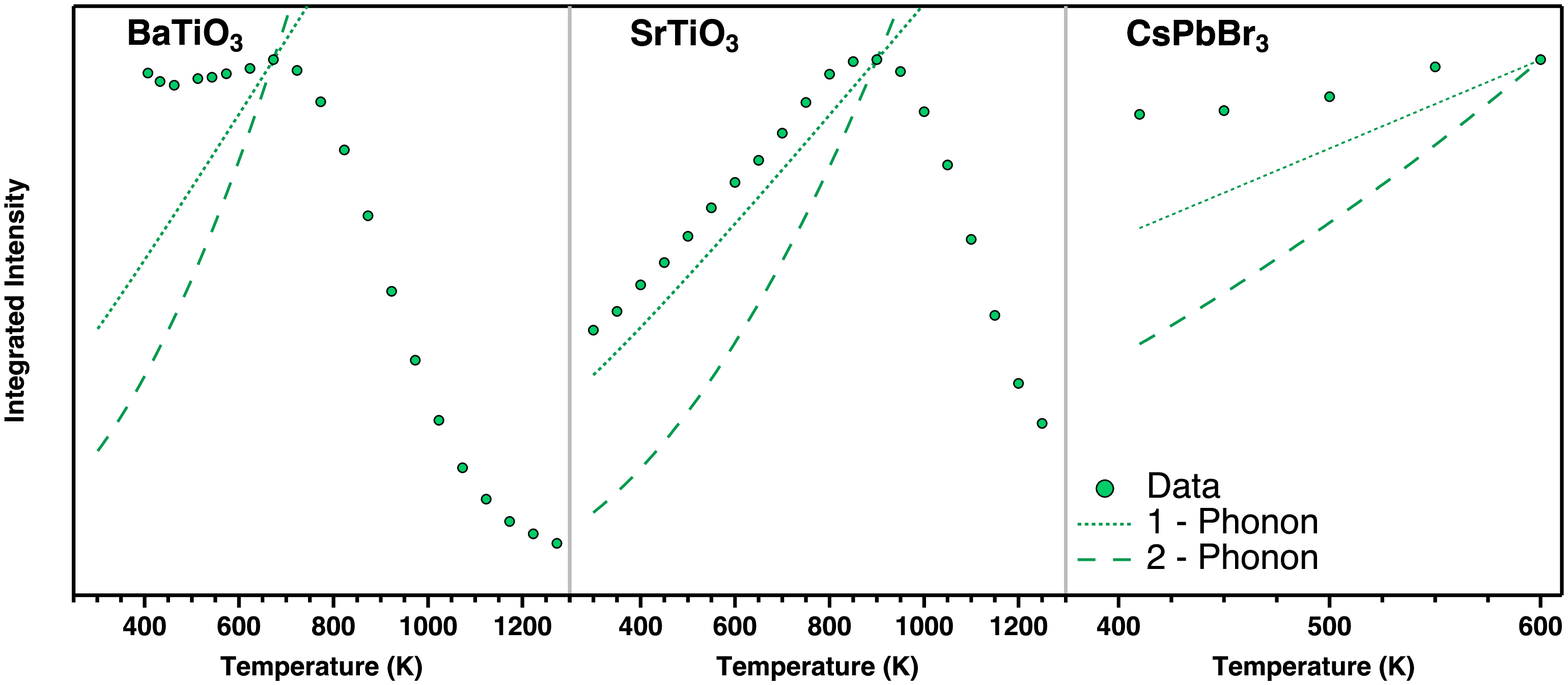}
        \caption{Temperature-dependent Stokes integrated intensity (circles) of BaTiO$_{\vec{3}}$, SrTiO$_{\vec{3}}$ and CsPbBr$_{\vec{3}}$ in the cubic phase compared with the expected integrated intensity from 1-phonon (dense-dash) and 2-phonon (wide-dash) scattering, normalized to cross the maximum intensity point.
        The integrated intensity of \BTO\ and \CPB\ do not follow any of the calculated dependencies.
        The data of \STO\ does follow the 1-phonon line below 900~K, which is violation of the Raman selection rules for the crystal structure.
        }
        \label{fig:SI_expII}
    \end{figure}

\section{\label{secSI_NumerSim}Numerical Simulation}


To test the feasibility of our hypothesis, that a double-well (DW) potential could give rise to a non-monotonic trend in Raman scattering, we numerically solved the time-independent Schr{\"o}dinger equation,
\begin{equation}
\label{SI_eq:scrodinger}
    \left(\frac{-\hbar^{2}}{2m}\frac{\partial^{2}}{\partial x^{2}}+V(x)\right)\psi(x)=E\psi(x)
\end{equation}
with $\hbar$ the reduced Planck constant and $m$ the particle mass.
Because the solution is numerical we are able to solve for any random potential $V(x)$.
We used MATLAB software and a 1D grid of 2401 points, yielding 2401 eigenstates and corresponding eigenvalues.
Specifically, Eq.~(S13) was solved for $m=\hbar=1$, and
\begin{equation}
    V_{\tt{Har}}(x)=\frac{1}{2} m \omega^2 x^2
\end{equation}
for a perfect harmonic potential, and
\begin{equation}
    V_{\tt{DW}}(x)= V_{\tt{Har}}(x) + B \mathcal{G}(x,0,1)
\end{equation}
for the DW potential. 
$\mathcal{G}(x,0,1)$ is a Gaussian barrier centered at $x=0$ with a full-with-half-maximum of 1.
The $x$ range for the solution was chosen in an iterative process, making sure that at the highest temperature $x_{\tt{max}} >> \Delta x$ and the probability density to find the particle decays to zero at $\pm x_{\tt{max}}$.
The finite nature of the calculation necessarily gives a maximum value for the potential surface, and therefore not all calculated states are bound.
This created non-physical solutions for high enough eigenvalues, but these were verified to have negligible contributions in all pertinent temperatures.
Moreover, to avoid additional non-physical errors due to the numerical nature of the simulation, the solution for $V_{\tt{Har}}$ was compared to the analytical solution by evaluating:
\begin{equation}
    \frac{E_n}{\omega (n+1/2)}=\hbar=1~(\pm 0.1 \%)
\end{equation}
where $E_n$ is the $n^{th}$ eigenvalue and all eigenvalues deviated by more than 0.1\% were rejected.

The set of eigenvalues $\left\{E_{n}\right\}$ and eigenstates $\left\{\psi_{n}(x)\right\}$ determines the temperature-dependent probability density and Raman spectra.
If the canonical ensemble density operator is given by
\begin{equation}
    \rho(T)=\sum_{n}p_{n}(T)\left|\psi_{n}\rangle \langle \psi_{n}\right|
    =\sum_{n}\frac{e^{-E_{n}\beta}}{Z_{T}}\left|\psi_{n}\rangle \langle \psi_{n}\right|
\end{equation}
with $Z_{T}=\sum_{n}e^{-E_{n}\beta}$ the canonical partition function, and $\beta=(k_{B}T)^{-1}$,
the probability density is given by

\begin{equation}
\begin{split}
    \mathcal{P} (x) & = \tt{tr} \left( \rho (T) | x \rangle \langle x | \right) = \tt{tr} \left( \sum_{n} p_{n}(T) | \psi_{n} \rangle \langle \psi_{n} | x \rangle \langle x | \right) \\
    & = \int\tt{d}x'~ \left\langle x' \left| \sum_{n} p_{n} (T) \right| \psi_{n} \right\rangle \langle \psi_{n} |x \rangle \langle x|x' \rangle \\
    & = \int \tt{d}x'~ \left\langle x' \left| \sum_{n} p_{n} (T) \right| \psi_{n} \right\rangle \langle \psi_{n} | x \rangle \delta \left( x-x' \right) \\
    & = \sum_{n} p_{n} (T) \langle x | \psi_{n} \rangle \langle \psi_{n} | x \rangle  = \sum_{n} \frac{e^{-E_{n}\beta}}{Z_{T}} | \psi_{n} (x) |^{2}
\end{split}
\end{equation}

To calculate the Raman spectra we start from~\cite{Cowley1964}:
\begin{equation}
\label{SI_eq:raman}
    \mathcal{S}(\omega)\propto\int_{-\infty}^\infty\tt{d}\tau~ e^{-i\omega\tau}\langle \chi(\tau)^{*}\chi(0) \rangle_T
\end{equation}
with $\chi(\tau)$ the time-dependent polarizability operator and $\langle\cdot\rangle_T$ denoting a thermal average.
We have no information about the polarizability, but assume its time-dependence originates in the ionic motion $x(t)$. 
We therefore Taylor expand the polarizability in atomic displacement $x$:
\begin{equation}
    \chi(x)=\chi_{0}+\left(\frac{\tt{d}\chi}{\tt{d}x}\right)_{x=0}x+\left(\frac{\tt{d}^{2}\chi}{\tt{d}x^{2}}\right)_{x=0}x^{2}+\cdots.
\end{equation}
Plugging the expansion back into Eq.~(S19) we can isolate any scattering order we desire.
Due to the selection rules for cubic perovskite structure, we isolate the \nd -order contributions, to get:
\begin{equation}
\label{SI_eq:cross-section}
    \mathcal{S}_2(\omega) \propto \int \tt{d}t~ e^{-i\omega t} \avg{ x(t)x(t)x(0)x(0)}_{T}
\end{equation}
where the auto-correlation is now for displacements instead of polarizabilities.
Although our solutions are stationary, frequency-dependent spectra may be obtained through the constraint of energy conservation and the Fourier decomposition of a delta-function.
Using the Heisenberg representation for the location operator, $\hat{x}\left(\tau\right)=e^{i\frac{H\tau}{\hbar}}\hat{x}e^{-i\frac{H\tau}{\hbar}}$, Eq.~(S21) becomes:

\pagebreak

\begin{figure}[ht]
    \centering
    \includegraphics[width = 12.5 cm]{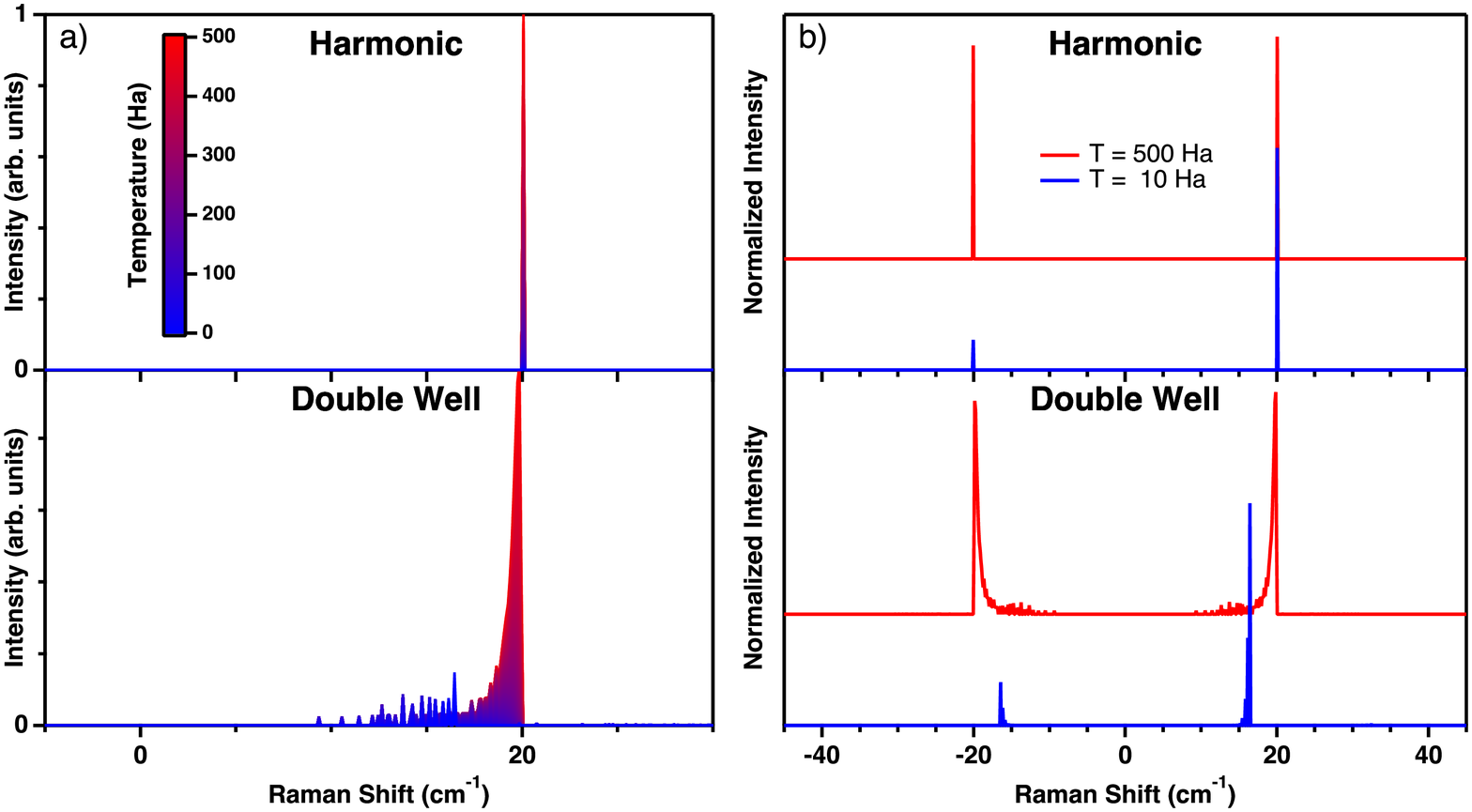}
    \caption{Simulated inelastic light scattering (Raman) spectra from a particle in a an harmonic (top panels) and a double-well (bottom panels) potential. a) All calculated spectra. b) Normalized spectra at $T=10~\tt{Ha}$ and $T=500~\tt{Ha}$ showing the difference in dominating features due to thermal population.}
    \label{fig:SI_NumerSimRaman}
\end{figure}

\begin{equation}
\begin{split}
    \int\tt{d}\tau e^{-i\omega\tau} & \avg{x(\tau)x(\tau)x(0)x(0)} _{T} = \\
    & = \int\tt{d}\tau e^{-i\omega\tau}\sum_{n}p_{n}(T)\left\langle n\left| U^\dagger (\tau) \hat{x} U(\tau) U^\dagger (\tau) \hat{x} U(\tau) \hat{x}\hat{x} \right|n\right\rangle \\
    & = \int\tt{d}\tau e^{-i\omega\tau}\sum_{n}p_{n}(T)\left\langle n\left| U^\dagger (\tau) \hat{x}\hat{x} U(\tau) \hat{x}\hat{x} \right|n\right\rangle \\
    & = \int\tt{d}\tau e^{-i\omega\tau}\sum_{n}p_{n}(T)\left\langle n\left| e^{i\frac{H\tau}{\hbar}}\hat{x}\ket{l} \braxket{l}{\hat{x}}{m} \bra{m}e^{-i\frac{H\tau}{\hbar}}\hat{x}\ket{k} \bra{k}\hat{x}\right|n\right\rangle \\
    & = \int\tt{d}\tau e^{-i\omega\tau}\sum_{n}p_{n}(T)\left\langle n\left| e^{i\frac{E_n\tau}{\hbar}}\hat{x}\ket{l} \braxket{l}{\hat{x}}{m} \bra{m}e^{-i\frac{E_m\tau}{\hbar}}\hat{x}\ket{k} \bra{k}\hat{x}\right|n\right\rangle \\
    & = \sum_{n,m,l,k}p_{n}(T) \braxket{n}{\hat{x}}{l} \braxket{l}{\hat{x}}{m} \braxket{m}{\hat{x}}{k} \braxket{k}{\hat{x}}{n} \int\tt{d}\tau e^{i\left(\omega_{nm}-\omega\right)\tau}\\
    & = 2\pi\sum_{n,m,l,k}p_{n}(T) \braxket{n}{\hat{x}}{l} \braxket{l}{\hat{x}}{m} \braxket{m}{\hat{x}}{k} \braxket{k}{\hat{x}}{n} \delta\left(\omega_{nm}-\omega\right)\\
    & = 2\pi\sum_{n,m}p_{n}(T) \braxket{n}{\hat{x}^2}{m} \braxket{m}{\hat{x}^2}{n}  \delta\left(\omega_{nm}-\omega\right)
\end{split}
\end{equation}
with $\omega_{nm}\equiv\frac{E_{n}-E_{m}}{\hbar}$.

The matrix elements can now be directly evaluated by
\begin{equation}
    \braxket{n}{\hat{x}^2}{m} = \int_{x_{\tt{min}}}^{x_{\tt{max}}}\tt{d}x~ \psi_m^{*}(x)\cdot x^2 \cdot \psi_n(x)
\end{equation}
and the delta function is numerically approximated by a square function of a single pixel in our $x$-axis.

Finally, the integrated intensities are given by 
\begin{equation}
    I=\int_0^{\omega_{\tt{max}}}\mathcal{S}_2(\omega)~\tt{d}\omega.
\end{equation}

    \begin{figure}[t]
        \centering
        \includegraphics[width = 14 cm]{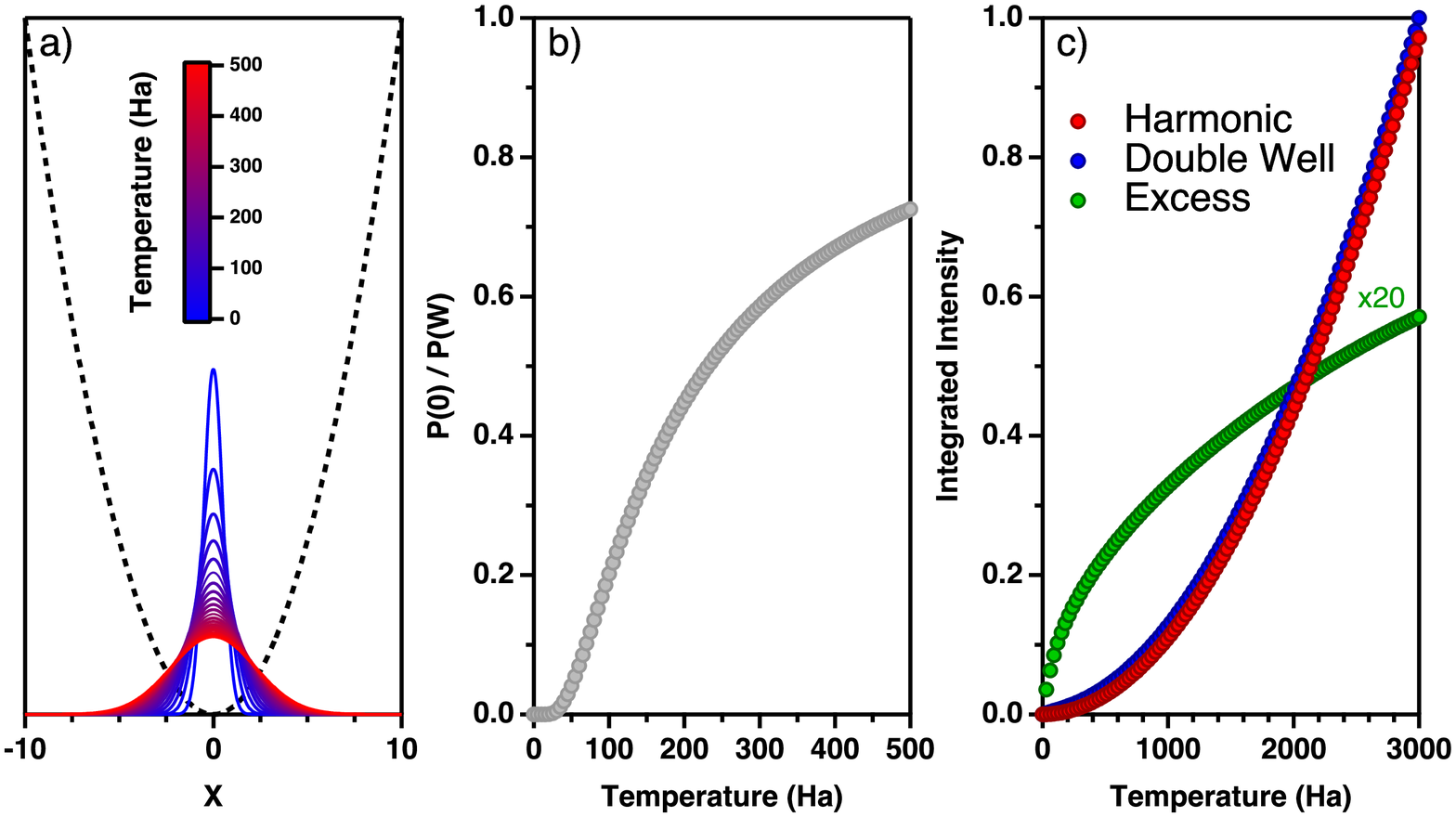}
        \caption{a) Probability densities of finding the particle in a parabolic potential as function of temperature showing the expected Guassian probability ($x_{\tt{max}}=16$). 
        b) The ratio between the probability density at the barrier ($P(x=0)$) and the probability density at the bottom of the well ($P(W)$) under a double-well potential. The ratio is small at low temperatures, when the particle is confined to the wells, and increases with temperature as the probability density broadens.
        c) Integrated intensity of the simulated \nd -order Raman spectra as function of temperature, at higher temperature range, under parabolic (red) and double-well (blue) potentials, showing an increase in the excess intensity (green) with temperature.}
        \label{fig:SI_NumerSimExtra}
    \end{figure}

Figure~S5(a) shows the simulated \nd -order Raman spectra as a function of temperature for the harmonic (top panel) and DW (bottom panel) potential. 
The Raman spectra of the harmonic potential exhibit a single peak at $2\omega = 20~cm^{-1}$ while the Raman spectra of the DW potential exhibit many transitions around $2\omega$ as well as low frequency features.
Figure~S5(b) shows the normalized simulated \nd -order Raman spectra at $T=10~\tt{Ha}$ and $T=500~\tt{Ha}$, emphasizing the dominant features in each spectrum.
Clearly, in the harmonic potential, the dominant feature remains the $2\omega$ peak at all temperature range, while in the DW potential the low frequency features dominate at low temperatures.

Figure~S6(a) shows the probability density to find the particle in the harmonic potential.
As expected, the probability density is a Gaussian distribution around the potential well which decays at the edges ($x_{\tt{max}}=16$) and broadens with temperatures.
The ratio between the probability density at the top of the barrier and the bottom of the well ($r_\mathcal{P}=\mathcal{P}(0)/\mathcal{P}(W)$), as function of temperature,  is depicted in Fig.~S6(b).
$r_\mathcal{P}$ correlates very well with the DW integrated intensity, as depicted in Fig.~2 in the main text.
At low temperatures, $r_\mathcal{P}$ is very low and $\mathcal{P}(0)$ is negligible, and increases monotonically with temperature, approaching a constant value.
At the same low temperatures range, the DW integrated intensity exhibits a sharp increase, and aligns parallel to the harmonic integrated intensity with increasing temperature.
This can be understood in the context of spontaneous symmetry breaking: at low temperatures, when the particle in confined to the wells, only transitions which are forbidden in the harmonic case are activated and contribute to the integrated intensity. 
As temperature increases, the particle feels an harmonic-like potential.
Harmonically allowed transitions are thermally activated and their relative contribution increases rapidly, resulting in an harmonic-like integrated intensity.
    
The monotonic increase in integrated intensity continues at higher temperatures.
Figure~S6(c) shows the temperature dependent integrated intensity of the simulated \nd -order Raman spectra under an harmonic (red trace) and DW (blue trace) potential and their difference (green trace), at a higher temperatures.
Clearly, the excess intensity increases monotonically with temperature.
The higher temperatures simulation was performed with a 1D grid of 12001 points, to keep the nominal thermal contribution of high-energy eigenstates negligible and a similar resolution in $x$.

  \begin{figure}[ht]
        \centering
        \includegraphics[width = 14 cm]{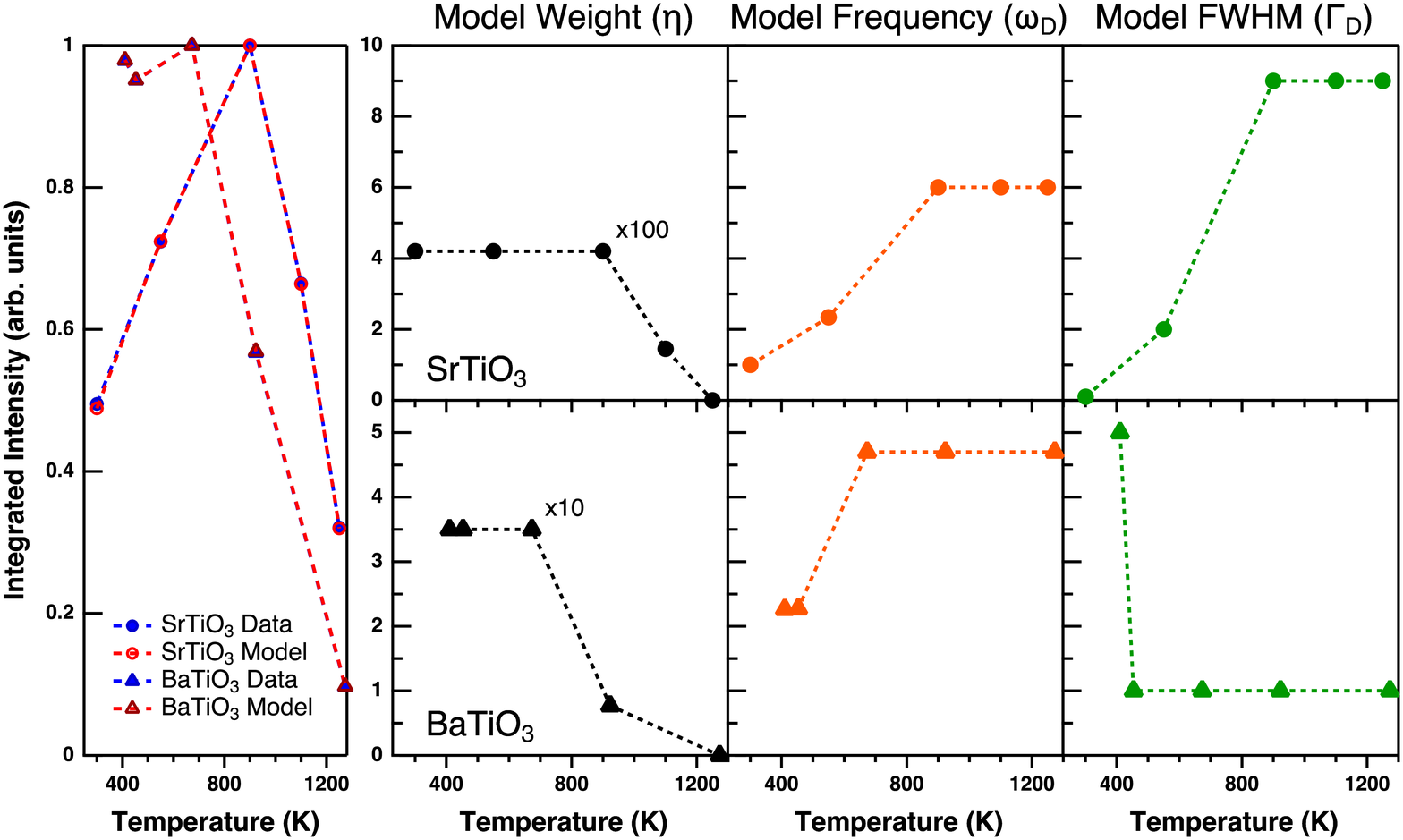}
        \caption{a) Integrated intensity of \STO\ and \BTO\ as function of temperature compared to the integrated intensity from the fit. A good fit was defined according to the match of the integrated intensities at all temperatures. b) Fit result parameters of \STO\ (top) and \BTO\ (bottom) as function of temperature.
        The dashed lines are guides to the eye.}
        \label{fig:SI_SBTOTrends}
    \end{figure}

\section{\label{secSI_Fitting}Fit results}

Figures~S7 and S8 show the parameters from fitting the temperature dependent Raman spectra of \BTO, \STO\ and \CPB\ (Fig.~1 in the main text) to Eq.~(14) in the main text.
The fit was done by matching both the integrated intensity and the spectral line-shape.
The results reflect the chosen constraints to the fit, where $\eta$ was kept constant in the range of $T_c < T \leq T_m$, and was the only free fit parameter for $T>T_m$.

   \begin{figure}[h!]
        \centering
        \includegraphics[width = 15 cm]{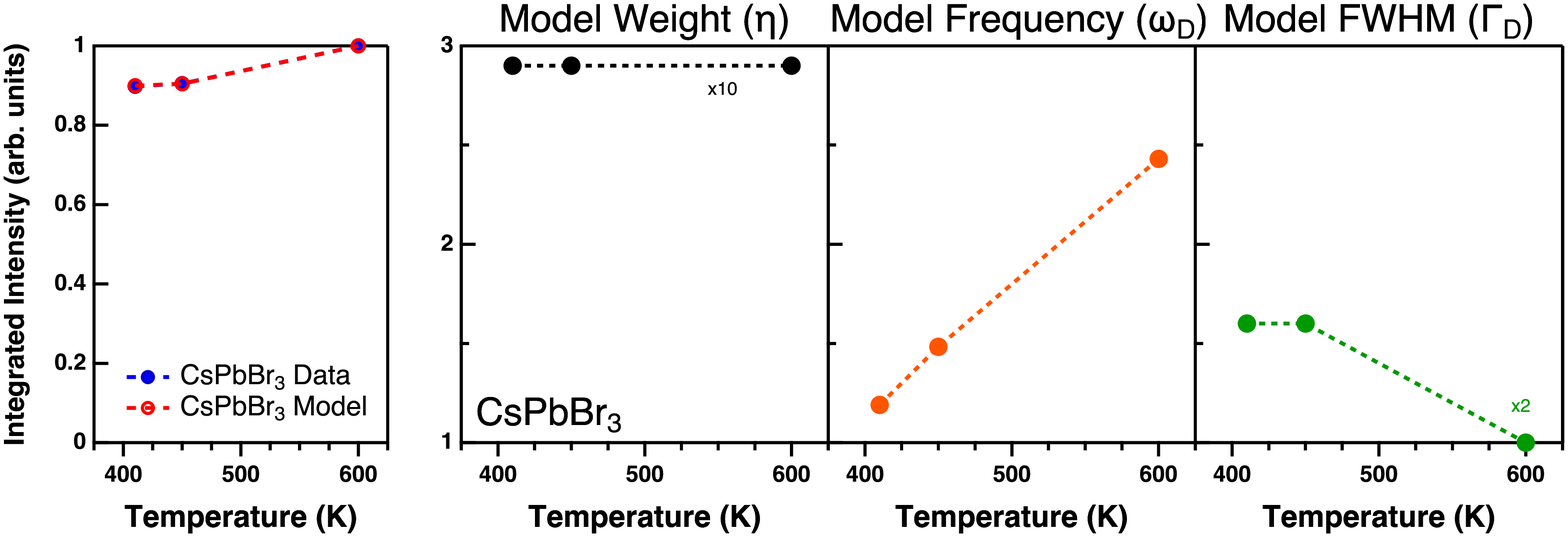}
        \caption{a) Integrated intensity of \CPB\ as function of temperature compared to the integrated intensity from the fit. A good fit was defined according to the match of the integrated intensities at all temperatures. b) Fit result parameters of \CPB\ as function of temperature.
        The dashed lines are guides to the eye.}
        \label{fig:SI_CPBTrends}
    \end{figure}


\twocolumngrid

\clearpage
\bibliography{Bibliography.bib}

\begin{thebibliography}{96}%
\makeatletter
\providecommand \@ifxundefined [1]{%
 \@ifx{#1\undefined}
}%
\providecommand \@ifnum [1]{%
 \ifnum #1\expandafter \@firstoftwo
 \else \expandafter \@secondoftwo
 \fi
}%
\providecommand \@ifx [1]{%
 \ifx #1\expandafter \@firstoftwo
 \else \expandafter \@secondoftwo
 \fi
}%
\providecommand \natexlab [1]{#1}%
\providecommand \enquote  [1]{``#1''}%
\providecommand \bibnamefont  [1]{#1}%
\providecommand \bibfnamefont [1]{#1}%
\providecommand \citenamefont [1]{#1}%
\providecommand \href@noop [0]{\@secondoftwo}%
\providecommand \href [0]{\begingroup \@sanitize@url \@href}%
\providecommand \@href[1]{\@@startlink{#1}\@@href}%
\providecommand \@@href[1]{\endgroup#1\@@endlink}%
\providecommand \@sanitize@url [0]{\catcode `\\12\catcode `\$12\catcode
  `\&12\catcode `\#12\catcode `\^12\catcode `\_12\catcode `\%12\relax}%
\providecommand \@@startlink[1]{}%
\providecommand \@@endlink[0]{}%
\providecommand \url  [0]{\begingroup\@sanitize@url \@url }%
\providecommand \@url [1]{\endgroup\@href {#1}{\urlprefix }}%
\providecommand \urlprefix  [0]{URL }%
\providecommand \Eprint [0]{\href }%
\providecommand \doibase [0]{https://doi.org/}%
\providecommand \selectlanguage [0]{\@gobble}%
\providecommand \bibinfo  [0]{\@secondoftwo}%
\providecommand \bibfield  [0]{\@secondoftwo}%
\providecommand \translation [1]{[#1]}%
\providecommand \BibitemOpen [0]{}%
\providecommand \bibitemStop [0]{}%
\providecommand \bibitemNoStop [0]{.\EOS\space}%
\providecommand \EOS [0]{\spacefactor3000\relax}%
\providecommand \BibitemShut  [1]{\csname bibitem#1\endcsname}%
\let\auto@bib@innerbib\@empty
\bibitem [{\citenamefont {Kamba}(2021)}]{Kamba2021}%
  \BibitemOpen
  \bibfield  {author} {\bibinfo {author} {\bibfnamefont {S.}~\bibnamefont
  {Kamba}},\ }\bibfield  {title} {\bibinfo {title} {{Soft-mode spectroscopy of
  ferroelectrics and multiferroics: A review}},\ }\href
  {https://doi.org/10.1063/5.0036066} {\bibfield  {journal} {\bibinfo
  {journal} {APL Materials}\ }\textbf {\bibinfo {volume} {9}},\ \bibinfo
  {pages} {20704} (\bibinfo {year} {2021})}\BibitemShut {NoStop}%
\bibitem [{\citenamefont {Cowley}\ and\ \citenamefont
  {Shapiro}(2006)}]{Cowley2006}%
  \BibitemOpen
  \bibfield  {author} {\bibinfo {author} {\bibfnamefont {R.~A.}\ \bibnamefont
  {Cowley}}\ and\ \bibinfo {author} {\bibfnamefont {S.~M.}\ \bibnamefont
  {Shapiro}},\ }\bibfield  {title} {\bibinfo {title} {{Structural phase
  transitions}},\ }\bibfield  {journal} {\bibinfo  {journal} {Journal of the
  Physical Society of Japan}\ }\textbf {\bibinfo {volume} {75}},\ \href
  {https://doi.org/10.1143/JPSJ.75.111001} {10.1143/JPSJ.75.111001} (\bibinfo
  {year} {2006})\BibitemShut {NoStop}%
\bibitem [{\citenamefont {Miyata}\ \emph {et~al.}(2017)\citenamefont {Miyata},
  \citenamefont {Atallah},\ and\ \citenamefont {Zhu}}]{Miyata2017}%
  \BibitemOpen
  \bibfield  {author} {\bibinfo {author} {\bibfnamefont {K.}~\bibnamefont
  {Miyata}}, \bibinfo {author} {\bibfnamefont {T.~L.}\ \bibnamefont
  {Atallah}},\ and\ \bibinfo {author} {\bibfnamefont {X.~Y.}\ \bibnamefont
  {Zhu}},\ }\bibfield  {title} {\bibinfo {title} {{Lead halide perovskites:
  Crystal-liquid duality, phonon glass electron crystals, and large polaron
  formation}},\ }\href {https://doi.org/10.1126/sciadv.1701469} {\bibfield
  {journal} {\bibinfo  {journal} {Science Advances}\ }\textbf {\bibinfo
  {volume} {3}},\ \bibinfo {pages} {1} (\bibinfo {year} {2017})}\BibitemShut
  {NoStop}%
\bibitem [{\citenamefont {Mitzi}(1999)}]{Mitzi1999}%
  \BibitemOpen
  \bibfield  {author} {\bibinfo {author} {\bibfnamefont {D.~B.}\ \bibnamefont
  {Mitzi}},\ }\href {https://doi.org/10.1002/9780470166499.ch1} {\emph
  {\bibinfo {title} {{Synthesis, Structure, and Properties of Organic-Inorganic
  Perovskites and Related Materials}}}}\ (\bibinfo  {publisher} {John Wiley \&
  Sons, Ltd},\ \bibinfo {year} {1999})\ pp.\ \bibinfo {pages}
  {1--121}\BibitemShut {NoStop}%
\bibitem [{\citenamefont {Lanigan-Atkins}\ \emph {et~al.}(2021)\citenamefont
  {Lanigan-Atkins}, \citenamefont {He}, \citenamefont {Krogstad}, \citenamefont
  {Pajerowski}, \citenamefont {Abernathy}, \citenamefont {Xu}, \citenamefont
  {Xu}, \citenamefont {Chung}, \citenamefont {Kanatzidis}, \citenamefont
  {Rosenkranz}, \citenamefont {Osborn},\ and\ \citenamefont
  {Delaire}}]{Lanigan-Atkins2021}%
  \BibitemOpen
  \bibfield  {author} {\bibinfo {author} {\bibfnamefont {T.}~\bibnamefont
  {Lanigan-Atkins}}, \bibinfo {author} {\bibfnamefont {X.}~\bibnamefont {He}},
  \bibinfo {author} {\bibfnamefont {M.~J.}\ \bibnamefont {Krogstad}}, \bibinfo
  {author} {\bibfnamefont {D.~M.}\ \bibnamefont {Pajerowski}}, \bibinfo
  {author} {\bibfnamefont {D.~L.}\ \bibnamefont {Abernathy}}, \bibinfo {author}
  {\bibfnamefont {G.~N.}\ \bibnamefont {Xu}}, \bibinfo {author} {\bibfnamefont
  {Z.}~\bibnamefont {Xu}}, \bibinfo {author} {\bibfnamefont {D.~Y.}\
  \bibnamefont {Chung}}, \bibinfo {author} {\bibfnamefont {M.~G.}\ \bibnamefont
  {Kanatzidis}}, \bibinfo {author} {\bibfnamefont {S.}~\bibnamefont
  {Rosenkranz}}, \bibinfo {author} {\bibfnamefont {R.}~\bibnamefont {Osborn}},\
  and\ \bibinfo {author} {\bibfnamefont {O.}~\bibnamefont {Delaire}},\
  }\bibfield  {title} {\bibinfo {title} {{Two-dimensional overdamped
  fluctuations of the soft perovskite lattice in CsPbBr$_3$}},\ }\href
  {https://doi.org/10.1038/s41563-021-00947-y} {\bibfield  {journal} {\bibinfo
  {journal} {Nature Materials}\ }\textbf {\bibinfo {volume} {20}},\ \bibinfo
  {pages} {977} (\bibinfo {year} {2021})}\BibitemShut {NoStop}%
\bibitem [{\citenamefont {Wright}\ \emph {et~al.}(2016)\citenamefont {Wright},
  \citenamefont {Verdi}, \citenamefont {Milot}, \citenamefont {Eperon},
  \citenamefont {P{\'{e}}rez-Osorio}, \citenamefont {Snaith}, \citenamefont
  {Giustino}, \citenamefont {Johnston},\ and\ \citenamefont
  {Herz}}]{Wright2016}%
  \BibitemOpen
  \bibfield  {author} {\bibinfo {author} {\bibfnamefont {A.~D.}\ \bibnamefont
  {Wright}}, \bibinfo {author} {\bibfnamefont {C.}~\bibnamefont {Verdi}},
  \bibinfo {author} {\bibfnamefont {R.~L.}\ \bibnamefont {Milot}}, \bibinfo
  {author} {\bibfnamefont {G.~E.}\ \bibnamefont {Eperon}}, \bibinfo {author}
  {\bibfnamefont {M.~A.}\ \bibnamefont {P{\'{e}}rez-Osorio}}, \bibinfo {author}
  {\bibfnamefont {H.~J.}\ \bibnamefont {Snaith}}, \bibinfo {author}
  {\bibfnamefont {F.}~\bibnamefont {Giustino}}, \bibinfo {author}
  {\bibfnamefont {M.~B.}\ \bibnamefont {Johnston}},\ and\ \bibinfo {author}
  {\bibfnamefont {L.~M.}\ \bibnamefont {Herz}},\ }\bibfield  {title} {\bibinfo
  {title} {{Electron–phonon coupling in hybrid lead halide perovskites}},\
  }\bibfield  {journal} {\bibinfo  {journal} {Nature Communications}\ }\textbf
  {\bibinfo {volume} {7}},\ \href {https://doi.org/10.1038/ncomms11755}
  {10.1038/ncomms11755} (\bibinfo {year} {2016})\BibitemShut {NoStop}%
\bibitem [{\citenamefont {Munson}\ \emph {et~al.}(2018)\citenamefont {Munson},
  \citenamefont {Kennehan}, \citenamefont {Doucette},\ and\ \citenamefont
  {Asbury}}]{Munson2018}%
  \BibitemOpen
  \bibfield  {author} {\bibinfo {author} {\bibfnamefont {K.~T.}\ \bibnamefont
  {Munson}}, \bibinfo {author} {\bibfnamefont {E.~R.}\ \bibnamefont
  {Kennehan}}, \bibinfo {author} {\bibfnamefont {G.~S.}\ \bibnamefont
  {Doucette}},\ and\ \bibinfo {author} {\bibfnamefont {J.~B.}\ \bibnamefont
  {Asbury}},\ }\bibfield  {title} {\bibinfo {title} {{Dynamic Disorder
  Dominates Delocalization, Transport, and Recombination in Halide
  Perovskites}},\ }\href {https://doi.org/10.1016/j.chempr.2018.09.001}
  {\bibfield  {journal} {\bibinfo  {journal} {Chem}\ }\textbf {\bibinfo
  {volume} {4}},\ \bibinfo {pages} {2826} (\bibinfo {year} {2018})}\BibitemShut
  {NoStop}%
\bibitem [{\citenamefont {Migoni}\ \emph {et~al.}(1976)\citenamefont {Migoni},
  \citenamefont {Bilz},\ and\ \citenamefont {B{\"{a}}uerle}}]{Migoni1976}%
  \BibitemOpen
  \bibfield  {author} {\bibinfo {author} {\bibfnamefont {R.}~\bibnamefont
  {Migoni}}, \bibinfo {author} {\bibfnamefont {H.}~\bibnamefont {Bilz}},\ and\
  \bibinfo {author} {\bibfnamefont {D.}~\bibnamefont {B{\"{a}}uerle}},\
  }\bibfield  {title} {\bibinfo {title} {{Origin of Raman scattering and
  ferroelectricity in oxidic perovskites}},\ }\href
  {https://doi.org/10.1103/PhysRevLett.37.1155} {\bibfield  {journal} {\bibinfo
   {journal} {Physical Review Letters}\ }\textbf {\bibinfo {volume} {37}},\
  \bibinfo {pages} {1155} (\bibinfo {year} {1976})}\BibitemShut {NoStop}%
\bibitem [{\citenamefont {Fontana}\ \emph {et~al.}(1991)\citenamefont
  {Fontana}, \citenamefont {Idrissi}, \citenamefont {Kugel},\ and\
  \citenamefont {Wojcik}}]{Fontana1991}%
  \BibitemOpen
  \bibfield  {author} {\bibinfo {author} {\bibfnamefont {M.~D.}\ \bibnamefont
  {Fontana}}, \bibinfo {author} {\bibfnamefont {H.}~\bibnamefont {Idrissi}},
  \bibinfo {author} {\bibfnamefont {G.~E.}\ \bibnamefont {Kugel}},\ and\
  \bibinfo {author} {\bibfnamefont {K.}~\bibnamefont {Wojcik}},\ }\bibfield
  {title} {\bibinfo {title} {{Raman spectrum in PbTiO$_3$ re-examined: Dynamics
  of the soft phonon and the central peak}},\ }\href
  {https://doi.org/10.1088/0953-8984/3/44/014} {\bibfield  {journal} {\bibinfo
  {journal} {Journal of Physics: Condensed Matter}\ }\textbf {\bibinfo {volume}
  {3}},\ \bibinfo {pages} {8695} (\bibinfo {year} {1991})}\BibitemShut
  {NoStop}%
\bibitem [{\citenamefont {Menahem}\ \emph {et~al.}(2021)\citenamefont
  {Menahem}, \citenamefont {Dai}, \citenamefont {Aharon}, \citenamefont
  {Sharma}, \citenamefont {Asher}, \citenamefont {Diskin-Posner}, \citenamefont
  {Korobko}, \citenamefont {Rappe},\ and\ \citenamefont {Yaffe}}]{Menahem2021}%
  \BibitemOpen
  \bibfield  {author} {\bibinfo {author} {\bibfnamefont {M.}~\bibnamefont
  {Menahem}}, \bibinfo {author} {\bibfnamefont {Z.}~\bibnamefont {Dai}},
  \bibinfo {author} {\bibfnamefont {S.}~\bibnamefont {Aharon}}, \bibinfo
  {author} {\bibfnamefont {R.}~\bibnamefont {Sharma}}, \bibinfo {author}
  {\bibfnamefont {M.}~\bibnamefont {Asher}}, \bibinfo {author} {\bibfnamefont
  {Y.}~\bibnamefont {Diskin-Posner}}, \bibinfo {author} {\bibfnamefont
  {R.}~\bibnamefont {Korobko}}, \bibinfo {author} {\bibfnamefont {A.~M.}\
  \bibnamefont {Rappe}},\ and\ \bibinfo {author} {\bibfnamefont
  {O.}~\bibnamefont {Yaffe}},\ }\bibfield  {title} {\bibinfo {title} {{Strongly
  Anharmonic Octahedral Tilting in Two-Dimensional Hybrid Halide
  Perovskites}},\ }\href {https://doi.org/10.1021/acsnano.1c02022} {\bibfield
  {journal} {\bibinfo  {journal} {ACS Nano}\ }\textbf {\bibinfo {volume}
  {15}},\ \bibinfo {pages} {10153} (\bibinfo {year} {2021})}\BibitemShut
  {NoStop}%
\bibitem [{\citenamefont {Sharma}\ \emph {et~al.}(2020)\citenamefont {Sharma},
  \citenamefont {Menahem}, \citenamefont {Dai}, \citenamefont {Gao},
  \citenamefont {Korobko}, \citenamefont {Pinkas}, \citenamefont {Rappe},
  \citenamefont {Yaffe}, \citenamefont {Brenner}, \citenamefont {Yadgarov},
  \citenamefont {Zhang}, \citenamefont {Rakita}, \citenamefont {Korobko},
  \citenamefont {Pinkas}, \citenamefont {Rappe},\ and\ \citenamefont
  {Yaffe}}]{SharmaMAPI2}%
  \BibitemOpen
  \bibfield  {author} {\bibinfo {author} {\bibfnamefont {R.}~\bibnamefont
  {Sharma}}, \bibinfo {author} {\bibfnamefont {M.}~\bibnamefont {Menahem}},
  \bibinfo {author} {\bibfnamefont {Z.}~\bibnamefont {Dai}}, \bibinfo {author}
  {\bibfnamefont {L.}~\bibnamefont {Gao}}, \bibinfo {author} {\bibfnamefont
  {R.}~\bibnamefont {Korobko}}, \bibinfo {author} {\bibfnamefont
  {I.}~\bibnamefont {Pinkas}}, \bibinfo {author} {\bibfnamefont {A.~M.}\
  \bibnamefont {Rappe}}, \bibinfo {author} {\bibfnamefont {O.}~\bibnamefont
  {Yaffe}}, \bibinfo {author} {\bibfnamefont {T.~M.}\ \bibnamefont {Brenner}},
  \bibinfo {author} {\bibfnamefont {L.}~\bibnamefont {Yadgarov}}, \bibinfo
  {author} {\bibfnamefont {J.}~\bibnamefont {Zhang}}, \bibinfo {author}
  {\bibfnamefont {Y.}~\bibnamefont {Rakita}}, \bibinfo {author} {\bibfnamefont
  {R.}~\bibnamefont {Korobko}}, \bibinfo {author} {\bibfnamefont
  {I.}~\bibnamefont {Pinkas}}, \bibinfo {author} {\bibfnamefont {A.~M.}\
  \bibnamefont {Rappe}},\ and\ \bibinfo {author} {\bibfnamefont
  {O.}~\bibnamefont {Yaffe}},\ }\bibfield  {title} {\bibinfo {title} {{Lattice
  mode symmetry analysis of the orthorhombic phase of methylammonium lead
  iodide using polarized Raman}},\ }\href
  {https://doi.org/10.1103/PhysRevMaterials.4.051601} {\bibfield  {journal}
  {\bibinfo  {journal} {Physical Review Materials}\ }\textbf {\bibinfo {volume}
  {4}},\ \bibinfo {pages} {1} (\bibinfo {year} {2020})},\ \Eprint
  {https://arxiv.org/abs/1912.00363} {arXiv:1912.00363} \BibitemShut {NoStop}%
\bibitem [{\citenamefont {Cohen}\ \emph {et~al.}(2022)\citenamefont {Cohen},
  \citenamefont {Brenner}, \citenamefont {Klarbring}, \citenamefont {Sharma},
  \citenamefont {Fabini}, \citenamefont {Korobko}, \citenamefont {Nayak},
  \citenamefont {Hellman},\ and\ \citenamefont {Yaffe}}]{Cohen2022}%
  \BibitemOpen
  \bibfield  {author} {\bibinfo {author} {\bibfnamefont {A.}~\bibnamefont
  {Cohen}}, \bibinfo {author} {\bibfnamefont {T.~M.}\ \bibnamefont {Brenner}},
  \bibinfo {author} {\bibfnamefont {J.}~\bibnamefont {Klarbring}}, \bibinfo
  {author} {\bibfnamefont {R.}~\bibnamefont {Sharma}}, \bibinfo {author}
  {\bibfnamefont {D.~H.}\ \bibnamefont {Fabini}}, \bibinfo {author}
  {\bibfnamefont {R.}~\bibnamefont {Korobko}}, \bibinfo {author} {\bibfnamefont
  {P.~K.}\ \bibnamefont {Nayak}}, \bibinfo {author} {\bibfnamefont
  {O.}~\bibnamefont {Hellman}},\ and\ \bibinfo {author} {\bibfnamefont
  {O.}~\bibnamefont {Yaffe}},\ }\bibfield  {title} {\bibinfo {title} {Diverging
  expressions of anharmonicity in halide perovskites},\ }\href
  {https://doi.org/https://doi.org/10.1002/adma.202107932} {\bibfield
  {journal} {\bibinfo  {journal} {Advanced Materials}\ }\textbf {\bibinfo
  {volume} {34}},\ \bibinfo {pages} {2107932} (\bibinfo {year}
  {2022})}\BibitemShut {NoStop}%
\bibitem [{\citenamefont {Bhagavantam}\ and\ \citenamefont
  {Venkatarayudu}(1931)}]{Bhagavantam1931}%
  \BibitemOpen
  \bibfield  {author} {\bibinfo {author} {\bibfnamefont {S.}~\bibnamefont
  {Bhagavantam}}\ and\ \bibinfo {author} {\bibfnamefont {T.}~\bibnamefont
  {Venkatarayudu}},\ }\bibfield  {title} {\bibinfo {title} {{Raman effect in
  relation to crystal structure}},\ }\href {https://doi.org/10.1007/BF03172556}
  {\bibfield  {journal} {\bibinfo  {journal} {Proc. Indian Acad. Sci. - Sect.
  A}\ }\textbf {\bibinfo {volume} {9}},\ \bibinfo {pages} {224} (\bibinfo
  {year} {1931})}\BibitemShut {NoStop}%
\bibitem [{\citenamefont {Cowley1968}(1968)}]{Cowley1968}%
  \BibitemOpen
  \bibfield  {author} {\bibinfo {author} {\bibnamefont {Cowley1968}},\
  }\bibfield  {title} {\bibinfo {title} {{Anharmonic crystals}},\ }\href
  {https://doi.org/10.1088/0034-4885/31/1/303} {\bibfield  {journal} {\bibinfo
  {journal} {Reports Prog. Phys.}\ }\textbf {\bibinfo {volume} {31}},\ \bibinfo
  {pages} {123} (\bibinfo {year} {1968})}\BibitemShut {NoStop}%
\bibitem [{\citenamefont {Saksena}(1940)}]{Saksena1940}%
  \BibitemOpen
  \bibfield  {author} {\bibinfo {author} {\bibfnamefont {B.~D.}\ \bibnamefont
  {Saksena}},\ }\bibfield  {title} {\bibinfo {title} {{Raman effect and crystal
  symmetry}},\ }\href {https://doi.org/10.1007/BF03046551} {\bibfield
  {journal} {\bibinfo  {journal} {Proc. Indian Acad. Sci. - Sect. A}\ }\textbf
  {\bibinfo {volume} {11}},\ \bibinfo {pages} {229} (\bibinfo {year}
  {1940})}\BibitemShut {NoStop}%
\bibitem [{\citenamefont {Fontana}\ \emph {et~al.}(2020)\citenamefont
  {Fontana}, \citenamefont {Kokanyan},\ and\ \citenamefont
  {Kauffmann}}]{Fontana2020}%
  \BibitemOpen
  \bibfield  {author} {\bibinfo {author} {\bibfnamefont {M.~D.}\ \bibnamefont
  {Fontana}}, \bibinfo {author} {\bibfnamefont {N.}~\bibnamefont {Kokanyan}},\
  and\ \bibinfo {author} {\bibfnamefont {T.~H.}\ \bibnamefont {Kauffmann}},\
  }\bibfield  {title} {\bibinfo {title} {{Sub-THz Raman response in BaTiO$_3$
  and link with structural phase transition}},\ }\href
  {https://doi.org/10.1088/1361-648X/ab808e} {\bibfield  {journal} {\bibinfo
  {journal} {Journal of Physics Condensed Matter}\ }\textbf {\bibinfo {volume}
  {32}},\ \bibinfo {pages} {285403} (\bibinfo {year} {2020})}\BibitemShut
  {NoStop}%
\bibitem [{\citenamefont {Perry}\ \emph {et~al.}(1967)\citenamefont {Perry},
  \citenamefont {Fertel},\ and\ \citenamefont {McNelly}}]{Perry1967}%
  \BibitemOpen
  \bibfield  {author} {\bibinfo {author} {\bibfnamefont {C.~H.}\ \bibnamefont
  {Perry}}, \bibinfo {author} {\bibfnamefont {J.~H.}\ \bibnamefont {Fertel}},\
  and\ \bibinfo {author} {\bibfnamefont {T.~F.}\ \bibnamefont {McNelly}},\
  }\bibfield  {title} {\bibinfo {title} {{Temperature dependence of the Raman
  spectrum of SrTiO$_3$ and KTaO$_3$}},\ }\href
  {https://doi.org/10.1063/1.1712142} {\bibfield  {journal} {\bibinfo
  {journal} {The Journal of Chemical Physics}\ }\textbf {\bibinfo {volume}
  {47}},\ \bibinfo {pages} {1619} (\bibinfo {year} {1967})}\BibitemShut
  {NoStop}%
\bibitem [{\citenamefont {Cowley}(1964{\natexlab{a}})}]{Cowley1964b}%
  \BibitemOpen
  \bibfield  {author} {\bibinfo {author} {\bibfnamefont {R.~A.}\ \bibnamefont
  {Cowley}},\ }\bibfield  {title} {\bibinfo {title} {{Lattice Dynamics and
  Phase Transitions of Strontium Titanate}},\ }\href
  {https://doi.org/10.1103/PhysRev.134.A981} {\bibfield  {journal} {\bibinfo
  {journal} {Phys. Rev.}\ }\textbf {\bibinfo {volume} {134}},\ \bibinfo {pages}
  {401} (\bibinfo {year} {1964}{\natexlab{a}})}\BibitemShut {NoStop}%
\bibitem [{\citenamefont {Fontana}\ \emph {et~al.}(1990)\citenamefont
  {Fontana}, \citenamefont {Idrissi},\ and\ \citenamefont
  {Wojcik}}]{Fontana1990}%
  \BibitemOpen
  \bibfield  {author} {\bibinfo {author} {\bibfnamefont {M.~D.}\ \bibnamefont
  {Fontana}}, \bibinfo {author} {\bibfnamefont {H.}~\bibnamefont {Idrissi}},\
  and\ \bibinfo {author} {\bibfnamefont {K.}~\bibnamefont {Wojcik}},\
  }\bibfield  {title} {\bibinfo {title} {{Displacive to order disorder
  crossover in the cubic tetragonal phase transition of PbTiO$_3$}},\ }\href
  {https://doi.org/10.1209/0295-5075/11/5/006} {\bibfield  {journal} {\bibinfo
  {journal} {EPL}\ }\textbf {\bibinfo {volume} {11}},\ \bibinfo {pages} {419}
  (\bibinfo {year} {1990})}\BibitemShut {NoStop}%
\bibitem [{\citenamefont {Born}\ and\ \citenamefont
  {Bradburn}(1947)}]{Born1947}%
  \BibitemOpen
  \bibfield  {author} {\bibinfo {author} {\bibfnamefont {M.}~\bibnamefont
  {Born}}\ and\ \bibinfo {author} {\bibfnamefont {M.}~\bibnamefont
  {Bradburn}},\ }\bibfield  {title} {\bibinfo {title} {{The theory of the Raman
  effect in crystals, in particular rocksalt.}},\ }\href
  {https://doi.org/10.1098/rspa.1947.0002} {\bibfield  {journal} {\bibinfo
  {journal} {Proceedings of the Royal Society A}\ }\textbf {\bibinfo {volume}
  {188}},\ \bibinfo {pages} {161} (\bibinfo {year} {1947})}\BibitemShut
  {NoStop}%
\bibitem [{\citenamefont {Johnson}\ and\ \citenamefont
  {Loudon}(1964)}]{Johnson1964}%
  \BibitemOpen
  \bibfield  {author} {\bibinfo {author} {\bibfnamefont {F.~A.}\ \bibnamefont
  {Johnson}}\ and\ \bibinfo {author} {\bibfnamefont {R.}~\bibnamefont
  {Loudon}},\ }\bibfield  {title} {\bibinfo {title} {{Critical-point analysis
  of the phonon spectra of diamond, silicon and germanium}},\ }\href
  {https://doi.org/https://doi.org/10.1098/rspa.1964.0182} {\bibfield
  {journal} {\bibinfo  {journal} {Proceedings of the Royal Society A}\ }\textbf
  {\bibinfo {volume} {281}},\ \bibinfo {pages} {274} (\bibinfo {year}
  {1964})}\BibitemShut {NoStop}%
\bibitem [{\citenamefont {Burstein}\ \emph {et~al.}(1965)\citenamefont
  {Burstein}, \citenamefont {Johnson},\ and\ \citenamefont
  {Loudon}}]{Burstein1965}%
  \BibitemOpen
  \bibfield  {author} {\bibinfo {author} {\bibfnamefont {E.}~\bibnamefont
  {Burstein}}, \bibinfo {author} {\bibfnamefont {F.~A.}\ \bibnamefont
  {Johnson}},\ and\ \bibinfo {author} {\bibfnamefont {R.}~\bibnamefont
  {Loudon}},\ }\bibfield  {title} {\bibinfo {title} {{Selection rules for
  second-order infrared and raman processes in the rocksalt structure and
  interpretation of the Raman spectra of NaCl, KBr, and NaI}},\ }\bibfield
  {journal} {\bibinfo  {journal} {Physical Review}\ }\textbf {\bibinfo {volume}
  {139}},\ \href {https://doi.org/10.1103/PhysRev.139.A1239}
  {10.1103/PhysRev.139.A1239} (\bibinfo {year} {1965})\BibitemShut {NoStop}%
\bibitem [{\citenamefont {Fontana}\ and\ \citenamefont
  {Lambert}(1972)}]{Fontana1972}%
  \BibitemOpen
  \bibfield  {author} {\bibinfo {author} {\bibfnamefont {M.~P.}\ \bibnamefont
  {Fontana}}\ and\ \bibinfo {author} {\bibfnamefont {M.}~\bibnamefont
  {Lambert}},\ }\bibfield  {title} {\bibinfo {title} {{Linear disorder and
  temperature dependence of Raman scattering in BaTiO$_3$}},\ }\href
  {https://doi.org/10.1016/0038-1098(72)90334-1} {\bibfield  {journal}
  {\bibinfo  {journal} {Solid State Communications}\ }\textbf {\bibinfo
  {volume} {10}},\ \bibinfo {pages} {1} (\bibinfo {year} {1972})}\BibitemShut
  {NoStop}%
\bibitem [{\citenamefont {H{\"{u}}ller}(1969)}]{Huller1969}%
  \BibitemOpen
  \bibfield  {author} {\bibinfo {author} {\bibfnamefont {A.}~\bibnamefont
  {H{\"{u}}ller}},\ }\bibfield  {title} {\bibinfo {title} {{Soft Phonon
  Dispersion in BaTiO$_3$}},\ }\href@noop {} {\bibfield  {journal} {\bibinfo
  {journal} {Z. Physik}\ }\textbf {\bibinfo {volume} {220}},\ \bibinfo {pages}
  {145} (\bibinfo {year} {1969})}\BibitemShut {NoStop}%
\bibitem [{\citenamefont {Khatib}\ \emph {et~al.}(1989)\citenamefont {Khatib},
  \citenamefont {Migoni}, \citenamefont {Kugel},\ and\ \citenamefont
  {Godefroy}}]{Khatib1989}%
  \BibitemOpen
  \bibfield  {author} {\bibinfo {author} {\bibfnamefont {D.}~\bibnamefont
  {Khatib}}, \bibinfo {author} {\bibfnamefont {R.}~\bibnamefont {Migoni}},
  \bibinfo {author} {\bibfnamefont {G.~E.}\ \bibnamefont {Kugel}},\ and\
  \bibinfo {author} {\bibfnamefont {L.}~\bibnamefont {Godefroy}},\ }\bibfield
  {title} {\bibinfo {title} {{Lattice dynamics of BaTiO$_3$ in the cubic
  phase}},\ }\href {https://doi.org/10.1088/0953-8984/1/49/002} {\bibfield
  {journal} {\bibinfo  {journal} {Journal of Physics: Condensed Matter}\
  }\textbf {\bibinfo {volume} {1}},\ \bibinfo {pages} {9811} (\bibinfo {year}
  {1989})}\BibitemShut {NoStop}%
\bibitem [{\citenamefont {Stachiotti}\ \emph {et~al.}(1993)\citenamefont
  {Stachiotti}, \citenamefont {Dobry}, \citenamefont {Migoni},\ and\
  \citenamefont {Bussmann-Holder}}]{Stachiotti1993}%
  \BibitemOpen
  \bibfield  {author} {\bibinfo {author} {\bibfnamefont {M.}~\bibnamefont
  {Stachiotti}}, \bibinfo {author} {\bibfnamefont {A.}~\bibnamefont {Dobry}},
  \bibinfo {author} {\bibfnamefont {R.}~\bibnamefont {Migoni}},\ and\ \bibinfo
  {author} {\bibfnamefont {A.}~\bibnamefont {Bussmann-Holder}},\ }\bibfield
  {title} {\bibinfo {title} {{Crossover from a displacive to an order-disorder
  transition in the nonlinear-polarizability model}},\ }\href
  {https://doi.org/10.1103/PhysRevB.47.2473} {\bibfield  {journal} {\bibinfo
  {journal} {Physical Review B}\ }\textbf {\bibinfo {volume} {47}},\ \bibinfo
  {pages} {2473} (\bibinfo {year} {1993})}\BibitemShut {NoStop}%
\bibitem [{\citenamefont {Guo}\ \emph {et~al.}(2017{\natexlab{a}})\citenamefont
  {Guo}, \citenamefont {Yaffe}, \citenamefont {Paley}, \citenamefont {Beecher},
  \citenamefont {Hull}, \citenamefont {Szpak}, \citenamefont {Owen},
  \citenamefont {Brus},\ and\ \citenamefont {Pimenta}}]{Guo2017b}%
  \BibitemOpen
  \bibfield  {author} {\bibinfo {author} {\bibfnamefont {Y.}~\bibnamefont
  {Guo}}, \bibinfo {author} {\bibfnamefont {O.}~\bibnamefont {Yaffe}}, \bibinfo
  {author} {\bibfnamefont {D.~W.}\ \bibnamefont {Paley}}, \bibinfo {author}
  {\bibfnamefont {A.~N.}\ \bibnamefont {Beecher}}, \bibinfo {author}
  {\bibfnamefont {T.~D.}\ \bibnamefont {Hull}}, \bibinfo {author}
  {\bibfnamefont {G.}~\bibnamefont {Szpak}}, \bibinfo {author} {\bibfnamefont
  {J.~S.}\ \bibnamefont {Owen}}, \bibinfo {author} {\bibfnamefont {L.~E.}\
  \bibnamefont {Brus}},\ and\ \bibinfo {author} {\bibfnamefont {M.~A.}\
  \bibnamefont {Pimenta}},\ }\bibfield  {title} {\bibinfo {title} {{Interplay
  between organic cations and inorganic framework and incommensurability in
  hybrid lead-halide perovskite CH$_{3}$NH$ _{3}$PbBr$_{3}$}},\ }\href
  {https://doi.org/10.1103/PhysRevMaterials.1.042401} {\bibfield  {journal}
  {\bibinfo  {journal} {Phys. Rev. Mater.}\ }\textbf {\bibinfo {volume} {1}},\
  \bibinfo {pages} {042401} (\bibinfo {year} {2017}{\natexlab{a}})}\BibitemShut
  {NoStop}%
\bibitem [{\citenamefont {Gao}\ \emph {et~al.}(2021)\citenamefont {Gao},
  \citenamefont {Yadgarov}, \citenamefont {Sharma}, \citenamefont {Korobko},
  \citenamefont {McCall}, \citenamefont {Fabini}, \citenamefont {Stoumpos},
  \citenamefont {Kanatzidis}, \citenamefont {Rappe},\ and\ \citenamefont
  {Yaffe}}]{Gao2021}%
  \BibitemOpen
  \bibfield  {author} {\bibinfo {author} {\bibfnamefont {L.}~\bibnamefont
  {Gao}}, \bibinfo {author} {\bibfnamefont {L.}~\bibnamefont {Yadgarov}},
  \bibinfo {author} {\bibfnamefont {R.}~\bibnamefont {Sharma}}, \bibinfo
  {author} {\bibfnamefont {R.}~\bibnamefont {Korobko}}, \bibinfo {author}
  {\bibfnamefont {K.~M.}\ \bibnamefont {McCall}}, \bibinfo {author}
  {\bibfnamefont {D.~H.}\ \bibnamefont {Fabini}}, \bibinfo {author}
  {\bibfnamefont {C.~C.}\ \bibnamefont {Stoumpos}}, \bibinfo {author}
  {\bibfnamefont {M.~G.}\ \bibnamefont {Kanatzidis}}, \bibinfo {author}
  {\bibfnamefont {A.~M.}\ \bibnamefont {Rappe}},\ and\ \bibinfo {author}
  {\bibfnamefont {O.}~\bibnamefont {Yaffe}},\ }\bibfield  {title} {\bibinfo
  {title} {{Metal cation s lone-pairs increase octahedral tilting instabilities
  in halide perovskites}},\ }\href {https://doi.org/10.1039/d1ma00288k}
  {\bibfield  {journal} {\bibinfo  {journal} {Materials Advances}\ }\textbf
  {\bibinfo {volume} {2}},\ \bibinfo {pages} {4610} (\bibinfo {year}
  {2021})}\BibitemShut {NoStop}%
\bibitem [{\citenamefont {Yaffe}\ \emph {et~al.}(2017)\citenamefont {Yaffe},
  \citenamefont {Guo}, \citenamefont {Tan}, \citenamefont {Egger},
  \citenamefont {Hull}, \citenamefont {Stoumpos}, \citenamefont {Zheng},
  \citenamefont {Heinz}, \citenamefont {Kronik}, \citenamefont {Kanatzidis},
  \citenamefont {Owen}, \citenamefont {Rappe}, \citenamefont {Pimenta},\ and\
  \citenamefont {Brus}}]{YaffePRL2017}%
  \BibitemOpen
  \bibfield  {author} {\bibinfo {author} {\bibfnamefont {O.}~\bibnamefont
  {Yaffe}}, \bibinfo {author} {\bibfnamefont {Y.}~\bibnamefont {Guo}}, \bibinfo
  {author} {\bibfnamefont {L.~Z.}\ \bibnamefont {Tan}}, \bibinfo {author}
  {\bibfnamefont {D.~A.}\ \bibnamefont {Egger}}, \bibinfo {author}
  {\bibfnamefont {T.~D.}\ \bibnamefont {Hull}}, \bibinfo {author}
  {\bibfnamefont {C.~C.}\ \bibnamefont {Stoumpos}}, \bibinfo {author}
  {\bibfnamefont {F.}~\bibnamefont {Zheng}}, \bibinfo {author} {\bibfnamefont
  {T.~F.}\ \bibnamefont {Heinz}}, \bibinfo {author} {\bibfnamefont
  {L.}~\bibnamefont {Kronik}}, \bibinfo {author} {\bibfnamefont {M.~G.}\
  \bibnamefont {Kanatzidis}}, \bibinfo {author} {\bibfnamefont {J.~S.}\
  \bibnamefont {Owen}}, \bibinfo {author} {\bibfnamefont {A.~M.}\ \bibnamefont
  {Rappe}}, \bibinfo {author} {\bibfnamefont {M.~A.}\ \bibnamefont {Pimenta}},\
  and\ \bibinfo {author} {\bibfnamefont {L.~E.}\ \bibnamefont {Brus}},\
  }\bibfield  {title} {\bibinfo {title} {{Local Polar Fluctuations in Lead
  Halide Perovskite Crystals}},\ }\href
  {https://doi.org/10.1103/PhysRevLett.118.136001} {\bibfield  {journal}
  {\bibinfo  {journal} {Physical Review Letters}\ }\textbf {\bibinfo {volume}
  {118}},\ \bibinfo {pages} {1} (\bibinfo {year} {2017})},\ \Eprint
  {https://arxiv.org/abs/1604.08107} {arXiv:1604.08107} \BibitemShut {NoStop}%
\bibitem [{\citenamefont {Bechtel}\ \emph {et~al.}(2019)\citenamefont
  {Bechtel}, \citenamefont {Thomas},\ and\ \citenamefont {{Van Der
  Ven}}}]{BechtelPRM2019}%
  \BibitemOpen
  \bibfield  {author} {\bibinfo {author} {\bibfnamefont {J.~S.}\ \bibnamefont
  {Bechtel}}, \bibinfo {author} {\bibfnamefont {J.~C.}\ \bibnamefont
  {Thomas}},\ and\ \bibinfo {author} {\bibfnamefont {A.}~\bibnamefont {{Van Der
  Ven}}},\ }\bibfield  {title} {\bibinfo {title} {{Finite-temperature
  simulation of anharmonicity and octahedral tilting transitions in halide
  perovskites}},\ }\href {https://doi.org/10.1103/PhysRevMaterials.3.113605}
  {\bibfield  {journal} {\bibinfo  {journal} {Physical Review Materials}\
  }\textbf {\bibinfo {volume} {3}},\ \bibinfo {pages} {113605} (\bibinfo {year}
  {2019})}\BibitemShut {NoStop}%
\bibitem [{\citenamefont {Beecher}\ \emph {et~al.}(2016)\citenamefont
  {Beecher}, \citenamefont {Semonin}, \citenamefont {Skelton}, \citenamefont
  {Frost}, \citenamefont {Terban}, \citenamefont {Zhai}, \citenamefont
  {Alatas}, \citenamefont {Owen}, \citenamefont {Walsh},\ and\ \citenamefont
  {Billinge}}]{Beecher2016}%
  \BibitemOpen
  \bibfield  {author} {\bibinfo {author} {\bibfnamefont {A.~N.}\ \bibnamefont
  {Beecher}}, \bibinfo {author} {\bibfnamefont {O.~E.}\ \bibnamefont
  {Semonin}}, \bibinfo {author} {\bibfnamefont {J.~M.}\ \bibnamefont
  {Skelton}}, \bibinfo {author} {\bibfnamefont {J.~M.}\ \bibnamefont {Frost}},
  \bibinfo {author} {\bibfnamefont {M.~W.}\ \bibnamefont {Terban}}, \bibinfo
  {author} {\bibfnamefont {H.}~\bibnamefont {Zhai}}, \bibinfo {author}
  {\bibfnamefont {A.}~\bibnamefont {Alatas}}, \bibinfo {author} {\bibfnamefont
  {J.~S.}\ \bibnamefont {Owen}}, \bibinfo {author} {\bibfnamefont
  {A.}~\bibnamefont {Walsh}},\ and\ \bibinfo {author} {\bibfnamefont
  {S.~J.~L.}\ \bibnamefont {Billinge}},\ }\bibfield  {title} {\bibinfo {title}
  {{Direct Observation of Dynamic Symmetry Breaking above Room Temperature in
  Methylammonium Lead Iodide Perovskite}},\ }\href
  {https://doi.org/10.1021/acsenergylett.6b00381} {\bibfield  {journal}
  {\bibinfo  {journal} {ACS Energy Letters}\ }\textbf {\bibinfo {volume} {1}},\
  \bibinfo {pages} {880} (\bibinfo {year} {2016})},\ \Eprint
  {https://arxiv.org/abs/arXiv:1606.09267v1} {arXiv:arXiv:1606.09267v1}
  \BibitemShut {NoStop}%
\bibitem [{\citenamefont {Whalley}\ \emph {et~al.}(2016)\citenamefont
  {Whalley}, \citenamefont {Skelton}, \citenamefont {Frost},\ and\
  \citenamefont {Walsh}}]{Whalley2016b}%
  \BibitemOpen
  \bibfield  {author} {\bibinfo {author} {\bibfnamefont {L.~D.}\ \bibnamefont
  {Whalley}}, \bibinfo {author} {\bibfnamefont {J.~M.}\ \bibnamefont
  {Skelton}}, \bibinfo {author} {\bibfnamefont {J.~M.}\ \bibnamefont {Frost}},\
  and\ \bibinfo {author} {\bibfnamefont {A.}~\bibnamefont {Walsh}},\ }\bibfield
   {title} {\bibinfo {title} {{Phonon anharmonicity, lifetimes, and thermal
  transport in CH$ _{3} $NH$ _{3} $PbI$ _{3} $ from many-body perturbation
  theory}},\ }\href {https://doi.org/10.1103/PhysRevB.94.220301} {\bibfield
  {journal} {\bibinfo  {journal} {Phys. Rev. B}\ }\textbf {\bibinfo {volume}
  {94}},\ \bibinfo {pages} {220301} (\bibinfo {year} {2016})}\BibitemShut
  {NoStop}%
\bibitem [{\citenamefont {Zhu}\ and\ \citenamefont {Ertekin}(2019)}]{Zhu2019}%
  \BibitemOpen
  \bibfield  {author} {\bibinfo {author} {\bibfnamefont {T.}~\bibnamefont
  {Zhu}}\ and\ \bibinfo {author} {\bibfnamefont {E.}~\bibnamefont {Ertekin}},\
  }\bibfield  {title} {\bibinfo {title} {{Mixed phononic and non-phononic
  transport in hybrid lead halide perovskites: Glass-crystal duality, dynamical
  disorder, and anharmonicity}},\ }\href {https://doi.org/10.1039/c8ee02820f}
  {\bibfield  {journal} {\bibinfo  {journal} {Energy Environ. Sci.}\ }\textbf
  {\bibinfo {volume} {12}},\ \bibinfo {pages} {216} (\bibinfo {year}
  {2019})}\BibitemShut {NoStop}%
\bibitem [{\citenamefont {Gehrmann}\ and\ \citenamefont
  {Egger}(2019)}]{Gehrmann2019}%
  \BibitemOpen
  \bibfield  {author} {\bibinfo {author} {\bibfnamefont {C.}~\bibnamefont
  {Gehrmann}}\ and\ \bibinfo {author} {\bibfnamefont {D.~A.}\ \bibnamefont
  {Egger}},\ }\bibfield  {title} {\bibinfo {title} {{Dynamic shortening of
  disorder potentials in anharmonic halide perovskites}},\ }\bibfield
  {journal} {\bibinfo  {journal} {Nature Communications}\ }\textbf {\bibinfo
  {volume} {10}},\ \href {https://doi.org/10.1038/s41467-019-11087-y}
  {10.1038/s41467-019-11087-y} (\bibinfo {year} {2019})\BibitemShut {NoStop}%
\bibitem [{\citenamefont {{De Wolf}}\ \emph {et~al.}(2014)\citenamefont {{De
  Wolf}}, \citenamefont {Holovsky}, \citenamefont {Moon}, \citenamefont
  {L{\"{o}}per}, \citenamefont {Niesen}, \citenamefont {Ledinsky},
  \citenamefont {Haug}, \citenamefont {Yum},\ and\ \citenamefont
  {Ballif}}]{DeWolf2014}%
  \BibitemOpen
  \bibfield  {author} {\bibinfo {author} {\bibfnamefont {S.}~\bibnamefont {{De
  Wolf}}}, \bibinfo {author} {\bibfnamefont {J.}~\bibnamefont {Holovsky}},
  \bibinfo {author} {\bibfnamefont {S.~J.}\ \bibnamefont {Moon}}, \bibinfo
  {author} {\bibfnamefont {P.}~\bibnamefont {L{\"{o}}per}}, \bibinfo {author}
  {\bibfnamefont {B.}~\bibnamefont {Niesen}}, \bibinfo {author} {\bibfnamefont
  {M.}~\bibnamefont {Ledinsky}}, \bibinfo {author} {\bibfnamefont {F.~J.}\
  \bibnamefont {Haug}}, \bibinfo {author} {\bibfnamefont {J.~H.}\ \bibnamefont
  {Yum}},\ and\ \bibinfo {author} {\bibfnamefont {C.}~\bibnamefont {Ballif}},\
  }\bibfield  {title} {\bibinfo {title} {{Organometallic halide perovskites:
  Sharp optical absorption edge and its relation to photovoltaic
  performance}},\ }\href {https://doi.org/10.1021/jz500279b} {\bibfield
  {journal} {\bibinfo  {journal} {J. Phys. Chem. Lett.}\ }\textbf {\bibinfo
  {volume} {5}},\ \bibinfo {pages} {1035} (\bibinfo {year} {2014})}\BibitemShut
  {NoStop}%
\bibitem [{\citenamefont {Maeng}\ \emph {et~al.}(2019)\citenamefont {Maeng},
  \citenamefont {Lee}, \citenamefont {Park}, \citenamefont {Raga},
  \citenamefont {Kang}, \citenamefont {Kee}, \citenamefont {Yu}, \citenamefont
  {Hong}, \citenamefont {Ono}, \citenamefont {Qi}, \citenamefont {Jung},\ and\
  \citenamefont {Nakamura}}]{Maeng2019}%
  \BibitemOpen
  \bibfield  {author} {\bibinfo {author} {\bibfnamefont {I.}~\bibnamefont
  {Maeng}}, \bibinfo {author} {\bibfnamefont {Y.~M.}\ \bibnamefont {Lee}},
  \bibinfo {author} {\bibfnamefont {J.}~\bibnamefont {Park}}, \bibinfo {author}
  {\bibfnamefont {S.~R.}\ \bibnamefont {Raga}}, \bibinfo {author}
  {\bibfnamefont {C.}~\bibnamefont {Kang}}, \bibinfo {author} {\bibfnamefont
  {C.~S.}\ \bibnamefont {Kee}}, \bibinfo {author} {\bibfnamefont {B.~D.}\
  \bibnamefont {Yu}}, \bibinfo {author} {\bibfnamefont {S.}~\bibnamefont
  {Hong}}, \bibinfo {author} {\bibfnamefont {L.~K.}\ \bibnamefont {Ono}},
  \bibinfo {author} {\bibfnamefont {Y.}~\bibnamefont {Qi}}, \bibinfo {author}
  {\bibfnamefont {M.~C.}\ \bibnamefont {Jung}},\ and\ \bibinfo {author}
  {\bibfnamefont {M.}~\bibnamefont {Nakamura}},\ }\bibfield  {title} {\bibinfo
  {title} {{Significant THz absorption in CH 3 NH 2 molecular
  defect-incorporated organic-inorganic hybrid perovskite thin film}},\
  }\bibfield  {journal} {\bibinfo  {journal} {Scientific Reports}\ }\textbf
  {\bibinfo {volume} {9}},\ \href {https://doi.org/10.1038/s41598-019-42359-8}
  {10.1038/s41598-019-42359-8} (\bibinfo {year} {2019})\BibitemShut {NoStop}%
\bibitem [{\citenamefont {Xia}\ \emph {et~al.}(2021)\citenamefont {Xia},
  \citenamefont {Ponc{\'{e}}}, \citenamefont {Peng}, \citenamefont {Ulatowski},
  \citenamefont {Patel}, \citenamefont {Wright}, \citenamefont {Milot},
  \citenamefont {Kraus}, \citenamefont {Lin}, \citenamefont {Herz},
  \citenamefont {Giustino},\ and\ \citenamefont {Johnston}}]{Xia2021}%
  \BibitemOpen
  \bibfield  {author} {\bibinfo {author} {\bibfnamefont {C.~Q.}\ \bibnamefont
  {Xia}}, \bibinfo {author} {\bibfnamefont {S.}~\bibnamefont {Ponc{\'{e}}}},
  \bibinfo {author} {\bibfnamefont {J.}~\bibnamefont {Peng}}, \bibinfo {author}
  {\bibfnamefont {A.~M.}\ \bibnamefont {Ulatowski}}, \bibinfo {author}
  {\bibfnamefont {J.~B.}\ \bibnamefont {Patel}}, \bibinfo {author}
  {\bibfnamefont {A.~D.}\ \bibnamefont {Wright}}, \bibinfo {author}
  {\bibfnamefont {R.~L.}\ \bibnamefont {Milot}}, \bibinfo {author}
  {\bibfnamefont {H.}~\bibnamefont {Kraus}}, \bibinfo {author} {\bibfnamefont
  {Q.}~\bibnamefont {Lin}}, \bibinfo {author} {\bibfnamefont {L.~M.}\
  \bibnamefont {Herz}}, \bibinfo {author} {\bibfnamefont {F.}~\bibnamefont
  {Giustino}},\ and\ \bibinfo {author} {\bibfnamefont {M.~B.}\ \bibnamefont
  {Johnston}},\ }\bibfield  {title} {\bibinfo {title} {{Ultrafast photo-induced
  phonon hardening due to Pauli blocking in MAPbI3single-crystal and
  polycrystalline perovskites}},\ }\bibfield  {journal} {\bibinfo  {journal}
  {JPhys Materials}\ }\textbf {\bibinfo {volume} {4}},\ \href
  {https://doi.org/10.1088/2515-7639/ac22b9} {10.1088/2515-7639/ac22b9}
  (\bibinfo {year} {2021})\BibitemShut {NoStop}%
\bibitem [{\citenamefont {La-O-Vorakiat}\ \emph {et~al.}(2016)\citenamefont
  {La-O-Vorakiat}, \citenamefont {Xia}, \citenamefont {Kadro}, \citenamefont
  {Salim}, \citenamefont {Zhao}, \citenamefont {Ahmed}, \citenamefont {Lam},
  \citenamefont {Zhu}, \citenamefont {Marcus}, \citenamefont {Michel-Beyerle},\
  and\ \citenamefont {Chia}}]{VorakiatJPhysChemLett2016}%
  \BibitemOpen
  \bibfield  {author} {\bibinfo {author} {\bibfnamefont {C.}~\bibnamefont
  {La-O-Vorakiat}}, \bibinfo {author} {\bibfnamefont {H.}~\bibnamefont {Xia}},
  \bibinfo {author} {\bibfnamefont {J.}~\bibnamefont {Kadro}}, \bibinfo
  {author} {\bibfnamefont {T.}~\bibnamefont {Salim}}, \bibinfo {author}
  {\bibfnamefont {D.}~\bibnamefont {Zhao}}, \bibinfo {author} {\bibfnamefont
  {T.}~\bibnamefont {Ahmed}}, \bibinfo {author} {\bibfnamefont {Y.~M.}\
  \bibnamefont {Lam}}, \bibinfo {author} {\bibfnamefont {J.~X.}\ \bibnamefont
  {Zhu}}, \bibinfo {author} {\bibfnamefont {R.~A.}\ \bibnamefont {Marcus}},
  \bibinfo {author} {\bibfnamefont {M.~E.}\ \bibnamefont {Michel-Beyerle}},\
  and\ \bibinfo {author} {\bibfnamefont {E.~E.}\ \bibnamefont {Chia}},\
  }\bibfield  {title} {\bibinfo {title} {{Phonon Mode Transformation Across the
  Orthohombic-Tetragonal Phase Transition in a Lead Iodide Perovskite
  CH3NH3PbI3: A Terahertz Time-Domain Spectroscopy Approach}},\ }\href
  {https://doi.org/10.1021/acs.jpclett.5b02223} {\bibfield  {journal} {\bibinfo
   {journal} {Journal of Physical Chemistry Letters}\ }\textbf {\bibinfo
  {volume} {7}},\ \bibinfo {pages} {1} (\bibinfo {year} {2016})}\BibitemShut
  {NoStop}%
\bibitem [{\citenamefont {Yang}\ \emph {et~al.}(2020)\citenamefont {Yang},
  \citenamefont {Skelton}, \citenamefont {{Da Silva}}, \citenamefont {Frost},\
  and\ \citenamefont {Walsh}}]{Yang2020}%
  \BibitemOpen
  \bibfield  {author} {\bibinfo {author} {\bibfnamefont {R.~X.}\ \bibnamefont
  {Yang}}, \bibinfo {author} {\bibfnamefont {J.~M.}\ \bibnamefont {Skelton}},
  \bibinfo {author} {\bibfnamefont {E.~L.}\ \bibnamefont {{Da Silva}}},
  \bibinfo {author} {\bibfnamefont {J.~M.}\ \bibnamefont {Frost}},\ and\
  \bibinfo {author} {\bibfnamefont {A.}~\bibnamefont {Walsh}},\ }\bibfield
  {title} {\bibinfo {title} {{Assessment of dynamic structural instabilities
  across 24 cubic inorganic halide perovskites}},\ }\href
  {https://doi.org/10.1063/1.5131575} {\bibfield  {journal} {\bibinfo
  {journal} {J. Chem. Phys}\ }\textbf {\bibinfo {volume} {152}},\ \bibinfo
  {pages} {24703} (\bibinfo {year} {2020})}\BibitemShut {NoStop}%
\bibitem [{He2(2020)}]{He2020}%
  \BibitemOpen
  \bibfield  {title} {\bibinfo {title} {{Anharmonic Eigenvectors and Acoustic
  Phonon Disappearance in Quantum Paraelectric SrTiO$_3$}},\ }\href
  {https://doi.org/10.1103/PhysRevLett.124.145901} {\bibfield  {journal}
  {\bibinfo  {journal} {Physical Review Letters}\ }\textbf {\bibinfo {volume}
  {124}},\ \bibinfo {pages} {145901} (\bibinfo {year} {2020})}\BibitemShut
  {NoStop}%
\bibitem [{\citenamefont {Laabidi}\ \emph {et~al.}(1991)\citenamefont
  {Laabidi}, \citenamefont {Fontana},\ and\ \citenamefont
  {Jannot}}]{Laabidi1991}%
  \BibitemOpen
  \bibfield  {author} {\bibinfo {author} {\bibfnamefont {K.}~\bibnamefont
  {Laabidi}}, \bibinfo {author} {\bibfnamefont {M.}~\bibnamefont {Fontana}},\
  and\ \bibinfo {author} {\bibfnamefont {B.}~\bibnamefont {Jannot}},\
  }\bibfield  {title} {\bibinfo {title} {{Existence of two time scales in the
  phase transitions of BaTiO$_3$}},\ }\href
  {https://doi.org/10.1080/00150199108209438} {\bibfield  {journal} {\bibinfo
  {journal} {Ferroelectrics}\ }\textbf {\bibinfo {volume} {124}},\ \bibinfo
  {pages} {201} (\bibinfo {year} {1991})}\BibitemShut {NoStop}%
\bibitem [{\citenamefont {Cardona}\ and\ \citenamefont
  {Guntherodt}(1982)}]{Cardona1982}%
  \BibitemOpen
  \bibfield  {author} {\bibinfo {author} {\bibfnamefont {M.}~\bibnamefont
  {Cardona}}\ and\ \bibinfo {author} {\bibfnamefont {G.}~\bibnamefont
  {Guntherodt}},\ }\href {https://doi.org/10.1016/0030-3992(77)90116-5} {\emph
  {\bibinfo {title} {{Light Scattering in Solids II - Basic Concepts and
  Instrumentation}}}}\ (\bibinfo  {publisher} {Springer-Verlag},\ \bibinfo
  {year} {1982})\ pp.\ \bibinfo {pages} {1--252}\BibitemShut {NoStop}%
\bibitem [{\citenamefont {Egger}\ \emph {et~al.}(2018)\citenamefont {Egger},
  \citenamefont {Bera}, \citenamefont {Cahen}, \citenamefont {Hodes},
  \citenamefont {Kirchartz}, \citenamefont {Kronik}, \citenamefont {Lovrincic},
  \citenamefont {Rappe}, \citenamefont {Reichman},\ and\ \citenamefont
  {Yaffe}}]{BigReview}%
  \BibitemOpen
  \bibfield  {author} {\bibinfo {author} {\bibfnamefont {D.~A.}\ \bibnamefont
  {Egger}}, \bibinfo {author} {\bibfnamefont {A.}~\bibnamefont {Bera}},
  \bibinfo {author} {\bibfnamefont {D.}~\bibnamefont {Cahen}}, \bibinfo
  {author} {\bibfnamefont {G.}~\bibnamefont {Hodes}}, \bibinfo {author}
  {\bibfnamefont {T.}~\bibnamefont {Kirchartz}}, \bibinfo {author}
  {\bibfnamefont {L.}~\bibnamefont {Kronik}}, \bibinfo {author} {\bibfnamefont
  {R.}~\bibnamefont {Lovrincic}}, \bibinfo {author} {\bibfnamefont {A.~M.}\
  \bibnamefont {Rappe}}, \bibinfo {author} {\bibfnamefont {D.~R.}\ \bibnamefont
  {Reichman}},\ and\ \bibinfo {author} {\bibfnamefont {O.}~\bibnamefont
  {Yaffe}},\ }\bibfield  {title} {\bibinfo {title} {What remains unexplained
  about the properties of halide perovskites?},\ }\href
  {https://doi.org/https://doi.org/10.1002/adma.201800691} {\bibfield
  {journal} {\bibinfo  {journal} {Advanced Materials}\ }\textbf {\bibinfo
  {volume} {30}},\ \bibinfo {pages} {1800691} (\bibinfo {year}
  {2018})}\BibitemShut {NoStop}%
\bibitem [{\citenamefont {Marronnier}\ \emph {et~al.}(2017)\citenamefont
  {Marronnier}, \citenamefont {Lee}, \citenamefont {Geffroy}, \citenamefont
  {Even}, \citenamefont {Bonnassieux},\ and\ \citenamefont
  {Roma}}]{Marronnier2017}%
  \BibitemOpen
  \bibfield  {author} {\bibinfo {author} {\bibfnamefont {A.}~\bibnamefont
  {Marronnier}}, \bibinfo {author} {\bibfnamefont {H.}~\bibnamefont {Lee}},
  \bibinfo {author} {\bibfnamefont {B.}~\bibnamefont {Geffroy}}, \bibinfo
  {author} {\bibfnamefont {J.}~\bibnamefont {Even}}, \bibinfo {author}
  {\bibfnamefont {Y.}~\bibnamefont {Bonnassieux}},\ and\ \bibinfo {author}
  {\bibfnamefont {G.}~\bibnamefont {Roma}},\ }\bibfield  {title} {\bibinfo
  {title} {{Structural Instabilities Related to Highly Anharmonic Phonons in
  Halide Perovskites}},\ }\href {https://doi.org/10.1021/acs.jpclett.7b00807}
  {\bibfield  {journal} {\bibinfo  {journal} {Journal of Physical Chemistry
  Letters}\ }\textbf {\bibinfo {volume} {8}},\ \bibinfo {pages} {2659}
  (\bibinfo {year} {2017})}\BibitemShut {NoStop}%
\bibitem [{\citenamefont {Mayteevarunyoo}\ \emph {et~al.}(2008)\citenamefont
  {Mayteevarunyoo}, \citenamefont {Malomed},\ and\ \citenamefont
  {Dong}}]{Mayteevarunyoo2008}%
  \BibitemOpen
  \bibfield  {author} {\bibinfo {author} {\bibfnamefont {T.}~\bibnamefont
  {Mayteevarunyoo}}, \bibinfo {author} {\bibfnamefont {B.~A.}\ \bibnamefont
  {Malomed}},\ and\ \bibinfo {author} {\bibfnamefont {G.}~\bibnamefont
  {Dong}},\ }\bibfield  {title} {\bibinfo {title} {{Spontaneous symmetry
  breaking in a nonlinear double-well structure}},\ }\href
  {https://doi.org/10.1103/PhysRevA.78.053601} {\bibfield  {journal} {\bibinfo
  {journal} {Physical Review A}\ }\textbf {\bibinfo {volume} {78}},\ \bibinfo
  {pages} {053601} (\bibinfo {year} {2008})},\ \Eprint
  {https://arxiv.org/abs/0810.0859} {arXiv:0810.0859} \BibitemShut {NoStop}%
\bibitem [{\citenamefont {Wang}\ \emph {et~al.}(2021)\citenamefont {Wang},
  \citenamefont {Malyi}, \citenamefont {Zhao},\ and\ \citenamefont
  {Zunger}}]{Wang2021}%
  \BibitemOpen
  \bibfield  {author} {\bibinfo {author} {\bibfnamefont {Z.}~\bibnamefont
  {Wang}}, \bibinfo {author} {\bibfnamefont {O.~I.}\ \bibnamefont {Malyi}},
  \bibinfo {author} {\bibfnamefont {X.}~\bibnamefont {Zhao}},\ and\ \bibinfo
  {author} {\bibfnamefont {A.}~\bibnamefont {Zunger}},\ }\bibfield  {title}
  {\bibinfo {title} {{Mass enhancement in 3d and s-p perovskites from symmetry
  breaking}},\ }\href {https://doi.org/10.1103/PhysRevB.103.165110} {\bibfield
  {journal} {\bibinfo  {journal} {Phys. Rev. B}\ }\textbf {\bibinfo {volume}
  {103}},\ \bibinfo {pages} {165110} (\bibinfo {year} {2021})}\BibitemShut
  {NoStop}%
\bibitem [{\citenamefont {Zhao}\ \emph {et~al.}(2020)\citenamefont {Zhao},
  \citenamefont {Dalpian}, \citenamefont {Wang},\ and\ \citenamefont
  {Zunger}}]{Zhao2020}%
  \BibitemOpen
  \bibfield  {author} {\bibinfo {author} {\bibfnamefont {X.~G.}\ \bibnamefont
  {Zhao}}, \bibinfo {author} {\bibfnamefont {G.~M.}\ \bibnamefont {Dalpian}},
  \bibinfo {author} {\bibfnamefont {Z.}~\bibnamefont {Wang}},\ and\ \bibinfo
  {author} {\bibfnamefont {A.}~\bibnamefont {Zunger}},\ }\bibfield  {title}
  {\bibinfo {title} {{Polymorphous nature of cubic halide perovskites}},\
  }\href {https://doi.org/10.1103/PhysRevB.101.155137} {\bibfield  {journal}
  {\bibinfo  {journal} {Phys. Rev. B}\ }\textbf {\bibinfo {volume} {101}},\
  \bibinfo {pages} {155137} (\bibinfo {year} {2020})}\BibitemShut {NoStop}%
\bibitem [{\citenamefont {Zhao}\ \emph {et~al.}(2021)\citenamefont {Zhao},
  \citenamefont {Wang}, \citenamefont {Malyi},\ and\ \citenamefont
  {Zunger}}]{Zhao2021}%
  \BibitemOpen
  \bibfield  {author} {\bibinfo {author} {\bibfnamefont {X.~G.}\ \bibnamefont
  {Zhao}}, \bibinfo {author} {\bibfnamefont {Z.}~\bibnamefont {Wang}}, \bibinfo
  {author} {\bibfnamefont {O.~I.}\ \bibnamefont {Malyi}},\ and\ \bibinfo
  {author} {\bibfnamefont {A.}~\bibnamefont {Zunger}},\ }\bibfield  {title}
  {\bibinfo {title} {{Effect of static local distortions vs. dynamic motions on
  the stability and band gaps of cubic oxide and halide perovskites}},\ }\href
  {https://doi.org/10.1016/j.mattod.2021.05.021} {\bibfield  {journal}
  {\bibinfo  {journal} {Mater. Today}\ }\textbf {\bibinfo {volume} {49}},\
  \bibinfo {pages} {107} (\bibinfo {year} {2021})}\BibitemShut {NoStop}%
\bibitem [{\citenamefont {Zhao}\ \emph {et~al.}(2022)\citenamefont {Zhao},
  \citenamefont {Malyi}, \citenamefont {Billinge},\ and\ \citenamefont
  {Zunger}}]{Zhao2022}%
  \BibitemOpen
  \bibfield  {author} {\bibinfo {author} {\bibfnamefont {X.-g.}\ \bibnamefont
  {Zhao}}, \bibinfo {author} {\bibfnamefont {O.~I.}\ \bibnamefont {Malyi}},
  \bibinfo {author} {\bibfnamefont {S.~J.~L.}\ \bibnamefont {Billinge}},\ and\
  \bibinfo {author} {\bibfnamefont {A.}~\bibnamefont {Zunger}},\ }\bibfield
  {title} {\bibinfo {title} {{Intrinsic local symmetry breaking in nominally
  cubic paraelectric BaTiO$_\text{3}$}},\ }\href
  {https://doi.org/10.1103/PhysRevB.105.224108} {\bibfield  {journal} {\bibinfo
   {journal} {Phys. Rev. B}\ }\textbf {\bibinfo {volume} {105}},\ \bibinfo
  {pages} {224108} (\bibinfo {year} {2022})}\BibitemShut {NoStop}%
\bibitem [{\citenamefont {Benshalom}\ \emph {et~al.}(2022)\citenamefont
  {Benshalom}, \citenamefont {Reuveni}, \citenamefont {Korobko}, \citenamefont
  {Yaffe},\ and\ \citenamefont {Hellman}}]{Benshalom2022}%
  \BibitemOpen
  \bibfield  {author} {\bibinfo {author} {\bibfnamefont {N.}~\bibnamefont
  {Benshalom}}, \bibinfo {author} {\bibfnamefont {G.}~\bibnamefont {Reuveni}},
  \bibinfo {author} {\bibfnamefont {R.}~\bibnamefont {Korobko}}, \bibinfo
  {author} {\bibfnamefont {O.}~\bibnamefont {Yaffe}},\ and\ \bibinfo {author}
  {\bibfnamefont {O.}~\bibnamefont {Hellman}},\ }\bibfield  {title} {\bibinfo
  {title} {{Dielectric response of rock-salt crystals at finite temperatures
  from first principles}},\ }\href
  {https://doi.org/10.1103/PhysRevMaterials.6.033607} {\bibfield  {journal}
  {\bibinfo  {journal} {Physical Review Materials}\ }\textbf {\bibinfo {volume}
  {6}},\ \bibinfo {pages} {033607} (\bibinfo {year} {2022})}\BibitemShut
  {NoStop}%
\bibitem [{\citenamefont {Kwok}(1968)}]{Kwok1968}%
  \BibitemOpen
  \bibfield  {author} {\bibinfo {author} {\bibfnamefont {P.~C.}\ \bibnamefont
  {Kwok}},\ }\bibfield  {title} {\bibinfo {title} {{Green's Function Method in
  Lattice Dynamics}},\ }in\ \href
  {https://doi.org/10.1016/S0081-1947(08)60219-2} {\emph {\bibinfo {booktitle}
  {Solid State Phys.}}}\ (\bibinfo {year} {1968})\ pp.\ \bibinfo {pages}
  {213--303}\BibitemShut {NoStop}%
\bibitem [{\citenamefont {Andrews}\ \emph {et~al.}(1982)\citenamefont
  {Andrews}, \citenamefont {Harley}, \citenamefont {Jahn},\ and\ \citenamefont
  {Sherman}}]{Andrews1982}%
  \BibitemOpen
  \bibfield  {author} {\bibinfo {author} {\bibfnamefont {S.~R.}\ \bibnamefont
  {Andrews}}, \bibinfo {author} {\bibfnamefont {R.~T.}\ \bibnamefont {Harley}},
  \bibinfo {author} {\bibfnamefont {I.~R.}\ \bibnamefont {Jahn}},\ and\
  \bibinfo {author} {\bibfnamefont {W.~F.}\ \bibnamefont {Sherman}},\
  }\bibfield  {title} {\bibinfo {title} {{Quasi-elastic light scattering at the
  order-disorder phase transitions of NH4Cl and NH4Br}},\ }\href
  {https://doi.org/10.1088/0022-3719/15/22/014} {\bibfield  {journal} {\bibinfo
   {journal} {J. Phys. C Solid State Phys.}\ }\textbf {\bibinfo {volume}
  {15}},\ \bibinfo {pages} {4679} (\bibinfo {year} {1982})}\BibitemShut
  {NoStop}%
\bibitem [{\citenamefont {Krumhansl}\ and\ \citenamefont
  {Schrieffer}(1975)}]{Krumhansl1975}%
  \BibitemOpen
  \bibfield  {author} {\bibinfo {author} {\bibfnamefont {J.~A.}\ \bibnamefont
  {Krumhansl}}\ and\ \bibinfo {author} {\bibfnamefont {J.~R.}\ \bibnamefont
  {Schrieffer}},\ }\bibfield  {title} {\bibinfo {title} {{Dynamics and
  statistical mechanics of a one-dimensional model Hamiltonian for structural
  phase transitions}},\ }\href {https://doi.org/10.1103/PhysRevB.11.3535}
  {\bibfield  {journal} {\bibinfo  {journal} {Phys. Rev. B}\ }\textbf {\bibinfo
  {volume} {11}},\ \bibinfo {pages} {3535} (\bibinfo {year}
  {1975})}\BibitemShut {NoStop}%
\bibitem [{\citenamefont {Porter}\ \emph {et~al.}(2009)\citenamefont {Porter},
  \citenamefont {Zabusky}, \citenamefont {Hu},\ and\ \citenamefont
  {Campbell}}]{Porter2009}%
  \BibitemOpen
  \bibfield  {author} {\bibinfo {author} {\bibfnamefont {M.~A.}\ \bibnamefont
  {Porter}}, \bibinfo {author} {\bibfnamefont {N.~J.}\ \bibnamefont {Zabusky}},
  \bibinfo {author} {\bibfnamefont {B.}~\bibnamefont {Hu}},\ and\ \bibinfo
  {author} {\bibfnamefont {D.~K.}\ \bibnamefont {Campbell}},\ }\bibfield
  {title} {\bibinfo {title} {{Fermi, Pasta, Ulam and the Birth of Experimental
  Mathematics}},\ }\href {https://doi.org/10.1511/2009.78.214} {\bibfield
  {journal} {\bibinfo  {journal} {Am. Sci.}\ }\textbf {\bibinfo {volume}
  {97}},\ \bibinfo {pages} {214} (\bibinfo {year} {2009})}\BibitemShut
  {NoStop}%
\bibitem [{\citenamefont {Shirane}\ \emph {et~al.}(1993)\citenamefont
  {Shirane}, \citenamefont {Cowley}, \citenamefont {Matsuda},\ and\
  \citenamefont {Shapiro}}]{Shirane1993}%
  \BibitemOpen
  \bibfield  {author} {\bibinfo {author} {\bibfnamefont {G.}~\bibnamefont
  {Shirane}}, \bibinfo {author} {\bibfnamefont {R.~A.}\ \bibnamefont {Cowley}},
  \bibinfo {author} {\bibfnamefont {M.}~\bibnamefont {Matsuda}},\ and\ \bibinfo
  {author} {\bibfnamefont {S.~M.}\ \bibnamefont {Shapiro}},\ }\bibfield
  {title} {\bibinfo {title} {{Q dependence of the central peak in the
  inelastic-neutron-scattering spectrum of SrTiO$_3$}},\ }\href
  {https://doi.org/10.1103/PhysRevB.48.15595} {\bibfield  {journal} {\bibinfo
  {journal} {Physical Review B}\ }\textbf {\bibinfo {volume} {48}},\ \bibinfo
  {pages} {15595} (\bibinfo {year} {1993})}\BibitemShut {NoStop}%
\bibitem [{\citenamefont {Safran}\ \emph {et~al.}(1977)\citenamefont {Safran},
  \citenamefont {Dresselhaus},\ and\ \citenamefont {Lax}}]{Safran1977}%
  \BibitemOpen
  \bibfield  {author} {\bibinfo {author} {\bibfnamefont {S.~A.}\ \bibnamefont
  {Safran}}, \bibinfo {author} {\bibfnamefont {G.}~\bibnamefont
  {Dresselhaus}},\ and\ \bibinfo {author} {\bibfnamefont {B.}~\bibnamefont
  {Lax}},\ }\bibfield  {title} {\bibinfo {title} {{Theory of spin-disorder
  Raman scattering in magnetic semiconductors}},\ }\href
  {https://doi.org/https://doi.org/10.1103/PhysRevB.16.2749} {\bibfield
  {journal} {\bibinfo  {journal} {Phys. Rev. B}\ }\textbf {\bibinfo {volume}
  {16}},\ \bibinfo {pages} {2749} (\bibinfo {year} {1977})}\BibitemShut
  {NoStop}%
\bibitem [{\citenamefont {Dultz}\ and\ \citenamefont
  {Ihlefeld}(1973)}]{Dultz1973}%
  \BibitemOpen
  \bibfield  {author} {\bibinfo {author} {\bibfnamefont {W.}~\bibnamefont
  {Dultz}}\ and\ \bibinfo {author} {\bibfnamefont {H.}~\bibnamefont
  {Ihlefeld}},\ }\bibfield  {title} {\bibinfo {title} {{One‐phonon
  density‐of‐states and spatial correlation from the Raman spectrum of
  molecular crystals: NH 4 I}},\ }\href {https://doi.org/10.1063/1.1679663}
  {\bibfield  {journal} {\bibinfo  {journal} {J. Chem. Phys.}\ }\textbf
  {\bibinfo {volume} {58}},\ \bibinfo {pages} {3365} (\bibinfo {year}
  {1973})}\BibitemShut {NoStop}%
\bibitem [{\citenamefont {Dultz}(1976)}]{Dultz1976}%
  \BibitemOpen
  \bibfield  {author} {\bibinfo {author} {\bibfnamefont {W.}~\bibnamefont
  {Dultz}},\ }\bibfield  {title} {\bibinfo {title} {{Critical effects of the
  light scattering intensity at an order–disorder phase transition: KCN}},\
  }\href {https://doi.org/10.1063/1.433429} {\bibfield  {journal} {\bibinfo
  {journal} {J. Chem. Phys.}\ }\textbf {\bibinfo {volume} {65}},\ \bibinfo
  {pages} {2812} (\bibinfo {year} {1976})}\BibitemShut {NoStop}%
\bibitem [{\citenamefont {Briganti}\ \emph {et~al.}(1981)\citenamefont
  {Briganti}, \citenamefont {Mazzacurati}, \citenamefont {Signorelli},\ and\
  \citenamefont {Nardone}}]{Briganti1981}%
  \BibitemOpen
  \bibfield  {author} {\bibinfo {author} {\bibfnamefont {G.}~\bibnamefont
  {Briganti}}, \bibinfo {author} {\bibfnamefont {V.}~\bibnamefont
  {Mazzacurati}}, \bibinfo {author} {\bibfnamefont {G.}~\bibnamefont
  {Signorelli}},\ and\ \bibinfo {author} {\bibfnamefont {M.}~\bibnamefont
  {Nardone}},\ }\bibfield  {title} {\bibinfo {title} {{Interaction induced
  light scattering in orientationally disordered crystals: the translational
  phonon region}},\ }\href {https://doi.org/10.1080/00268978100102111}
  {\bibfield  {journal} {\bibinfo  {journal} {Mol. Phys.}\ }\textbf {\bibinfo
  {volume} {43}},\ \bibinfo {pages} {1347} (\bibinfo {year}
  {1981})}\BibitemShut {NoStop}%
\bibitem [{\citenamefont {Sanyal}\ and\ \citenamefont
  {Sharma}(1982)}]{Sanyal1982}%
  \BibitemOpen
  \bibfield  {author} {\bibinfo {author} {\bibfnamefont {S.~P.}\ \bibnamefont
  {Sanyal}}\ and\ \bibinfo {author} {\bibfnamefont {T.}~\bibnamefont
  {Sharma}},\ }\bibfield  {title} {\bibinfo {title} {{Two phonon density of
  states of partially disordered crystal NH4I}},\ }\href
  {https://doi.org/10.1016/0022-3697(82)90127-5} {\bibfield  {journal}
  {\bibinfo  {journal} {J. Phys. Chem. Solids}\ }\textbf {\bibinfo {volume}
  {43}},\ \bibinfo {pages} {111} (\bibinfo {year} {1982})}\BibitemShut
  {NoStop}%
\bibitem [{\citenamefont {Cowley}(1996)}]{Cowley1996}%
  \BibitemOpen
  \bibfield  {author} {\bibinfo {author} {\bibfnamefont {R.~A.}\ \bibnamefont
  {Cowley}},\ }\bibfield  {title} {\bibinfo {title} {{Are there two length
  scales at phase transitions?}},\ }\href
  {https://doi.org/10.1088/0031-8949/1996/t66/003} {\bibfield  {journal}
  {\bibinfo  {journal} {Physica Scripta}\ }\textbf {\bibinfo {volume} {T66}},\
  \bibinfo {pages} {24} (\bibinfo {year} {1996})}\BibitemShut {NoStop}%
\bibitem [{\citenamefont {Papon}\ \emph {et~al.}(2002)\citenamefont {Papon},
  \citenamefont {Leblond},\ and\ \citenamefont {Meijer}}]{Papon2002}%
  \BibitemOpen
  \bibfield  {author} {\bibinfo {author} {\bibfnamefont {P.}~\bibnamefont
  {Papon}}, \bibinfo {author} {\bibfnamefont {J.}~\bibnamefont {Leblond}},\
  and\ \bibinfo {author} {\bibfnamefont {P.~H.}\ \bibnamefont {Meijer}},\
  }\bibinfo {title} {{Thermodynamics and Statistical Mechanics of Phase
  Transitions}},\ in\ \href@noop {} {\emph {\bibinfo {booktitle} {{Physics of
  Phase Transitions - Concepts and Applications}}}}\ (\bibinfo  {publisher}
  {Springer-Verlag},\ \bibinfo {address} {Berlin Heidelberg},\ \bibinfo {year}
  {2002})\ pp.\ \bibinfo {pages} {1--34}\BibitemShut {NoStop}%
\bibitem [{\citenamefont {DiAntonio}\ \emph {et~al.}(1993)\citenamefont
  {DiAntonio}, \citenamefont {Vugmeister}, \citenamefont {Toulouse},\ and\
  \citenamefont {Boatner}}]{DiAntonio1993}%
  \BibitemOpen
  \bibfield  {author} {\bibinfo {author} {\bibfnamefont {P.}~\bibnamefont
  {DiAntonio}}, \bibinfo {author} {\bibfnamefont {B.~E.}\ \bibnamefont
  {Vugmeister}}, \bibinfo {author} {\bibfnamefont {J.}~\bibnamefont
  {Toulouse}},\ and\ \bibinfo {author} {\bibfnamefont {L.~A.}\ \bibnamefont
  {Boatner}},\ }\bibfield  {title} {\bibinfo {title} {{Polar fluctuations and
  first-order Raman scattering in highly polarizable KTaO$_3$ crystals with
  off-center Li and Nb ions}},\ }\href
  {https://doi.org/10.1103/PhysRevB.47.5629} {\bibfield  {journal} {\bibinfo
  {journal} {Physical Review B}\ }\textbf {\bibinfo {volume} {47}},\ \bibinfo
  {pages} {5629} (\bibinfo {year} {1993})}\BibitemShut {NoStop}%
\bibitem [{\citenamefont {Vugmeister}\ \emph {et~al.}(1999)\citenamefont
  {Vugmeister}, \citenamefont {Yacoby}, \citenamefont {Toulouse},\ and\
  \citenamefont {Rabitz}}]{Vugmeister1999}%
  \BibitemOpen
  \bibfield  {author} {\bibinfo {author} {\bibfnamefont {B.~E.}\ \bibnamefont
  {Vugmeister}}, \bibinfo {author} {\bibfnamefont {Y.}~\bibnamefont {Yacoby}},
  \bibinfo {author} {\bibfnamefont {J.}~\bibnamefont {Toulouse}},\ and\
  \bibinfo {author} {\bibfnamefont {H.}~\bibnamefont {Rabitz}},\ }\bibfield
  {title} {\bibinfo {title} {{Second-order central peak in the Raman spectra of
  disordered ferroelectrics}},\ }\href
  {https://doi.org/10.1103/PhysRevB.59.8602} {\bibfield  {journal} {\bibinfo
  {journal} {Physical Review B}\ }\textbf {\bibinfo {volume} {59}},\ \bibinfo
  {pages} {8602} (\bibinfo {year} {1999})}\BibitemShut {NoStop}%
\bibitem [{\citenamefont {Stanley}(1971)}]{OZTheory}%
  \BibitemOpen
  \bibfield  {author} {\bibinfo {author} {\bibfnamefont {H.~E.}\ \bibnamefont
  {Stanley}},\ }\bibinfo {title} {{The Pair Correlation Function and The
  Ornstein-Zernike Theory}},\ in\ \href@noop {} {\emph {\bibinfo {booktitle}
  {{Introduction to Phase Transitions and Critical Phenomena}}}},\
  Vol.~\bibinfo {volume} {7}\ (\bibinfo  {publisher} {Clarendon Press,
  Oxford},\ \bibinfo {address} {London},\ \bibinfo {year} {1971})\ pp.\
  \bibinfo {pages} {94--108}\BibitemShut {NoStop}%
\bibitem [{\citenamefont {Safran}\ \emph {et~al.}(1976)\citenamefont {Safran},
  \citenamefont {Lax},\ and\ \citenamefont {Dresselhaus}}]{Safran1976}%
  \BibitemOpen
  \bibfield  {author} {\bibinfo {author} {\bibfnamefont {S.}~\bibnamefont
  {Safran}}, \bibinfo {author} {\bibfnamefont {B.}~\bibnamefont {Lax}},\ and\
  \bibinfo {author} {\bibfnamefont {G.}~\bibnamefont {Dresselhaus}},\
  }\bibfield  {title} {\bibinfo {title} {{Phenomenological theory of Raman
  scattering in europium chalcogenides}},\ }\href
  {https://doi.org/10.1016/0038-1098(76)90823-1} {\bibfield  {journal}
  {\bibinfo  {journal} {Solid State Commun.}\ }\textbf {\bibinfo {volume}
  {19}},\ \bibinfo {pages} {1217} (\bibinfo {year} {1976})}\BibitemShut
  {NoStop}%
\bibitem [{\citenamefont {Schmutz}\ \emph {et~al.}(1979)\citenamefont
  {Schmutz}, \citenamefont {Dresselhaus},\ and\ \citenamefont
  {Dresselhaus}}]{Schmutz1979}%
  \BibitemOpen
  \bibfield  {author} {\bibinfo {author} {\bibfnamefont {L.~E.}\ \bibnamefont
  {Schmutz}}, \bibinfo {author} {\bibfnamefont {G.}~\bibnamefont
  {Dresselhaus}},\ and\ \bibinfo {author} {\bibfnamefont {M.~S.}\ \bibnamefont
  {Dresselhaus}},\ }\bibfield  {title} {\bibinfo {title} {{Raman scattering in
  magnetic europium telluride}},\ }\href
  {https://doi.org/10.1016/0304-8853(79)90301-9} {\bibfield  {journal}
  {\bibinfo  {journal} {Journal of Magnetism and Magnetic Materials}\ }\textbf
  {\bibinfo {volume} {11}},\ \bibinfo {pages} {412} (\bibinfo {year}
  {1979})}\BibitemShut {NoStop}%
\bibitem [{\citenamefont {Fleury}\ \emph {et~al.}(1968)\citenamefont {Fleury},
  \citenamefont {Scott},\ and\ \citenamefont {Worlock}}]{Fleury1968}%
  \BibitemOpen
  \bibfield  {author} {\bibinfo {author} {\bibfnamefont {P.~A.}\ \bibnamefont
  {Fleury}}, \bibinfo {author} {\bibfnamefont {J.~F.}\ \bibnamefont {Scott}},\
  and\ \bibinfo {author} {\bibfnamefont {J.~M.}\ \bibnamefont {Worlock}},\
  }\bibfield  {title} {\bibinfo {title} {{Soft Phonon Modes and the 110°K
  Phase Transition in SrTiO$_3$}},\ }\href
  {https://doi.org/10.1103/PhysRevLett.21.16} {\bibfield  {journal} {\bibinfo
  {journal} {Phys. Rev. Lett.}\ }\textbf {\bibinfo {volume} {21}},\ \bibinfo
  {pages} {16} (\bibinfo {year} {1968})}\BibitemShut {NoStop}%
\bibitem [{\citenamefont {Shirane}\ and\ \citenamefont
  {Yamada}(1969)}]{Shirane1969}%
  \BibitemOpen
  \bibfield  {author} {\bibinfo {author} {\bibfnamefont {G.}~\bibnamefont
  {Shirane}}\ and\ \bibinfo {author} {\bibfnamefont {Y.}~\bibnamefont
  {Yamada}},\ }\bibfield  {title} {\bibinfo {title} {{Lattice-Dynamical Study
  of the 110°K Phase Transition in SrTiO$_3$}},\ }\href
  {https://doi.org/10.1103/PhysRev.177.858} {\bibfield  {journal} {\bibinfo
  {journal} {Phys. Rev.}\ }\textbf {\bibinfo {volume} {177}},\ \bibinfo {pages}
  {858} (\bibinfo {year} {1969})}\BibitemShut {NoStop}%
\bibitem [{\citenamefont {{Y. Yu}}\ and\ \citenamefont
  {Cardona}(2010{\natexlab{a}})}]{Y.Yu2010}%
  \BibitemOpen
  \bibfield  {author} {\bibinfo {author} {\bibfnamefont {P.}~\bibnamefont {{Y.
  Yu}}}\ and\ \bibinfo {author} {\bibfnamefont {M.}~\bibnamefont {Cardona}},\
  }\href@noop {} {\emph {\bibinfo {title} {{Fundamentals of Semiconductors,
  Physics and Materials Properties}}}},\ \bibinfo {edition} {4th}\ ed.\
  (\bibinfo  {publisher} {Springer},\ \bibinfo {year} {2010})\ p.\ \bibinfo
  {pages} {775}\BibitemShut {NoStop}%
\bibitem [{\citenamefont {Choudhury}\ \emph {et~al.}(2010)\citenamefont
  {Choudhury}, \citenamefont {Kolesnikov}, \citenamefont {Schober},
  \citenamefont {Walter}, \citenamefont {Johnson}, \citenamefont {Abernathy},\
  and\ \citenamefont {Lucas}}]{Choudhury2010}%
  \BibitemOpen
  \bibfield  {author} {\bibinfo {author} {\bibfnamefont {N.}~\bibnamefont
  {Choudhury}}, \bibinfo {author} {\bibfnamefont {A.~I.}\ \bibnamefont
  {Kolesnikov}}, \bibinfo {author} {\bibfnamefont {H.}~\bibnamefont {Schober}},
  \bibinfo {author} {\bibfnamefont {E.~J.}\ \bibnamefont {Walter}}, \bibinfo
  {author} {\bibfnamefont {M.}~\bibnamefont {Johnson}}, \bibinfo {author}
  {\bibfnamefont {D.~L.}\ \bibnamefont {Abernathy}},\ and\ \bibinfo {author}
  {\bibfnamefont {M.~S.}\ \bibnamefont {Lucas}},\ }\bibfield  {title} {\bibinfo
  {title} {{Phonon density of states of model ferroelectrics}},\ }\href
  {https://doi.org/10.1557/PROC-1262-W01-02} {\bibfield  {journal} {\bibinfo
  {journal} {MRS Online Proceedings Library 2010 1262:1}\ }\textbf {\bibinfo
  {volume} {1262}},\ \bibinfo {pages} {1} (\bibinfo {year} {2010})}\BibitemShut
  {NoStop}%
\bibitem [{\citenamefont {Scalabrin}\ \emph {et~al.}(1977)\citenamefont
  {Scalabrin}, \citenamefont {Chaves}, \citenamefont {Shim},\ and\
  \citenamefont {Porto}}]{Scalabrin1977}%
  \BibitemOpen
  \bibfield  {author} {\bibinfo {author} {\bibfnamefont {A.}~\bibnamefont
  {Scalabrin}}, \bibinfo {author} {\bibfnamefont {A.~S.}\ \bibnamefont
  {Chaves}}, \bibinfo {author} {\bibfnamefont {D.~S.}\ \bibnamefont {Shim}},\
  and\ \bibinfo {author} {\bibfnamefont {S.~P.}\ \bibnamefont {Porto}},\
  }\bibfield  {title} {\bibinfo {title} {{Temperature dependence of the A1 and
  E optical phonons in BaTiO$_3$}},\ }\href
  {https://doi.org/10.1002/pssb.2220790240} {\bibfield  {journal} {\bibinfo
  {journal} {Physica Status Solidi (B)}\ }\textbf {\bibinfo {volume} {79}},\
  \bibinfo {pages} {731} (\bibinfo {year} {1977})}\BibitemShut {NoStop}%
\bibitem [{Sha(1972)}]{Shapiro1972}%
  \BibitemOpen
  \bibfield  {title} {\bibinfo {title} {{Critical neutron scattering in
  SrTiO$_3$ and KMnF$_3$}},\ }\href {https://doi.org/10.1103/PhysRevB.6.4332}
  {\bibfield  {journal} {\bibinfo  {journal} {Physical Review B}\ }\textbf
  {\bibinfo {volume} {6}},\ \bibinfo {pages} {4332} (\bibinfo {year}
  {1972})}\BibitemShut {NoStop}%
\bibitem [{\citenamefont {Petzelt}\ \emph {et~al.}(1987)\citenamefont
  {Petzelt}, \citenamefont {Kozlov},\ and\ \citenamefont
  {Volkov}}]{Petzelt1987}%
  \BibitemOpen
  \bibfield  {author} {\bibinfo {author} {\bibfnamefont {J.}~\bibnamefont
  {Petzelt}}, \bibinfo {author} {\bibfnamefont {G.~V.}\ \bibnamefont
  {Kozlov}},\ and\ \bibinfo {author} {\bibfnamefont {A.~A.}\ \bibnamefont
  {Volkov}},\ }\bibfield  {title} {\bibinfo {title} {{Dielectric spectroscopy
  of paraelectric soft modes}},\ }\href
  {https://doi.org/10.1080/00150198708227912} {\bibfield  {journal} {\bibinfo
  {journal} {Ferroelectrics}\ }\textbf {\bibinfo {volume} {73}},\ \bibinfo
  {pages} {101} (\bibinfo {year} {1987})}\BibitemShut {NoStop}%
\bibitem [{\citenamefont {Weadock}\ \emph {et~al.}(2020)\citenamefont
  {Weadock}, \citenamefont {Gehring}, \citenamefont {Gold-Parker},
  \citenamefont {Smith}, \citenamefont {Karunadasa},\ and\ \citenamefont
  {Toney}}]{Weadock2020}%
  \BibitemOpen
  \bibfield  {author} {\bibinfo {author} {\bibfnamefont {N.~J.}\ \bibnamefont
  {Weadock}}, \bibinfo {author} {\bibfnamefont {P.~M.}\ \bibnamefont
  {Gehring}}, \bibinfo {author} {\bibfnamefont {A.}~\bibnamefont
  {Gold-Parker}}, \bibinfo {author} {\bibfnamefont {I.~C.}\ \bibnamefont
  {Smith}}, \bibinfo {author} {\bibfnamefont {H.~I.}\ \bibnamefont
  {Karunadasa}},\ and\ \bibinfo {author} {\bibfnamefont {M.~F.}\ \bibnamefont
  {Toney}},\ }\bibfield  {title} {\bibinfo {title} {{Test of the Dynamic-Domain
  and Critical Scattering Hypotheses in Cubic Methylammonium Lead Triiodide}},\
  }\bibfield  {journal} {\bibinfo  {journal} {Physical Review Letters}\
  }\textbf {\bibinfo {volume} {125}},\ \href
  {https://doi.org/10.1103/PhysRevLett.125.075701}
  {10.1103/PhysRevLett.125.075701} (\bibinfo {year} {2020})\BibitemShut
  {NoStop}%
\bibitem [{\citenamefont {Bruce}\ and\ \citenamefont
  {Cowley}(1980)}]{Bruce1980}%
  \BibitemOpen
  \bibfield  {author} {\bibinfo {author} {\bibfnamefont {A.~D.}\ \bibnamefont
  {Bruce}}\ and\ \bibinfo {author} {\bibfnamefont {R.~A.}\ \bibnamefont
  {Cowley}},\ }\bibfield  {title} {\bibinfo {title} {{Structural phase
  transitions III. Critical dynamics and quasi-elastic scattering}},\ }\href
  {https://doi.org/10.1080/00018738000101366} {\bibfield  {journal} {\bibinfo
  {journal} {Advances in Physics}\ }\textbf {\bibinfo {volume} {29}},\ \bibinfo
  {pages} {219} (\bibinfo {year} {1980})}\BibitemShut {NoStop}%
\bibitem [{\citenamefont {Leguy}\ \emph {et~al.}(2015)\citenamefont {Leguy},
  \citenamefont {Frost}, \citenamefont {McMahon}, \citenamefont {Sakai},
  \citenamefont {Kockelmann}, \citenamefont {Law}, \citenamefont {Li},
  \citenamefont {Foglia}, \citenamefont {Walsh}, \citenamefont {O'Regan},
  \citenamefont {Nelson}, \citenamefont {Cabral},\ and\ \citenamefont
  {Barnes}}]{LeguyNatComm2015}%
  \BibitemOpen
  \bibfield  {author} {\bibinfo {author} {\bibfnamefont {A.~M.~A.}\
  \bibnamefont {Leguy}}, \bibinfo {author} {\bibfnamefont {J.~M.}\ \bibnamefont
  {Frost}}, \bibinfo {author} {\bibfnamefont {A.~P.}\ \bibnamefont {McMahon}},
  \bibinfo {author} {\bibfnamefont {V.~G.}\ \bibnamefont {Sakai}}, \bibinfo
  {author} {\bibfnamefont {W.}~\bibnamefont {Kockelmann}}, \bibinfo {author}
  {\bibfnamefont {C.}~\bibnamefont {Law}}, \bibinfo {author} {\bibfnamefont
  {X.}~\bibnamefont {Li}}, \bibinfo {author} {\bibfnamefont {F.}~\bibnamefont
  {Foglia}}, \bibinfo {author} {\bibfnamefont {A.}~\bibnamefont {Walsh}},
  \bibinfo {author} {\bibfnamefont {B.~C.}\ \bibnamefont {O'Regan}}, \bibinfo
  {author} {\bibfnamefont {J.}~\bibnamefont {Nelson}}, \bibinfo {author}
  {\bibfnamefont {J.~T.}\ \bibnamefont {Cabral}},\ and\ \bibinfo {author}
  {\bibfnamefont {P.~R.~F.}\ \bibnamefont {Barnes}},\ }\bibfield  {title}
  {\bibinfo {title} {{The dynamics of methylammonium ions in hybrid
  organic–inorganic perovskite solar cells}},\ }\href
  {https://doi.org/10.1038/ncomms8780} {\bibfield  {journal} {\bibinfo
  {journal} {Nature Communications}\ }\textbf {\bibinfo {volume} {6}},\
  \bibinfo {pages} {7780} (\bibinfo {year} {2015})},\ \Eprint
  {https://arxiv.org/abs/1402.4980} {arXiv:1402.4980} \BibitemShut {NoStop}%
\bibitem [{\citenamefont {Scott}(1974)}]{Scott1974}%
  \BibitemOpen
  \bibfield  {author} {\bibinfo {author} {\bibfnamefont {J.~F.}\ \bibnamefont
  {Scott}},\ }\bibfield  {title} {\bibinfo {title} {{Soft-Mode Spectroscopy:
  Experimental Studies of Structural Phase Transitions}},\ }\href
  {https://doi.org/10.1103/RevModPhys.46.83} {\bibfield  {journal} {\bibinfo
  {journal} {Reviews of Modern Physics}\ }\textbf {\bibinfo {volume} {46}},\
  \bibinfo {pages} {83} (\bibinfo {year} {1974})}\BibitemShut {NoStop}%
\bibitem [{\citenamefont {Guo}\ \emph {et~al.}(2017{\natexlab{b}})\citenamefont
  {Guo}, \citenamefont {Xia}, \citenamefont {Gong}, \citenamefont {Stoumpos},
  \citenamefont {McCall}, \citenamefont {Alexander}, \citenamefont {Ma},
  \citenamefont {Zhou}, \citenamefont {Gosztola}, \citenamefont {Ketterson},
  \citenamefont {Kanatzidis}, \citenamefont {Xu}, \citenamefont {Chan},\ and\
  \citenamefont {Schaller}}]{Guo2017c}%
  \BibitemOpen
  \bibfield  {author} {\bibinfo {author} {\bibfnamefont {P.}~\bibnamefont
  {Guo}}, \bibinfo {author} {\bibfnamefont {Y.}~\bibnamefont {Xia}}, \bibinfo
  {author} {\bibfnamefont {J.}~\bibnamefont {Gong}}, \bibinfo {author}
  {\bibfnamefont {C.~C.}\ \bibnamefont {Stoumpos}}, \bibinfo {author}
  {\bibfnamefont {K.~M.}\ \bibnamefont {McCall}}, \bibinfo {author}
  {\bibfnamefont {G.~C.}\ \bibnamefont {Alexander}}, \bibinfo {author}
  {\bibfnamefont {Z.}~\bibnamefont {Ma}}, \bibinfo {author} {\bibfnamefont
  {H.}~\bibnamefont {Zhou}}, \bibinfo {author} {\bibfnamefont {D.~J.}\
  \bibnamefont {Gosztola}}, \bibinfo {author} {\bibfnamefont {J.~B.}\
  \bibnamefont {Ketterson}}, \bibinfo {author} {\bibfnamefont {M.~G.}\
  \bibnamefont {Kanatzidis}}, \bibinfo {author} {\bibfnamefont
  {T.}~\bibnamefont {Xu}}, \bibinfo {author} {\bibfnamefont {M.~K.}\
  \bibnamefont {Chan}},\ and\ \bibinfo {author} {\bibfnamefont {R.~D.}\
  \bibnamefont {Schaller}},\ }\bibfield  {title} {\bibinfo {title} {{Polar
  Fluctuations in Metal Halide Perovskites Uncovered by Acoustic Phonon
  Anomalies}},\ }\href {https://doi.org/10.1021/acsenergylett.7b00790}
  {\bibfield  {journal} {\bibinfo  {journal} {ACS Energy Letters}\ }\textbf
  {\bibinfo {volume} {2}},\ \bibinfo {pages} {2463} (\bibinfo {year}
  {2017}{\natexlab{b}})}\BibitemShut {NoStop}%
\bibitem [{\citenamefont {Songvilay}\ \emph {et~al.}(2019)\citenamefont
  {Songvilay}, \citenamefont {Giles-Donovan}, \citenamefont {Bari},
  \citenamefont {Ye}, \citenamefont {Minns}, \citenamefont {Green},
  \citenamefont {Xu}, \citenamefont {Gehring}, \citenamefont {Schmalzl},
  \citenamefont {Ratcliff}, \citenamefont {Brown}, \citenamefont {Chernyshov},
  \citenamefont {{Van Beek}}, \citenamefont {Cochran},\ and\ \citenamefont
  {Stock}}]{Songvilay2019}%
  \BibitemOpen
  \bibfield  {author} {\bibinfo {author} {\bibfnamefont {M.}~\bibnamefont
  {Songvilay}}, \bibinfo {author} {\bibfnamefont {N.}~\bibnamefont
  {Giles-Donovan}}, \bibinfo {author} {\bibfnamefont {M.}~\bibnamefont {Bari}},
  \bibinfo {author} {\bibfnamefont {Z.~G.}\ \bibnamefont {Ye}}, \bibinfo
  {author} {\bibfnamefont {J.~L.}\ \bibnamefont {Minns}}, \bibinfo {author}
  {\bibfnamefont {M.~A.}\ \bibnamefont {Green}}, \bibinfo {author}
  {\bibfnamefont {G.}~\bibnamefont {Xu}}, \bibinfo {author} {\bibfnamefont
  {P.~M.}\ \bibnamefont {Gehring}}, \bibinfo {author} {\bibfnamefont
  {K.}~\bibnamefont {Schmalzl}}, \bibinfo {author} {\bibfnamefont {W.~D.}\
  \bibnamefont {Ratcliff}}, \bibinfo {author} {\bibfnamefont {C.~M.}\
  \bibnamefont {Brown}}, \bibinfo {author} {\bibfnamefont {D.}~\bibnamefont
  {Chernyshov}}, \bibinfo {author} {\bibfnamefont {W.}~\bibnamefont {{Van
  Beek}}}, \bibinfo {author} {\bibfnamefont {S.}~\bibnamefont {Cochran}},\ and\
  \bibinfo {author} {\bibfnamefont {C.}~\bibnamefont {Stock}},\ }\bibfield
  {title} {\bibinfo {title} {{Common acoustic phonon lifetimes in inorganic and
  hybrid lead halide perovskites}},\ }\href
  {https://doi.org/10.1103/PhysRevMaterials.3.093602} {\bibfield  {journal}
  {\bibinfo  {journal} {Physical Review Materials}\ }\textbf {\bibinfo {volume}
  {3}},\ \bibinfo {pages} {93602} (\bibinfo {year} {2019})}\BibitemShut
  {NoStop}%
\bibitem [{\citenamefont {Mayers}\ \emph {et~al.}(2018)\citenamefont {Mayers},
  \citenamefont {Tan}, \citenamefont {Egger}, \citenamefont {Rappe},\ and\
  \citenamefont {Reichman}}]{Mayers2018}%
  \BibitemOpen
  \bibfield  {author} {\bibinfo {author} {\bibfnamefont {M.}~\bibnamefont
  {Mayers}}, \bibinfo {author} {\bibfnamefont {L.~Z.}\ \bibnamefont {Tan}},
  \bibinfo {author} {\bibfnamefont {D.~A.}\ \bibnamefont {Egger}}, \bibinfo
  {author} {\bibfnamefont {A.~M.}\ \bibnamefont {Rappe}},\ and\ \bibinfo
  {author} {\bibfnamefont {D.~R.}\ \bibnamefont {Reichman}},\ }\bibfield
  {title} {\bibinfo {title} {{How Lattice and Charge Fluctuations Control
  Carrier Dynamics in Halide Perovskites}},\ }\href
  {https://doi.org/10.1021/acs.nanolett.8b04276} {\bibfield  {journal}
  {\bibinfo  {journal} {Nano Letters}\ }\textbf {\bibinfo {volume} {18}},\
  \bibinfo {pages} {acs.nanolett.8b04276} (\bibinfo {year} {2018})}\BibitemShut
  {NoStop}%
\bibitem [{\citenamefont {Gold-Parker}\ \emph {et~al.}(2018)\citenamefont
  {Gold-Parker}, \citenamefont {Gehring}, \citenamefont {Skelton},
  \citenamefont {Smith}, \citenamefont {Parshall}, \citenamefont {Frost},
  \citenamefont {Karunadasa}, \citenamefont {Walsh},\ and\ \citenamefont
  {Toney}}]{Gold-Parker2018}%
  \BibitemOpen
  \bibfield  {author} {\bibinfo {author} {\bibfnamefont {A.}~\bibnamefont
  {Gold-Parker}}, \bibinfo {author} {\bibfnamefont {P.~M.}\ \bibnamefont
  {Gehring}}, \bibinfo {author} {\bibfnamefont {J.~M.}\ \bibnamefont
  {Skelton}}, \bibinfo {author} {\bibfnamefont {I.~C.}\ \bibnamefont {Smith}},
  \bibinfo {author} {\bibfnamefont {D.}~\bibnamefont {Parshall}}, \bibinfo
  {author} {\bibfnamefont {J.~M.}\ \bibnamefont {Frost}}, \bibinfo {author}
  {\bibfnamefont {H.~I.}\ \bibnamefont {Karunadasa}}, \bibinfo {author}
  {\bibfnamefont {A.}~\bibnamefont {Walsh}},\ and\ \bibinfo {author}
  {\bibfnamefont {M.~F.}\ \bibnamefont {Toney}},\ }\bibfield  {title} {\bibinfo
  {title} {{Acoustic phonon lifetimes limit thermal transport in methylammonium
  lead iodide}},\ }\href {https://doi.org/10.1073/pnas.1812227115} {\bibfield
  {journal} {\bibinfo  {journal} {Proceedings of the National Academy of
  Sciences of the United States of America}\ }\textbf {\bibinfo {volume}
  {115}},\ \bibinfo {pages} {11905} (\bibinfo {year} {2018})}\BibitemShut
  {NoStop}%
\bibitem [{\citenamefont {Filippone}\ \emph {et~al.}(2020)\citenamefont
  {Filippone}, \citenamefont {Zhao}, \citenamefont {Niu}, \citenamefont
  {Koocher}, \citenamefont {Silevitch}, \citenamefont {Fina}, \citenamefont
  {Rondinelli}, \citenamefont {Ravichandran},\ and\ \citenamefont
  {Jaramillo}}]{Filippone2020}%
  \BibitemOpen
  \bibfield  {author} {\bibinfo {author} {\bibfnamefont {S.}~\bibnamefont
  {Filippone}}, \bibinfo {author} {\bibfnamefont {B.}~\bibnamefont {Zhao}},
  \bibinfo {author} {\bibfnamefont {S.}~\bibnamefont {Niu}}, \bibinfo {author}
  {\bibfnamefont {N.~Z.}\ \bibnamefont {Koocher}}, \bibinfo {author}
  {\bibfnamefont {D.}~\bibnamefont {Silevitch}}, \bibinfo {author}
  {\bibfnamefont {I.}~\bibnamefont {Fina}}, \bibinfo {author} {\bibfnamefont
  {J.~M.}\ \bibnamefont {Rondinelli}}, \bibinfo {author} {\bibfnamefont
  {J.}~\bibnamefont {Ravichandran}},\ and\ \bibinfo {author} {\bibfnamefont
  {R.}~\bibnamefont {Jaramillo}},\ }\bibfield  {title} {\bibinfo {title}
  {{Discovery of highly polarizable semiconductors BaZrS3 and Ba3Zr2S7}},\
  }\bibfield  {journal} {\bibinfo  {journal} {Physical Review Materials}\
  }\textbf {\bibinfo {volume} {4}},\ \href
  {https://doi.org/10.1103/PhysRevMaterials.4.091601}
  {10.1103/PhysRevMaterials.4.091601} (\bibinfo {year} {2020})\BibitemShut
  {NoStop}%
\bibitem [{\citenamefont {Svirskas}\ \emph {et~al.}(2020)\citenamefont
  {Svirskas}, \citenamefont {Bal{\v{c}}iūnas}, \citenamefont {{\v{S}}imėnas},
  \citenamefont {Usevi{\v{c}}ius}, \citenamefont {Kinka}, \citenamefont
  {Veli{\v{c}}ka}, \citenamefont {Kubicki}, \citenamefont {Castillo},
  \citenamefont {Karabanov}, \citenamefont {Shvartsman}, \citenamefont {{De
  Ros{\'{a}}rio Soares}}, \citenamefont {{\v{S}}ablinskas}, \citenamefont
  {Salak}, \citenamefont {Lupascu},\ and\ \citenamefont
  {Banys}}]{Svirskas2020}%
  \BibitemOpen
  \bibfield  {author} {\bibinfo {author} {\bibfnamefont {{\v{S}}.}~\bibnamefont
  {Svirskas}}, \bibinfo {author} {\bibfnamefont {S.}~\bibnamefont
  {Bal{\v{c}}iūnas}}, \bibinfo {author} {\bibfnamefont {M.}~\bibnamefont
  {{\v{S}}imėnas}}, \bibinfo {author} {\bibfnamefont {G.}~\bibnamefont
  {Usevi{\v{c}}ius}}, \bibinfo {author} {\bibfnamefont {M.}~\bibnamefont
  {Kinka}}, \bibinfo {author} {\bibfnamefont {M.}~\bibnamefont
  {Veli{\v{c}}ka}}, \bibinfo {author} {\bibfnamefont {D.}~\bibnamefont
  {Kubicki}}, \bibinfo {author} {\bibfnamefont {M.~E.}\ \bibnamefont
  {Castillo}}, \bibinfo {author} {\bibfnamefont {A.}~\bibnamefont {Karabanov}},
  \bibinfo {author} {\bibfnamefont {V.~V.}\ \bibnamefont {Shvartsman}},
  \bibinfo {author} {\bibfnamefont {M.}~\bibnamefont {{De Ros{\'{a}}rio
  Soares}}}, \bibinfo {author} {\bibfnamefont {V.}~\bibnamefont
  {{\v{S}}ablinskas}}, \bibinfo {author} {\bibfnamefont {A.~N.}\ \bibnamefont
  {Salak}}, \bibinfo {author} {\bibfnamefont {D.~C.}\ \bibnamefont {Lupascu}},\
  and\ \bibinfo {author} {\bibfnamefont {J.}~\bibnamefont {Banys}},\ }\bibfield
   {title} {\bibinfo {title} {{Phase transitions, screening and dielectric
  response of CsPbBr$_3$}},\ }\href {https://doi.org/10.1039/d0ta04155f}
  {\bibfield  {journal} {\bibinfo  {journal} {Journal of Materials Chemistry
  A}\ }\textbf {\bibinfo {volume} {8}},\ \bibinfo {pages} {14015} (\bibinfo
  {year} {2020})}\BibitemShut {NoStop}%
\bibitem [{\citenamefont {Rakita}\ \emph {et~al.}(2016)\citenamefont {Rakita},
  \citenamefont {Kedem}, \citenamefont {Gupta}, \citenamefont {Sadhanala},
  \citenamefont {Kalchenko}, \citenamefont {B{\"{o}}hm}, \citenamefont
  {Kulbak}, \citenamefont {Friend}, \citenamefont {Cahen},\ and\ \citenamefont
  {Hodes}}]{Rakita2016}%
  \BibitemOpen
  \bibfield  {author} {\bibinfo {author} {\bibfnamefont {Y.}~\bibnamefont
  {Rakita}}, \bibinfo {author} {\bibfnamefont {N.}~\bibnamefont {Kedem}},
  \bibinfo {author} {\bibfnamefont {S.}~\bibnamefont {Gupta}}, \bibinfo
  {author} {\bibfnamefont {A.}~\bibnamefont {Sadhanala}}, \bibinfo {author}
  {\bibfnamefont {V.}~\bibnamefont {Kalchenko}}, \bibinfo {author}
  {\bibfnamefont {M.~L.}\ \bibnamefont {B{\"{o}}hm}}, \bibinfo {author}
  {\bibfnamefont {M.}~\bibnamefont {Kulbak}}, \bibinfo {author} {\bibfnamefont
  {R.~H.}\ \bibnamefont {Friend}}, \bibinfo {author} {\bibfnamefont
  {D.}~\bibnamefont {Cahen}},\ and\ \bibinfo {author} {\bibfnamefont
  {G.}~\bibnamefont {Hodes}},\ }\bibfield  {title} {\bibinfo {title}
  {{Low-Temperature Solution-Grown CsPbBr$_3$ Single Crystals and Their
  Characterization}},\ }\href {https://doi.org/10.1021/acs.cgd.6b00764}
  {\bibfield  {journal} {\bibinfo  {journal} {Crystal Growth and Design}\
  }\textbf {\bibinfo {volume} {16}},\ \bibinfo {pages} {5717} (\bibinfo {year}
  {2016})}\BibitemShut {NoStop}%
\bibitem [{\citenamefont {Asher}\ \emph {et~al.}(2020)\citenamefont {Asher},
  \citenamefont {Angerer}, \citenamefont {Korobko}, \citenamefont
  {Diskin-Posner}, \citenamefont {Egger},\ and\ \citenamefont
  {Yaffe}}]{Asher2020}%
  \BibitemOpen
  \bibfield  {author} {\bibinfo {author} {\bibfnamefont {M.}~\bibnamefont
  {Asher}}, \bibinfo {author} {\bibfnamefont {D.}~\bibnamefont {Angerer}},
  \bibinfo {author} {\bibfnamefont {R.}~\bibnamefont {Korobko}}, \bibinfo
  {author} {\bibfnamefont {Y.}~\bibnamefont {Diskin-Posner}}, \bibinfo {author}
  {\bibfnamefont {D.~A.}\ \bibnamefont {Egger}},\ and\ \bibinfo {author}
  {\bibfnamefont {O.}~\bibnamefont {Yaffe}},\ }\bibfield  {title} {\bibinfo
  {title} {{Anharmonic Lattice Vibrations in Small-Molecule Organic
  Semiconductors}},\ }\bibfield  {journal} {\bibinfo  {journal} {Advanced
  Materials}\ }\textbf {\bibinfo {volume} {32}},\ \href
  {https://doi.org/10.1002/adma.201908028} {10.1002/adma.201908028} (\bibinfo
  {year} {2020}),\ \Eprint {https://arxiv.org/abs/1912.03374}
  {arXiv:1912.03374} \BibitemShut {NoStop}%
\bibitem [{\citenamefont {Kranert}\ \emph
  {et~al.}(2016{\natexlab{a}})\citenamefont {Kranert}, \citenamefont {Sturm},
  \citenamefont {Schmidt-Grund},\ and\ \citenamefont
  {Grundmann}}]{KarnertSchiRep2016}%
  \BibitemOpen
  \bibfield  {author} {\bibinfo {author} {\bibfnamefont {C.}~\bibnamefont
  {Kranert}}, \bibinfo {author} {\bibfnamefont {C.}~\bibnamefont {Sturm}},
  \bibinfo {author} {\bibfnamefont {R.}~\bibnamefont {Schmidt-Grund}},\ and\
  \bibinfo {author} {\bibfnamefont {M.}~\bibnamefont {Grundmann}},\ }\bibfield
  {title} {\bibinfo {title} {{Raman tensor elements of $\beta$-Ga$_2$O$_3$}},\
  }\bibfield  {journal} {\bibinfo  {journal} {Scientific Reports}\ }\textbf
  {\bibinfo {volume} {6}},\ \href {https://doi.org/10.1038/srep35964}
  {10.1038/srep35964} (\bibinfo {year} {2016}{\natexlab{a}})\BibitemShut
  {NoStop}%
\bibitem [{\citenamefont {Kranert}\ \emph
  {et~al.}(2016{\natexlab{b}})\citenamefont {Kranert}, \citenamefont {Sturm},
  \citenamefont {Schmidt-Grund},\ and\ \citenamefont
  {Grundmann}}]{KranertPRL2016}%
  \BibitemOpen
  \bibfield  {author} {\bibinfo {author} {\bibfnamefont {C.}~\bibnamefont
  {Kranert}}, \bibinfo {author} {\bibfnamefont {C.}~\bibnamefont {Sturm}},
  \bibinfo {author} {\bibfnamefont {R.}~\bibnamefont {Schmidt-Grund}},\ and\
  \bibinfo {author} {\bibfnamefont {M.}~\bibnamefont {Grundmann}},\ }\bibfield
  {title} {\bibinfo {title} {{Raman Tensor Formalism for Optically Anisotropic
  Crystals}},\ }\href {https://doi.org/10.1103/PhysRevLett.116.127401}
  {\bibfield  {journal} {\bibinfo  {journal} {Physical Review Letters}\
  }\textbf {\bibinfo {volume} {116}},\ \bibinfo {pages} {127401} (\bibinfo
  {year} {2016}{\natexlab{b}})}\BibitemShut {NoStop}%
\bibitem [{\citenamefont {Fox}(2010)}]{FoxOPOS}%
  \BibitemOpen
  \bibfield  {author} {\bibinfo {author} {\bibfnamefont {M.}~\bibnamefont
  {Fox}},\ }\href {https://doi.org/10.1007/BF02751482} {\emph {\bibinfo {title}
  {{Optical properties of solids}}}},\ \bibinfo {edition} {2nd}\ ed.\ (\bibinfo
   {publisher} {Oxford University Press},\ \bibinfo {address} {New York, USA},\
  \bibinfo {year} {2010})\ p.~\bibinfo {pages} {49},\ \Eprint
  {https://arxiv.org/abs/arXiv:1011.1669v3} {arXiv:arXiv:1011.1669v3}
  \BibitemShut {NoStop}%
\bibitem [{\citenamefont {Guo}\ \emph {et~al.}(2018)\citenamefont {Guo},
  \citenamefont {Huang}, \citenamefont {Stoumpos}, \citenamefont {Mao},
  \citenamefont {Gong}, \citenamefont {Zeng}, \citenamefont {Diroll},
  \citenamefont {Xia}, \citenamefont {Ma}, \citenamefont {Gosztola},
  \citenamefont {Xu}, \citenamefont {Ketterson}, \citenamefont {Bedzyk},
  \citenamefont {Facchetti}, \citenamefont {Marks}, \citenamefont
  {Kanatzidis},\ and\ \citenamefont {Schaller}}]{GuoPRL2018}%
  \BibitemOpen
  \bibfield  {author} {\bibinfo {author} {\bibfnamefont {P.}~\bibnamefont
  {Guo}}, \bibinfo {author} {\bibfnamefont {W.}~\bibnamefont {Huang}}, \bibinfo
  {author} {\bibfnamefont {C.~C.}\ \bibnamefont {Stoumpos}}, \bibinfo {author}
  {\bibfnamefont {L.}~\bibnamefont {Mao}}, \bibinfo {author} {\bibfnamefont
  {J.}~\bibnamefont {Gong}}, \bibinfo {author} {\bibfnamefont {L.}~\bibnamefont
  {Zeng}}, \bibinfo {author} {\bibfnamefont {B.~T.}\ \bibnamefont {Diroll}},
  \bibinfo {author} {\bibfnamefont {Y.}~\bibnamefont {Xia}}, \bibinfo {author}
  {\bibfnamefont {X.}~\bibnamefont {Ma}}, \bibinfo {author} {\bibfnamefont
  {D.~J.}\ \bibnamefont {Gosztola}}, \bibinfo {author} {\bibfnamefont
  {T.}~\bibnamefont {Xu}}, \bibinfo {author} {\bibfnamefont {J.~B.}\
  \bibnamefont {Ketterson}}, \bibinfo {author} {\bibfnamefont {M.~J.}\
  \bibnamefont {Bedzyk}}, \bibinfo {author} {\bibfnamefont {A.}~\bibnamefont
  {Facchetti}}, \bibinfo {author} {\bibfnamefont {T.~J.}\ \bibnamefont
  {Marks}}, \bibinfo {author} {\bibfnamefont {M.~G.}\ \bibnamefont
  {Kanatzidis}},\ and\ \bibinfo {author} {\bibfnamefont {R.~D.}\ \bibnamefont
  {Schaller}},\ }\bibfield  {title} {\bibinfo {title} {{Hyperbolic Dispersion
  Arising from Anisotropic Excitons in Two-Dimensional Perovskites}},\
  }\bibfield  {journal} {\bibinfo  {journal} {Physical Review Letters}\
  }\textbf {\bibinfo {volume} {121}},\ \href
  {https://doi.org/10.1103/PhysRevLett.121.127401}
  {10.1103/PhysRevLett.121.127401} (\bibinfo {year} {2018})\BibitemShut
  {NoStop}%
\bibitem [{\citenamefont {Li}\ \emph {et~al.}(2020)\citenamefont {Li},
  \citenamefont {Ma}, \citenamefont {Cheng}, \citenamefont {Liu}, \citenamefont
  {Chen},\ and\ \citenamefont {Li}}]{Li2020}%
  \BibitemOpen
  \bibfield  {author} {\bibinfo {author} {\bibfnamefont {J.}~\bibnamefont
  {Li}}, \bibinfo {author} {\bibfnamefont {J.}~\bibnamefont {Ma}}, \bibinfo
  {author} {\bibfnamefont {X.}~\bibnamefont {Cheng}}, \bibinfo {author}
  {\bibfnamefont {Z.}~\bibnamefont {Liu}}, \bibinfo {author} {\bibfnamefont
  {Y.}~\bibnamefont {Chen}},\ and\ \bibinfo {author} {\bibfnamefont
  {D.}~\bibnamefont {Li}},\ }\bibfield  {title} {\bibinfo {title} {{Anisotropy
  of Excitons in Two-Dimensional Perovskite Crystals}},\ }\href
  {https://doi.org/10.1021/acsnano.9b08975} {\bibfield  {journal} {\bibinfo
  {journal} {ACS Nano}\ }\textbf {\bibinfo {volume} {14}},\ \bibinfo {pages}
  {2156} (\bibinfo {year} {2020})}\BibitemShut {NoStop}%
\bibitem [{\citenamefont {Schaufele}\ and\ \citenamefont
  {Weber}(1967)}]{Schaufele1967}%
  \BibitemOpen
  \bibfield  {author} {\bibinfo {author} {\bibfnamefont {R.~F.}\ \bibnamefont
  {Schaufele}}\ and\ \bibinfo {author} {\bibfnamefont {M.~J.}\ \bibnamefont
  {Weber}},\ }\bibfield  {title} {\bibinfo {title} {{First- and second-order
  raman scattering of SrTiO$_3$}},\ }\href {https://doi.org/10.1063/1.1841140}
  {\bibfield  {journal} {\bibinfo  {journal} {J. Chem. Phys.}\ }\textbf
  {\bibinfo {volume} {46}},\ \bibinfo {pages} {2859} (\bibinfo {year}
  {1967})}\BibitemShut {NoStop}%
\bibitem [{\citenamefont {Taylor}\ and\ \citenamefont
  {Murray}(1979)}]{Taylor1979}%
  \BibitemOpen
  \bibfield  {author} {\bibinfo {author} {\bibfnamefont {W.}~\bibnamefont
  {Taylor}}\ and\ \bibinfo {author} {\bibfnamefont {A.~F.}\ \bibnamefont
  {Murray}},\ }\bibfield  {title} {\bibinfo {title} {{Tetragonal SrTiO$_3$
  revisited: The effect of impurities on the Raman spectrum}},\ }\href
  {https://doi.org/10.1016/0038-1098(79)90005-X} {\bibfield  {journal}
  {\bibinfo  {journal} {Solid State Commun.}\ }\textbf {\bibinfo {volume}
  {31}},\ \bibinfo {pages} {937} (\bibinfo {year} {1979})}\BibitemShut
  {NoStop}%
\bibitem [{\citenamefont {Nilsen}\ and\ \citenamefont
  {Skinner}(1968)}]{Nilsen1968}%
  \BibitemOpen
  \bibfield  {author} {\bibinfo {author} {\bibfnamefont {W.~G.}\ \bibnamefont
  {Nilsen}}\ and\ \bibinfo {author} {\bibfnamefont {J.~G.}\ \bibnamefont
  {Skinner}},\ }\bibfield  {title} {\bibinfo {title} {{Raman spectrum of
  strontium titanate}},\ }\href {https://doi.org/10.1063/1.1669418} {\bibfield
  {journal} {\bibinfo  {journal} {The Journal of Chemical Physics}\ }\textbf
  {\bibinfo {volume} {48}},\ \bibinfo {pages} {2240} (\bibinfo {year}
  {1968})}\BibitemShut {NoStop}%
\bibitem [{\citenamefont {{Y. Yu}}\ and\ \citenamefont
  {Cardona}(2010{\natexlab{b}})}]{PortoNotationCardona}%
  \BibitemOpen
  \bibfield  {author} {\bibinfo {author} {\bibfnamefont {P.}~\bibnamefont {{Y.
  Yu}}}\ and\ \bibinfo {author} {\bibfnamefont {M.}~\bibnamefont {Cardona}},\
  }\bibinfo {title} {{Raman Rensors and Selection Rules}},\ in\ \href@noop {}
  {\emph {\bibinfo {booktitle} {{Fundamentals of Semiconductors, Physics and
  Materials Properties}}}}\ (\bibinfo  {publisher} {Springer},\ \bibinfo
  {address} {Berlin Heidelberg},\ \bibinfo {year} {2010})\ pp.\ \bibinfo
  {pages} {378--385},\ \bibinfo {edition} {4th}\ ed.\BibitemShut {Stop}%
\bibitem [{\citenamefont {Cowley}(1964{\natexlab{b}})}]{Cowley1964}%
  \BibitemOpen
  \bibfield  {author} {\bibinfo {author} {\bibfnamefont {R.~A.}\ \bibnamefont
  {Cowley}},\ }\bibfield  {title} {\bibinfo {title} {{The theory of Raman
  scattering from crystals}},\ }\href
  {https://doi.org/10.1088/0370-1328/84/2/311} {\bibfield  {journal} {\bibinfo
  {journal} {Proceedings of the Physical Society}\ }\textbf {\bibinfo {volume}
  {84}},\ \bibinfo {pages} {281} (\bibinfo {year}
  {1964}{\natexlab{b}})}\BibitemShut {NoStop}%
\end{thebibliography}%

\end{document}